\documentclass[article]{JHEP3} 
\pdfoutput=1

\JHEPspecialurl{http://jhep.sissa.it/JOURNAL/JHEP3.tar.gz}

\usepackage{epsfig,multicol,bbm}

\newcommand\fverb{\setbox\fverbbox=\hbox\bgroup\verb}
\newcommand\fverbdo{\egroup\medskip\noindent%
			\fbox{\unhbox\fverbbox}\ }
\newcommand\fverbit{\egroup\item[\fbox{\unhbox\fverbbox}]}
\newbox\fverbbox

\def\ap{\alpha'}
\def\fc{\frac}
\def\h{{1\over 2}}
\def\lsim{\mathrel{\rlap{\lower3pt\hbox{\hskip0pt$\sim$}}
    \raise1pt\hbox{$<$}}}         
\def\gsim{\mathrel{\rlap{\lower4pt\hbox{\hskip1pt$\sim$}}
    \raise1pt\hbox{$>$}}}         
\def\Re           {{\rm Re\hskip0.1em}}
\def\Im           {{\rm Im\hskip0.1em}}
\def\d{{\rm d}}
\def\text{\rm}
\def\beq{\begin{equation}}
\def\eeq{\end{equation}}

\title{Seeing through the String Landscape - a String Hunter's Companion in Particle Physics and
Cosmology\footnote{Review paper invited and accepted for publication by JHEP}}

\author{Dieter L\"ust\\
Arnold Sommerfeld Center for Theoretical Physics,
LMU-M\"unchen,
Theresienstr. 37,
80333 M\"unchen\\
and\\ 
Max-Planck-Institut f\"ur Physik, F\"ohringer Ring 6, 80805 M\"unchen, Germany\\
	E-mail: \email{dieter.luest@lmu.de, luest@mppmu.mpg.de}}

\received{\today} 		
\accepted{\today}		

\preprint{LMU-ASC 20/09\\MPP-2009-44}

\abstract{In this article we will overview 
several aspects of the string landscape, namely intersecting D-brane models and their 
statistics, possible model independent LHC signatures of intersecting brane models,
flux compactification, moduli stabilization in type II compactifications, domain wall solutions and brane inflation.
}


\begin{document} 


\section{Introduction}

Particle physics and cosmology are entering an unprecedentedly exciting epoch.
We are about to be confronted with new experimental data which will provide entirely new information
about the structure of matter, the fundamental interactions at short distances as well as about
the history of the early universe. Most notably the LHC experiment is expected to discover (or exclude) the Higgs
particle and hence will test our picture about the origin of mass in the Standard Model (SM).
In addition, the LHC might discover completely new particles in the TeV region as predicted e.g. by
supersymmetry. This would be a clear signal for physics beyond the SM. In astroparticle physics,
exciting new experiments such as the Planck satellite and others
will provide further data. This will allow to pin down the parameters of cosmic inflation, dark 
energy and dark matter with a much higher precision than before. Moreover laser interferometers
(LIGO and LISA) may for the first time discover  gravitational waves with possibly far reaching
consequences for our understanding of the early universe. 
As is clear by now, progress in particle
physics and cosmology go hand in hand. For example new particles discovered at LHC
may serve as dark matter candidates. On the other hand, to understand cosmic inflation and dark energy
at a microscopic level requires concrete underlying particle physics models for the physics of
the early universe.

In order to describe the physics at very high energies and during the very early universe, new theoretical
concepts are necessary which go beyond the SM of particle physics. Among various attempts in
this direction, string theory is perhaps the most successful and also the most ambitious approach since
besides the gauge interactions it includes also the gravitational force at the quantum level
(for some textbooks on string theory see
\cite{gsw87,Lust:1989tj,JP98, Kiritsis:2007zz,Becker:2007zj}).  Recently there has been achieved substantial progress in connecting
string theory with particle physics and cosmology. 
In addition, in spite of many open problems, it has become clear that string compactifications allow for a huge number
of possible ground states, nowadays referred  to as the landscape of string theory vacua.

The apparent existence of a string landscape is a double-edged sword.  Several
of these string vacua possess attractive phenomenological properties in that they come very close to
the SM of particle physics or to realistic models of cosmic inflation. Also the problem
of dark energy was recently addressed in string theory in an interesting way.
On the other hand, it was hoped  for a long time that 
the fundamental theory of nature would allow only for a single vacuum which would explain all physical phenomena and make unique predictions for future observations. As it stands at present, this hope may have been too naive. In fact, 
the existence of a large number of solutions of a physical theory is nothing specific to string theory. 
Every sufficiently complex gauge theory allows for a large number of at least meta-stable vacua whose phenomenological properties vary considerably.

Still, the emergence of the string landscape and the attempts for its
interpretation  may mark a shift in paradigm of how to treat the problem of unification and
also of uniqueness in a fundamental theory. As a totally new aspect that developed recently, the search for realistic string vacua
has lead to the application of 
statistical methods.  In a similar spirit it is argued (and heavily debated) that the problem of explaining the cosmological constant 
(and perhaps even other constants in nature) can be solved by some anthropic interpretation
of the string landscape.

The scope of the article is to explain recent developments in string theory with special emphasis on string compactifications from ten to four space-time dimensions and the associated landscape of string vacua. 
We are planning to discuss the following
two main aspects of the string landscape:
\begin{itemize}
\item
How string theory connects to the real world in particle physics and cosmology. Particular
   emphasis will be put on how to derive the SM from string compactifications and how
   to get viable models for cosmic inflation and for the description of dark energy. We will also discuss
   possible string signatures at the LHC collider experiment. 
\item How we describe and how we deal with the huge landscape in string theory.
   We will explain that statistical methods are useful for the search of the SM from string theory.
   In addition, we will discuss some general aspects of the landscape like transitions between
   different vacua and constraints from black hole decays.
\end{itemize}

One possible approach to string theory is the {\sl top-down} approach, which starts from the unification
of gravity and gauge interactions at very high energies, and then tries to deduce all
low energy observables from investigating the mathematical structures of the theory.
Although we do not yet know the typical string scale $M_{\rm string}$, where the
unification of gravity and gauge interaction takes places, one often assumes, at least
from a conservative point of view, that this happens at the Planck scale of about $M_{\rm Planck}
\simeq 10^{19}~{\rm GeV}$.
However no direct experiments
will guide us through the physics at such high energies, a fact which makes the top-down approach
very troublesome. But we like to emphasize already at this point that the string scale
$M_{\rm string}$ a free parameter, which a priori can take any value, especially in
type II orientifold models, where the SM lives on lower dimensional branes. In fact, as we will
discuss later, intersecting
brane models with all low string scale allow for model-independent predictions at collider
experiments, most notably at the LHC, which are generic for a large class of models inside the
string landscape. 

Let us be more precise what we actually mean by the string landscape. It is
 defined to be the space of all possible solutions
of the string equations of motion. In ten space-time dimensions, there exist just five different
formulations of string theory (two heterotic strings, type I type IIA and type IIB superstrings).
Exploring several kind of duality symmetries, it is conjectured that all these string theories
can be unified into M-theory, where also 11-dimensional supergravity is included.
However the number of lower-dimensional string solutions, i.e. lower dimensional string
ground states, which are obtained after compactification, is enormous.
This fact became clear
already in 1986 constructing heterotic
strings in four dimensions \cite{Kawai:1986va,Lerche:1986cx,Antoniadis:1986rn},
and within  the covariant lattice construction \cite{Lerche:1986cx}, the number of
possible four-dimensional string ground states was estimated to be of order $10^{1500}$.
More recently, the number  of discrete flux vacua of an effective supergravity potential
for type II compactifications on a generic Calabi-Yau manifold was shown
to be of order $10^{500}$ \cite{Bousso:2000xa,Douglas:2003um}.
Taken seriously, this vast landscape of distinct string vacua
really implies a big question
mark concerning the predictivity of string theory, since each point in the landscape essentially
corresponds to a different universe with different particle physics and cosmological properties.
To deal with such a huge number of possibilities, certain strategies are
required in order to proceed within the top-down approach.
One possible and legitimate approach is given by the investigation of the statistical properties
of the string landscape. I.e. one has to determine by statistical methods what is the fraction
of string vacua with good phenomenological properties. Possible statistical correlations
resp. anti-correlations would be especially worth to be discovered, like e.g. between the
number of families and the rank of the low-energy gauge groups, because they could
provide a step towards verifying or resp. falsifying string theory.
 Eventually, the statistical approach is likely to be merged with the anthropic principle
  \cite{Susskind:2003kw} (see also \cite{Schellekens:2008kg}).
Concerning the evolution of the universe (see e.g. \cite{Linde:2007fr}), the anthropic principle essentially requires
a multiverse with a huge number of bubbles, with each being  filled by one of the vacua of the landscape. The population of all possible bubbles in the universe
is possible in the context of eternal inflation, where transitions between different bubbles
due to quantum tunneling processes are going to happen.

Complementary to the top-down efforts, the {\sl bottom-up} approach is very important
 for connecting string theory with the real world. Here one tries to build consistent 
string models which contain as many SM features as possible. 
First one tries to build string models that contain as massless
states the particles of the SM, gauge bosons and three families of
quarks and leptons. Next, one has to derive the low-energy effective
action of the massless fields, in order to compute their couplings, like gauge couplings
and Yukawa couplings, which eventually can be compared with the experimentally known values.
Here another problem has to be solved, namely the problem of
moduli stabilization. String compactifications generically contain several massless
moduli fields with flat potential, which correspond to geometrical or other parameters
of the internal space. These have to be fixed, since the low-energy couplings of the
massless fields are functions of the moduli. In order to make predictions
one has to know the values of the moduli. In addition, massless moduli would
over-close the universe and also cause unobserved new forces \cite{de Carlos:1993jw}.

The bottom-up approach is especially useful in case some model independent and possibly
testable properties rise
just from the very fact that the SM has to be consistently embedded into string theory.
As we will discuss this happens for type II orientifold compactifications with a low string scale
around the TeV scale. In fact, the occurrence of Regge excitations of SM fields
is independent from the details of the internal geometry of the compactification.
If light enough, these Regge states can be possibly measured at the LHC
by scattering processes of quarks and gluons. The corresponding tree level string
cross sections are independent from in the internal geometry and hence independent from
the particular location of the model in the string landscape. This observation nullifies
in some sense the string landscape problem at the LHC.

As already mentioned, another window into new physics beyond the SM comes from astrophysics and cosmology.
Beautiful experiments, most notably COBE and WMAP, provided a precise
image of cosmic microwave background (CMB) radiation including its small
density variations. In this way the  inflationary scenario of the early universe is now
established as the standard model for cosmology. In addition, we know from
the astrophysical measurements that our universe is spatially flat. Its energy density
is dominated to about $74\%$ by a dark energy component, which behaves very similarly
to a positive cosmological constant. The explanation of this mysterious dark energy is one
of the biggest challenges for astroparticle physics, and hence also for string theory.
The remaining $26\%$ of the energy density is split into so far directly undiscovered dark matter particles (WIMPS), which account for $22\%$  of the total
energy density, and into a left-over $4\%$ component of visible SM
matter fields. Many properties of the dark matter fields are still unknown, although one very promising
candidate for dark matter is the lightest supersymmetric particle (LSP) in the MSSM.
So, strings also should be helpful to identify the nature of dark matter.
Hence the goal will be to use the data from the CMB, from dark matter and
from other astrophysical experiments to find or to probe the fundamental
theory of strings in the early universe.  
This will put further constraints on the allowed points in the string landscape,
often complementary to the particle physics constraints mentioned before.

In summary, successful  string model building must  take into account all these phenomenological
boundary conditions coming from the SM, from particle physics beyond the SM and also from cosmology.
The top-down constructions which  start from the geometry of the
compactification space must go hand in hand with the bottom-up approach, where
one is guided by  the phenomenological data. In this way, a (not necessarily one-to-one) map
between geometrical and topological properties of the compactifications spaces and
the particle physics observables will be provided. This dictionary between geometry/topology
and particle physics/cosmology is one of the most interesting aspects of string theory, and
will be demonstrated in this paper  by several examples.
Of course, it still has to be seen in the future if a string compactification can be found that matched
combined constraints from particle physics as well as from early time cosmology.

\section{Type II Intersecting brane models and their statistics}

\subsection{Overview over different classes of orientifold models}

In this section we will review how realistic string compactifications can be built from intersecting
D-brane orientifolds. They constitute a large class of models
in the string landscape. The orientifold region in the landscape  is complementary, and for type I strings  often dual, to heterotic string compactifications,
which will be left out here.
Specifically consider type II orientifold compactifications to four-dimensions on six-dimensional
manifolds ${\cal M}_6$, which were first discussed in       
\cite{Bachas:1995ik,Blumenhagen:2000wh, Angelantonj:2000hi,Aldazabal:2000dg,Cvetic:2001tj}
(for some reviews see \cite{Kiritsis:2003mc, Lust:2004ks, Blumenhagen:2005mu,Blumenhagen:2006ci,Marchesano:2007de}).
In order to incorporate non-Abelian gauge interactions and to obtain
massless fermions in non-trivial gauge representations, one has to introduce
D-branes in type II superstrings. Specifically there exist three classes of
four-dimensional models:

 \vskip0.3cm
 \noindent
 {\sl (i) Type I compactifications with D9/D5 branes:}
 
 \vskip0.3cm
 \noindent
This class of IIB models contain different stacks of D9-branes, which wrap the entire space ${\cal M}_6$, and which
also possess open string, magnetic,
Abelian gauge fields $F_{ab}$ on their world volumes (magnetized branes).
In other words, $F_{ab}$ corresponds to open string vector bundles, and this class of models
is string dual to heterotic string compactifications. For reasoning of Ramond tadpole cancellation,
one also needs an orientifold 9-plane (O9-plane). In addition one can also include D5-branes and
corresponding O5-planes. In the heterotic dual description the
D5/O5 open strings correspond to the non-perturbative sector of the theory. Since the open string gauge fields $F_{ab}$ induce mixed boundary conditions on the D-branes, the internal compact space can be regarded as a non-commutative
space.

\vskip0.3cm
 \noindent
 {\sl (ii) Type IIB compactifications with D7/D3 branes:}
 
 \vskip0.3cm
 \noindent
Here we are dealing with different stacks of D7-branes, which wrap different internal 4-cycles,
which intersect each other. The D7-branes can also carry non-vanishing open string
gauge flux $F_{ab}$. In addition, one can also allow for D3-branes, which are located at different
point of ${\cal M}_6$. In order to cancel all Ramond tadpoles one needs in general O3- and O7-planes.
Recently interesting GUT $SU(5)$ embeddings with D7-branes wrapped on Calabi-Yau 
cycles and $U(1)_Y$ background fluxes  were constructed {Blumenhagen:2008at}, which can be also formulated in F-theory \cite{Beasley:2008dc,Beasley:2008kw}.

\vskip0.3cm
 \noindent
 {\sl (iii) Type IIA compactifications with D6 branes:}
 
 \vskip0.3cm
 \noindent
 This class of models contains intersecting D6-branes, which are wrapped around 3-cycles
 of ${\cal M}_6$. Now, orientifold O6-planes are needed for Ramond tadpole 
 cancellation.
 One can show that
the cancellation of
the RR tadpoles implies absence of 
the non-Abelian anomalies in the effective 4D field theory.
However there can be still anomalous $U(1)$ gauge symmetries
in the effective 4D field theory.
These anomalies will be canceled by a Green-Schwarz mechanism
involving Ramond (pseudo)scalar field. As a result of these interactions
the corresponding $U(1)$ gauge boson will become massive.
Note that 
even an anomaly free $U(1)$ can become massive.
The massive $U(1)$ always remains as a global symmetry.
For SM engineering, we always have to require that the linear combination of $U(1)$'s that
corresponds to $U(1)_Y$ is anomaly free and massless.

\subsection{Intersecting D6-brane orientifolds}

D-brane models of these three different classes generically can be mapped onto each other by T-duality,
resp. IIA/IIB mirror symmetry including open strings and D-branes, and therefore are
essentially on equal footing. Hence, in the following
we will concentrate on IIA intersecting D6-brane models (class (iii)).

Let us therefore just summarize the main aspects of the intersecting D6-brane
models.
\begin{itemize}
\item 
We assume that six spatial directions are described
by a compact space  ${\cal M}_{6}$. 
In addition, a consistent orientifold projection is performed.
This yields O6-planes and in general changes the geometry.
The bulk space-time supersymmetry is reduced to ${\cal N}=1$ by the orientifold
projection.
To be more specific we will consider
a type IIA orientifold background of the form
\begin{eqnarray}
{\cal M}^{10} = ({\mathbb R}^{3,1} \times {\cal M}_{6})/
( \Omega\overline\sigma)\, ,\quad \Omega : {\rm world~ sheet ~parity.}
\end{eqnarray}
Here ${\cal M}_{6}$ is a Calabi-Yau 3-fold with a symmetry under 
$\overline\sigma$, the complex conjugation 
\begin{equation}{
\overline\sigma : z_i \mapsto \bar{z}_i,\ i=1,\, ...\, ,3,
}
\end{equation}
in local coordinates $z_i=x^i+iy^i$.
 It is combined with the world sheet parity $\Omega$ 
to form the orientifold projection $\Omega \overline\sigma$. 
This operation is actually a symmetry of the type IIA string 
on ${\cal M}_{6}$. 
Orientifold 6-planes are defined as the 
fixed locus 
\begin{center}
{ ${\mathbb R}^{3,1} \times{\rm Fix}(\overline\sigma)=
{\mathbb R}^{3,1} \times\pi_{O6}
$,} 
\end{center}
where   ${\rm Fix}(\overline\sigma)$ is a supersymmetric (sLag) 3-cycle on
$ {\cal M}^{6}$, denoted by $\pi_{O6}$.
It is special Lagrangian (sLag) and calibrated with 
respect to the real part of the holomorphic 3-form $\Omega_3$.

\vskip0.2cm
\noindent
Next we introduce {D6-branes}  with world-volume
\begin{center}
${\mathbb R}^{3,1} \times\pi_a$, 
\end{center}
i.e. they are
wrapped
around the {supersymmetric (sLag) 3-cycles $\pi_a$ 
and their $\Omega\overline\sigma$ images $\pi_a'$}
of ${\cal M}_6$,
which intersect  in ${\cal M}_6$.
Since the D-branes will
be wrapped around compact cycles
of the internal space,
multiple intersections will now be possible.
The chiral massless spectrum indeed is completely fixed
by the topological { intersection numbers $I$
of the 3-cycles} of the configuration.
\vskip 0.5cm
\begin{tabular}{lll}

\hspace{2cm} 
&
\hspace{2.5cm} 
&
\\

{\bf Sector} & {\bf Rep.} & {\bf Intersection number I} \\[0.2cm]
\hline
\hline
$a'\,a$
&
$A_a$
&
${1\over 2}\left(\pi'_a\circ\pi_a+\pi_{O6}\circ\pi_a\right)$
\\[0.2cm]

$a'\, a$
&
$S_a$
&
${1\over 2}\left(\pi'_a\circ\pi_a-\pi_{O6}\circ\pi_a\right)$
\\[0.2cm]

$a\,b$
&
$(\overline{N}_a,N_b)$
&
$\pi_{a}\circ \pi_b$
\\[0.2cm]

$a'\, b$
&
$({N}_a,{N}_b)$
&
$\pi'_{a}\circ \pi_b$
\\[0.2cm]
\hline
\end{tabular}
\vskip 0.4cm

Number of representations in each intersection sector in terms of the intersection numbers.

\vskip0.3cm

\item Since
the Ramond charges
of the space-time filling D-branes cannot `escape'
to infinite, the internal Ramond charges on compact space must cancel
(Gauss law). This is the issue of Ramond tadpole cancellation
which give some strong restrictions on the allowed D-brane
configurations. Specifically, the Ramond tadpole conditions follow from the equations
of motion for the gauge field $C_7$:
\begin{equation}
{ {1\over \kappa^2}\,
d\star d C_{7}=\mu_6\sum_a N_a\, \delta(\pi_a)+
                    \mu_6\sum_a N_a\, \delta(\pi'_a)
        + \mu_6 Q_6\,  \delta(\pi_{{\rm O}6}),
}
\end{equation}
where $\delta(\pi_a)$ denotes the Poincar\'e dual form of $\pi_a$, 
$\mu_p = 2\pi (4\pi^2\alpha')^{-(p+1)/2}$, and 
$2 \kappa^2= \mu_7^{-1}$.
Upon
integrating over ${\cal M}_6$ one obtains the RR-tadpole cancellation 
as equation
in homology:
\begin{equation}
\label{tadpoles}
\sum_a  N_a\, (\pi_a + \pi'_a)-4\pi_{{\rm O}6}=0.
\end{equation}
In principle it involves as many linear relations as there are
independent generators in $H_3({\cal M}_{6},R)$. But, of course,
the action of $\overline\sigma$ on ${\cal M}_{6}$ also induces an action
$[\overline\sigma]$ on the homology and cohomology. In particular,
$[\overline\sigma]$ swaps $H^{2,1}$ and $H^{1,2}$, and 
the number of conditions is halved.

\item Next, there
is the requirement of cancellation 
of the internal D-brane tensions, i.e
the forces between the D-branes must be balanced.
In terms of string amplitudes, it means that all NS tadpoles must
vanish, namely all NS tadpoles of the closed string moduli
fields and also of the dilaton field. Absence of these tadpoles means
that the potential of those fields is minimized. 
The disc level tension can be determined by 
integrating the Dirac-Born-Infeld effective
action. It is proportional to the
volume of the D-branes and the O-plane, so that the
disc level scalar potential reads
\begin{eqnarray}
\label{sp}
{\cal V}&=&T_6\, {e^{-\phi_{4}}
\over \sqrt{{\rm Vol({\cal M}_{6})}}}
               \left( \sum_a  N_a \left( {\rm Vol}({\rm D}6_a) + 
      {\rm Vol}({\rm D}6'_a) \right) -4 {\rm Vol}({\rm O}6)\right)\nonumber \\
     &=&T_6\, e^{-\phi_{4}} \left(
\sum_a{N_a \left| \int_{\pi_a} \widehat\Omega_3 \right|} +
 \sum_a{N_a \left| \int_{\pi'_a} \widehat\Omega_3 \right|} 
-4 \left| \int_{\pi_{{\rm O}6}} \widehat\Omega_3 \right |\right) .
\end{eqnarray}
The potential is easily seen to be positive semi-definite and its 
minimization imposes conditions on some of the moduli, freezing them 
to fixed values. Whenever the potential is positive, supersymmetry 
is broken and a classical vacuum energy generated by the net brane tension. 
It is easily demonstrated that the vanishing of ${\cal V}$ requires all 
the cycles wrapped by the D6-branes to be calibrated with respect to the same 
3-form as are the O6-planes.

\item
One can show that
the cancellation of
the RR tadpoles implies absence of 
the non-Abelian anomalies in the effective 4D field theory.
However there can be still anomalous $U(1)$ gauge symmetries
in the effective 4D field theory.
These anomalies will be canceled by a Green-Schwarz mechanism
involving Ramond (pseudo)scalar field. As a result of these interactions
the corresponding $U(1)$ gauge boson will become massive.
Considering the relevant triangle diagrams the condition for an anomaly free $U(1)_a$ is:
\begin{eqnarray}
N_a(\pi_a-\pi_a')\circ \pi_b=0\, .
\end{eqnarray}
Note that 
even an anomaly free $U(1)$ can become massive.
The massive $U(1)$ always remains as a global symmetry.
For SM engineering, we always have to require that the linear combination of $U(1)$'s that
corresponds to $U(1)_Y$ is anomaly free and massless.

\item
Besides the local triangle anomalies, field theoretical models can be plagued by global
$SU(2)$ gauge anomalies. In orientifold models this requirement can be deduced from a K-theory
analysis. In the case of our models, this condition requires an even amount 
of chiral matter  from $Sp(2)$ probe branes. In this case we obtain the following condition
for a model with $k$ stack of branes:
\begin{equation}
\sum_{a=1}^kN_a\pi_a\circ\pi_p\equiv 0\quad{\rm mod}\quad 2\, .
\end{equation}
This equation should hold for any probe brane $p$ invariant under the orientifold map.

\end{itemize}

\subsection{Getting the Standard Model}

Now we will discuss how the (supersymmetric) Standard Model (SM) can be obtained from
intersecting brane orientifolds. In general there will be two different brane sectors, namely one
local D-brane module, whose open string excitations correspond to the massless fields
of the SM. These SM branes are
wrapped around a subset of cycles in the internal space. The second module of D-branes constitute an Hidden Sector (HS) which interacts with
the SM only gravitationally or by some vector-like messenger fields. The HS D-branes are
wrapped around different internal cycles, and the HS is generically required
in order to satisfy all tadpole conditions, discussed above. (As we will discuss, there also exist
a few models without HS at all.) In addition, the HS often plays an important phenomenological role,
it can by responsible for spontaneous supersymmetry breaking in supersymmetric compactifications,
where the supersymmetry breakdown is transferred to the SM either by gravitational interactions
(gravity mediation) or by gauge interactions (gauge mediation). Moreover, the HS is responsible
for moduli stabilization (see section 4) and/or for cosmic inflation in the early universe (see section 5).
This scenario can be depicted in the figure 1.

\begin{figure}
\begin{center}
  \includegraphics[width=0.5\textwidth]{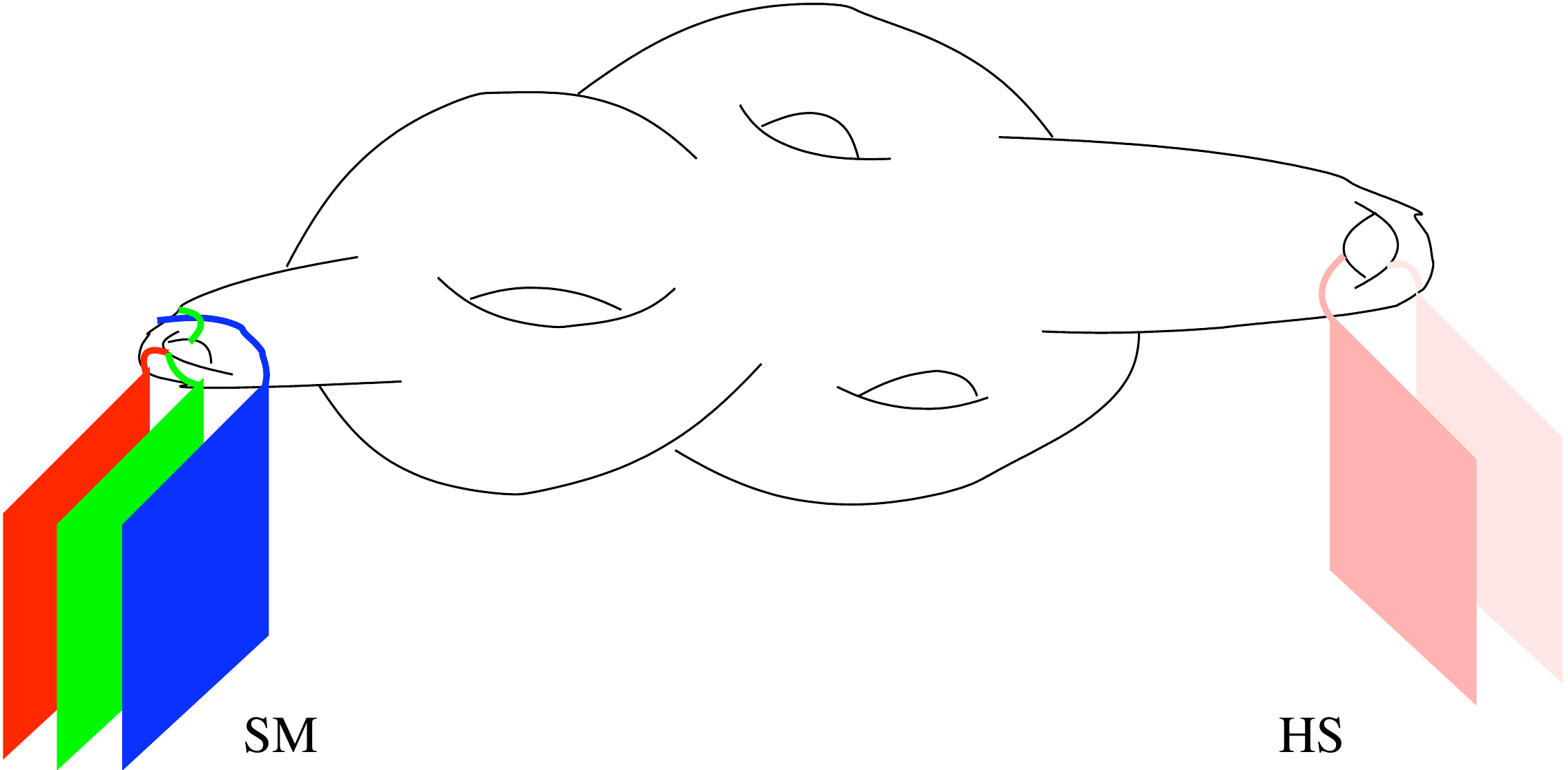}
\end{center}
\caption{Realization of the SM on a Calabi-Yau space by wrapped D-branes on the left side. The D-branes
on the right might be needed for tadpole cancellation and generate a hidden gauge sector.}
\end{figure}


For a realistic orientifold compactifications, we require that the following two conditions on the open string
spectrum are satisfied:
\begin{itemize}
\item
The open string on the SM branes lead fields of the SM with gauge bosons of the
group $SU(3)\times SU(2)_L\times U(1)_Y$ and three chiral families of quarks and leptons.
No other massless (chiral) states are allowed in the SM sector.
Hence the massless SM sector is rather model independent from the details of the internal geometry.
Each SM field possesses in general a tower of massive string excitations (Regge
excitations), which are also
independent from the internal geometry, and in addition massive Kaluza Klein (KK) and/or winding
states. This part of the spectrum does depend on the details of the internal geometry.
We will discuss in the next section, how SM scattering processes at the LHC can provide
rather model independent tests of the SM sector, in particular tests of the Regge spectrum.
\item The HS is depends to large extend on the details of the compactification. 
Its gauge symmetries and massless states are at the moment not further specified. As only constraint
on the massless spectrum of the HS we put the condition of {\it absence of chiral exotics} with SM
quantum numbers. This means that there must not be any chiral intersections of
the HS branes with the SM branes. However we can allow for vector-like states, which carry both SM
and HS quantum numbers. This vector-like states are expected to pair up, such that mass terms for them
are generated in the effective potential. Often these vector-like states are phenomenological 
attractive, since they can act as messenger fields for spontaneous supersymmetry breaking. As
a result of these interactions, soft SUSY breaking parameters are generated in the SM sector.
\end{itemize}

Let us now discuss more specifically the form of the SM brane sector. As emphasized already, one
can view this sector as a local D-brane module, which has to be implemented into a global
compactification model, i.e. the SM stack of D-branes has to be wrapped around some
cycles of the internal space. The massless part of the SM will then arise in a model independent way,
as well as its Regge excitations. So this part of the discussion covers a large part of the string landscape,
and its possible low energy signatures (see next section) are universal for a large class of point in the
landscape. In the following we will describe some local type IIA/IIB D-brane
configurations that lead to the SM
in a very economic way.

\subsubsection{Three stack D-brane models}

Here one starts with three stacks of D-branes with initial gauge symmetries:
\begin{equation}U(3)\times U(2) \times U(1)\times U(1)\ .
\end{equation}
The (left-handed) SM spectrum is shown in the table 1.

\begin{table}[htb]
\renewcommand{\arraystretch}{1.5}
\begin{center}
\begin{tabular}{|c|c|c|c|}
\hline
 matter  & $SU(3)\times SU(2)\times U(1)^3$ & $U(1)_Y$ & $U(1)_{B-L}$ \\
$ q$ & $({\bf 3},{\bf 2})_{(1,1,0)}$ &  ${1\over 3}$ &
${1\over 3}$ \\
  $\bar u$ & $(\overline{\bf 3},{\bf 1})_{(2,0,0)}$ &
      $-{4\over 3}$ &  $-{1\over 3}$  \\
$\bar d$ & $ (\overline{\bf 3},{\bf 1})_{(-1,0, 1)}$ &
     ${2\over 3}$ & $-{1\over 3}$   \\
     \hline
$l$ & $({\bf 1},{\bf 2})_{(0,-1,1)}$ & ${-1}$ &
         ${-1}$   \\
$\bar e$ & $({\bf 1},{\bf 1})_{(0,2,0)}$ & ${2}$ &
          ${1}$   \\
 $\bar\nu$ & $({\bf 1},{\bf 1})_{(0,0,-2)}$ & ${0}$ &
                ${1}$    \\
    \hline
\end{tabular}
\caption{\small Left-handed fermions for the 3 stack model.
}
\end{center}
\end{table}

The hypercharge $Q_Y$ is given as the following  linear combination
of the three $U(1)'$s:
\begin{equation}Q_Y=-{2\over 3}\ Q_a+{1\over 2} Q_b\ .
\end{equation}
Here one is forced to realize the left-handed $(\bar u,\bar c,\bar t)$-quarks
in the antisymmetric
representation of $U(3)$, which is the same as the anti-fundamental
representation $\overline 3$.
Note that  the three stack models with antisymmetric  matter  are dual to
the D3-brane quivers at CY
singularities \cite{Berenstein:2001nk,Verlinde:2005jr,Malyshev:2007zz}.
Alternative bottom-up constructions of the SM via D-branes can be found in 
\cite{Antoniadis:2000ena}.

\subsubsection{Four stack D-brane models}

One of the most common ways to realize the SM is by considering four stacks of
D-branes.
There are several simple ways to embed the SM
gauge group into products of unitary and symplectic gauge groups (see \cite{Blumenhagen:2006ci}).
For illustration and also
in the next section about LHC signatures, we will use as a prototype model four stacks of D-branes with
gauge symmetries:
\begin{equation}
U(3)_a\times U(2)_b \times U(1)_c\times U(1)_d\ .
\end{equation}
The intersection pattern of the four stacks of D6-branes can be
depicted as in figure 2.

\begin{figure}
\begin{center}
  \includegraphics[width=0.5\textwidth]{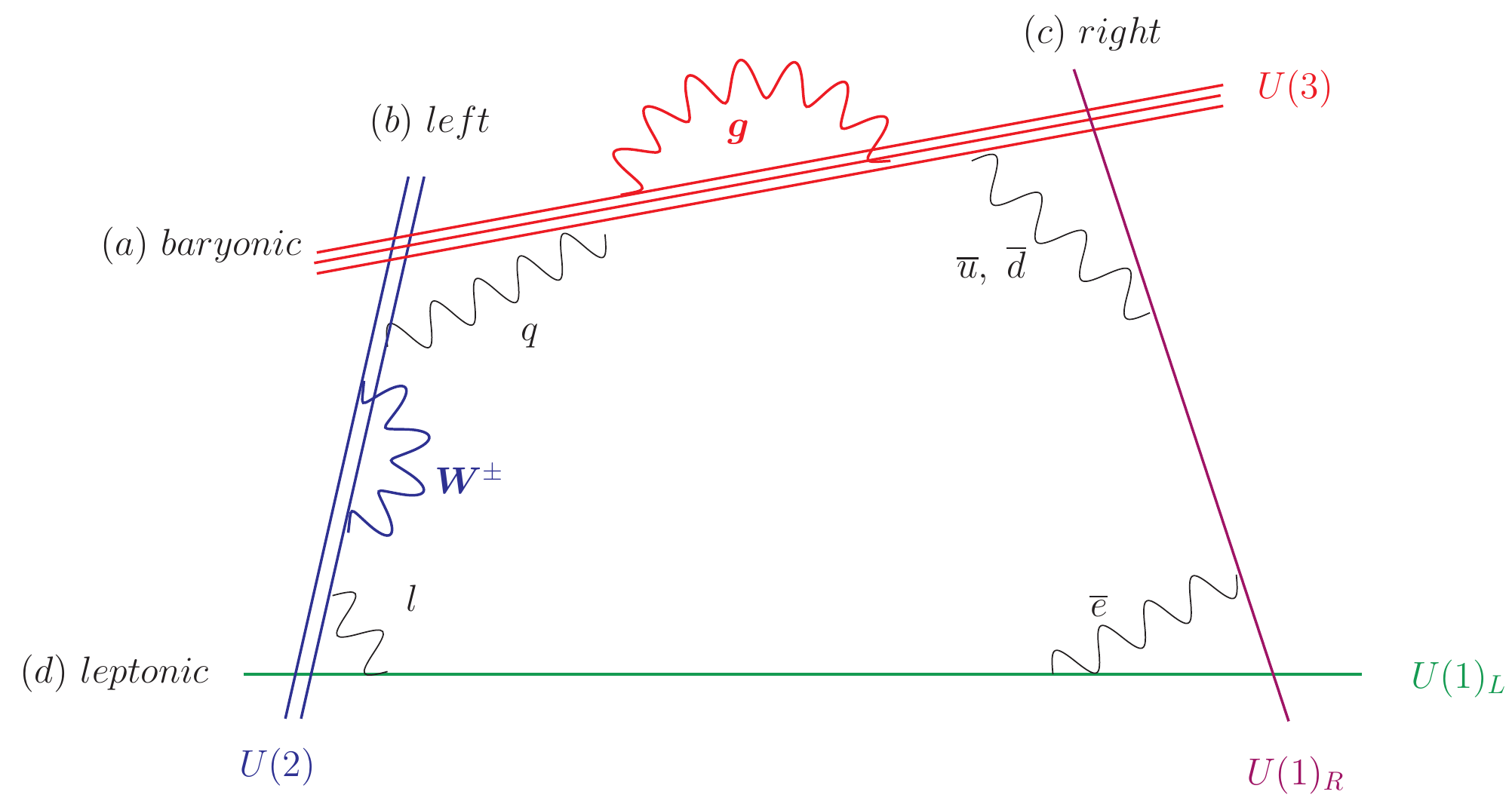}
\end{center}
\caption{A local module of four intersecting stacks of D-branes realizing the SM.}
\end{figure}


\noindent The chiral spectrum of the intersecting brane world
model should be identical to the chiral spectrum
of the SM particles.
In type IIA, this fixes uniquely the intersection numbers of the
3-cycles, $(\pi_a,\pi_b,\pi_c,\pi_d)$,
the four stacks of D6-branes are wrapped on.

There exist several ways to embed the hypercharge $Q_Y$ into the four $U(1)$
gauge symmetries.
The standard electroweak 
hypercharge $Q_Y^{(S)}$ is given as the following  linear combination
of three $U(1)'$s
\begin{equation}Q_Y^{(S)}={1\over 6}\ Q_a+{1\over 2}\ Q_c+{1\over 2}\ Q_d\ .\end{equation}
Therefore, in this case the gauge coupling of the hypercharge
is given as
\begin{equation}   {1\over \alpha_Y}={1\over 6}\ {1\over \alpha_a} +
                                {1\over 2}\ {1\over \alpha_c} +
                                {1\over 2}\ {1\over \alpha_d}\ .\end{equation}

Now we turn to the particle content of our prototype model.
In compact orientifold compactifications each stack of D-branes is 
accompanied by a orientifold mirror
stack of D$'$-branes. In the next Section about the amplitudes, we will not
make a difference between
the the D-brane and the mirror D$'$-branes. Hence we will use in the
following the indices
$a,b,c,d$ collectively for the D-branes as well as for their mirror branes. Then
self-intersections among D-branes include intersections between D-
and D$'$-branes. Furthermore, for simplicity, we will
suppress from the spectrum those open string states which one also gets from
intersections between
D-branes and orientifold planes. With these restrictions
the left-handed fermion spectrum
for our prototype model is presented in Table~2.

\def\1{\bf 1}
\def\2{\bf 2}
\def\3{\bf 3}
\def\ov{\overline}

\begin{table}[htb]
\renewcommand{\arraystretch}{1.5}
\begin{center}
\begin{tabular}{|c|c|c|}
\hline
particle & $U(3)_a\times U(2)_b \times U(1)_a\times U(1)_b\times U(1)_c \times U(1)_d$ & mult. \\
\hline
$q$ & $(\3,{\2})_{1,-1,0,0}+(\3,\2)_{1,1,0,0}$   & $I_{ab}$   \\
\hline
$\bar u$ &  $(\ov{\3},\1)_{-1,0,-1,0}+(\ov{\3},\1)_{-1,0,0,-1}$  & $I_{ac}+I_{ad}$ \\
\hline
$\bar d$ &  $(\ov{\3},\1)_{-1,0,1,0}+(\ov{\3},\1)_{-1,0,0,1}$  & $I_{ac}+I_{ad}$ \\
$\bar d'$ & $(\ov{\3}_A,\1)_{2,0,0,0}$  &  $\h I_{aa}$ \\
\hline 
$l$ & $(\1,\2)_{0,1,-1,0}+(\1,\2)_{0,1,0,-1}$  &         $I_{bc}+I_{bd}$       \\
& + $(\1,{\2})_{0,-1,-1,0}+(\1,{\2})_{0,-1,0,-1}$ &     \\
\hline
$\bar e$ & $(\1,\1)_{0,0,2,0}$ & $\h I_{cc}$ \\
$\bar e'$ & $(\1,\1)_{0,0,0,2}$ & $\h I_{dd}$ \\
$\bar e''$ & $(\1,\1)_{0,0,1,1}$ & $I_{cd}$\\
\hline
\end{tabular}
\caption{\small Chiral spectrum for the four stack model with $Q_Y^{(S)}$.
}
\end{center}
\end{table}

\noindent
To derive three generations of quark and leptons, the intersection
number in Table 4 must satisfy certain phenomenological restrictions:
We must have $I_{ab}=3$. From the left-handed anti u-quarks, we get that
$I_{ac}=3$, and likewise for the two types of left-handed anti d-quarks, we infer that
$I_{ac}+I_{ad}+{1\over 2}I_{aa}=3$.
In the lepton sector we require  that $I_{bc}+I_{bd}=3$ and
${1\over 2}(I_{cc}+I_{dd})+I_{cd}=3$.

\subsection{Intersecting D6-brane statistics}

In this section
we first want to count all different, consistent D-brane embeddings into a given closed string geometrical
background space.\footnote{For Standard Model searches and statistics of Gepner model and rational conformal field theory
orientifolds see  \cite{Dijkstra:2004ym,Dijkstra:2004cc,Anastasopoulos:2006da,Ibanez:2007rs,GatoRivera:2007yi,Kiritsis:2008ry,GatoRivera:2008zn}.
Orientifold based on free fermions were investigated in \cite{Kiritsis:2008mu}.
Statistics of the heterotic landscape were discussed in \cite{Dienes:2006ut,Lebedev:2006kn,Lebedev:2008un}.
Other aspects such as correlations in string statistics were discussed in \cite{Dienes:2007zz,Dienes:2008rm}.}
The aim is to find out how many of them lead to spectrum of the
supersymmetric SM.
To be specific, we now restrict ourselves on orbifold compactifications, i.e.
 ${\cal M}_6$ is a toroidal $Z_N$ resp. $Z_N\times Z_M$ orientifold. First, we
 consider the case  
${\cal M}_6=T^6/Z_2\times Z_2=\prod_{I=1}^3T^2_I/Z_2\times Z_2$. The D6-branes
are wrapping special Langrangian 3-cycles, which are products of 1-cycles
in each of the three subtori $T^2_I$. Hence they are characterized 
by three pairs of integer-valued wrapping numbers $X^I,Y^I$ ($I=0,\dots ,3)$.
The supersymmetry conditions, being equivalent to the vanishing of
the D-term scalar potential ${\cal V}$  have the form:
\begin{equation}\label{dterms}
\sum_{I=0}^3{Y^I\over U_I}=0\, ,\quad\sum_{I=0}^3X^IU_I>0\, .
\end{equation}
The $U_I$ are the three complex structure moduli of the three two-tori $T^2_I$.
The Ramond tadpole cancellation conditions for $k$ stacks
of $N_a$ D6-branes are given by
\begin{equation}\label{rtadpole}
\sum_{a=1}^kN_a\vec X_a=\vec L\, ,
\end{equation}
where the $L^I$ parametrize the orientifold charge. In addition there are 
some more constraints from K-theory.
Chiral matter in bifundamental representations originate from
open strings located at the intersection of two stacks of D6-branes
with a multiplicity  (generation) number
given by the intersection number
\begin{equation}\label{intersection}
I_{ab}=\sum_{I=0}^3(X^I_aY^I_b-X^I_bY^I_a)\, .
\end{equation}

\subsubsection{$Z_2\times Z_2$ orientifold}

Specifically, we first want to count all different, consistent D-brane embeddings
into the given $T^6/Z_2\times Z_2$ background geometry. I.e. we want to
count all possible solutions of the D-brane equations (\ref{dterms}) and (\ref{rtadpole}).
These set of equations are diophantic equations in the integer wrapping numbers $X^I,Y^I$,
and they contain as continuous parameters the complex structure moduli $U_I$.
First we want to know, if for any given tadpole charge
$\vec L$ there is a finite number of solutions of these equations.
Actually, based on a saddle point approximation, the total number of D-brane embeddings
can be estimated as follows \cite{Blumenhagen:2004xx}:
\begin{equation}
N_{\rm D-branes}(L)\simeq e^{2\sqrt{L\log L}}\, .
\end{equation}
For typical orientifold charges like $L=64$, one  obtains as estimate
that $N_{\rm D-branes}\simeq 2\times 10^{9}$.

Next, we explicitly count all possible solutions of the D-brane
equations (\ref{dterms}) and (\ref{rtadpole}) by running a computer program; this leads to a total of
$1.66\cdot 10^{8}$ supersymmetric  D-brane models
on the $Z_2\times Z_2$ orientifold \cite{Gmeiner:2005vz,Gmeiner:2005nh,Gmeiner:2006qw}. However  this computer count was limited
by the available CPU time of about $4\times 10^5$ hours, and hence it could be done
only for restricted, not too large values of the complex structure parameters $U_I$.
However, in \cite{Douglas:2006xy} an analytic proof was found that the number of solutions
for eqs.(\ref{dterms}) and (\ref{rtadpole}) is  indeed finite. Recently it was shown \cite{watinew} that for the $Z_2\times Z_2$ orientifold 
many models with standard model
like properties are also lying in the tail of the distribution with large complex structure parameters.

With this large sample of models we can ask the question which fraction of 
models satisfy several phenomenological constraints that gradually
approach the spectrum of the supersymmetric MSSM.

This is summarized
in the following table:

\vskip0.5cm

\begin{table}[htb!]
\begin{center}
\begin{tabular}{|l|r|}\hline
Restriction                    & Factor\\\hline
gauge factor $U(3)$            & $0.0816$\\
gauge factor $U(2)/Sp(2)$      & $0.992$\\
No symmetric representations   & $0.839$\\
Massless $U(1)_Y$	       & $0.423$\\
Three generations of quarks   ($I_{ab}^{\rm quarks}=3$) & $2.92\times10^{-5}$\\
Three generations of leptons  ($I_{ab}^{\rm leptons}=3$) & $1.62\times10^{-3}$\\\hline
\emph{Total}                   & $1.3\times10^{-9}$\\\hline
\end{tabular}
\end{center}
\end{table}

The total probability of $1.3\times 10^{-9}$ is simply obtained multiplying
each probability factors in the first six rows, since one can show that there
is little correlation between these individual probabilities.
We see that statistically  only one in a billion 
models give rise to an MSSM like D-brane vacuum.
Multiplying this result with the initial number of models, the chance to find the MSSM is
less then one. One can now compare this statistical result with the explicitly constructed
intersecting D6-brane models with MSSM like spectra. In fact, in \cite{Marchesano:2004xz} a 
$Z_2\times Z_2$ orientifold model with MSSM like spectrum was found that should be
contained in the statistical search discussed above. However unfortunately this model is
outside the range of complex structure moduli  covered by our computer scan.
Also note that all $Z_2\times Z_2$ MSSM like models found so far contain also chiral exotic particles, not
present in the MSSM. These chiral exotic particles can be avoided in the 
 $Z_6'$ orientifolds, as will now discuss in the next subsection.

\subsubsection{$Z_6$ and $Z_6'$ orientifolds}

To bypass the problem of getting always chiral exotic massless particles,
the statistical scan was extended in \cite{Gmeiner:2007we} to the case of the $Z_6$ orbifold geometry
and in \cite{Gmeiner:2007zz,Gmeiner:2008xq,Gmeiner:2008qq} to the $Z_6'$ orbifold
background geometry. For the first class of orbifold backgrounds explicit MSSM
like models were alrady constructed in \cite{Honecker:2004kb}.
Compared to the $Z_2\times Z_2$ orientifold, the $Z_6,Z_6'$ cases are more complex, because
it also contains exceptional, twisted (blowing-up) 3-cycles, besides the untwisted bulk 3-cycles.
The D6-branes wrapped around the exceptional 3-cycles correspond to fractional branes.
A general fractional cycle can be written with parameters  as
\begin{equation}\label{eq:brane}
 \Pi^{\rm frac} = \frac{1}{2} \Pi^{\rm bulk} +\frac{1}{2}\Pi^{\rm ex} \, ,
\end{equation}
where we introduced the notation bulk and exceptional cycles for the torus and $\mathbb{Z}_N$ cycles, respectively.

\vskip0.2cm
\noindent
$Z_6$-orientifold: First, it was possible to show that even in the presence of the exceptional cycles the
number of the D-brane solutions of the tadpole plus supersymmetry conditions is very large but
nevertheless finite.
Then, by extended computer scan it was found that there exist $3.4\times 10^{28}$ solutions
in total, of which $5.7\times 10^6$ contain the gauge group and the chiral
matter content of the MSSM. We therefore obtained a probability of $1.7\times 10^{-22}$ to
find MSSM like vacua, a number considerably lower than the value $10^{-9}$ for
the case of the  $Z_2\times Z_2$ orientifolds.
However still chiral exotics appear in all solutions.

\vskip0.2cm
\noindent
$Z_6'$: These are the so far best intersecting D6-brane models seen from the 
phenomenological point of view. Therefore let us describe the statistical results in more detail.

Table 3 from \cite{Gmeiner:2008xq}
 shows the total number of solutions of the tadpole and
supersymmetry conditions which is of order $10^{23}$.

\begin{figure}
\begin{center}
  \includegraphics[width=0.5\textwidth]{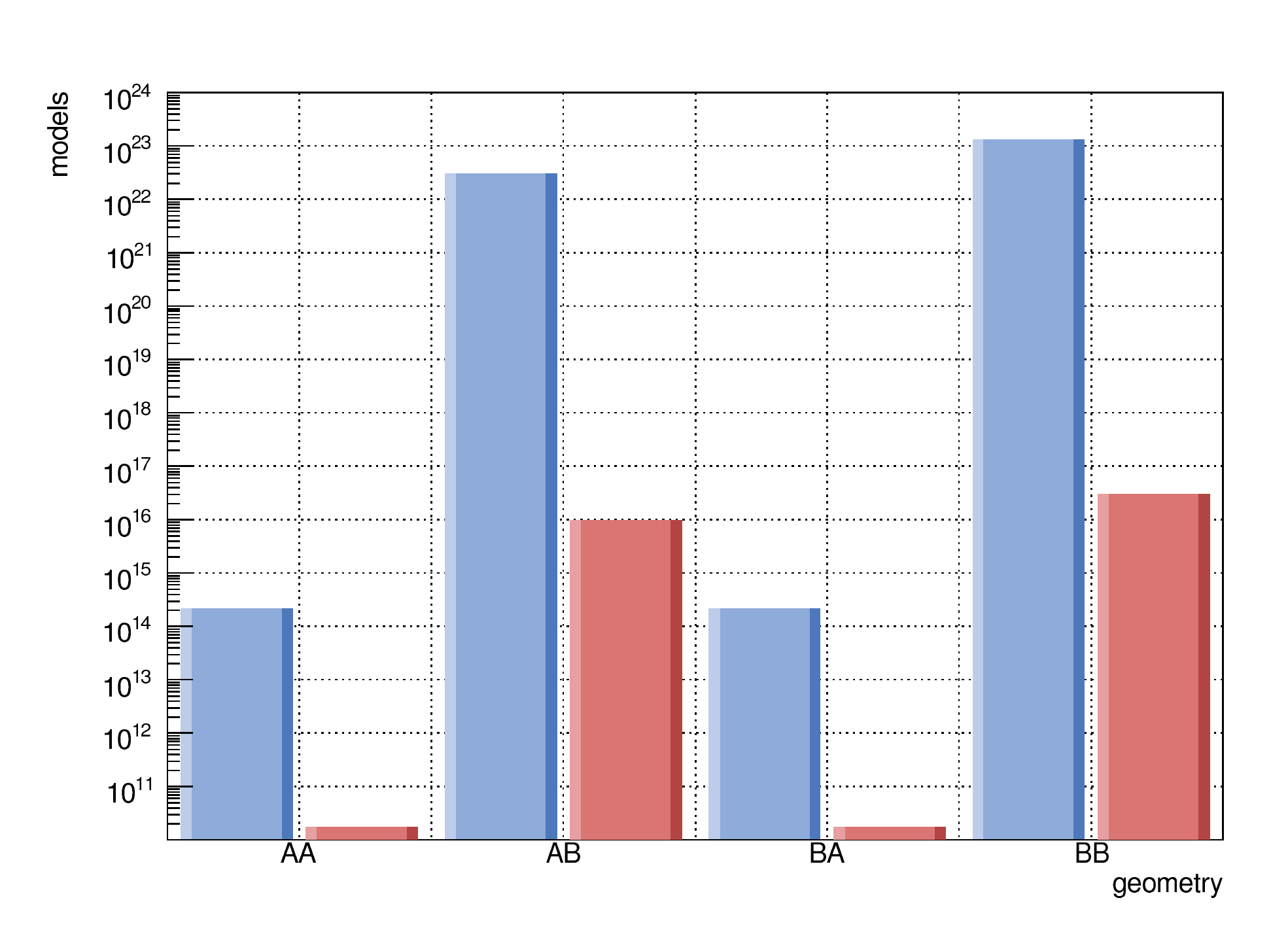}
\end{center}
\caption{The total number of consistent D6-brane embeddings on the $Z_6'$ orientifold, depending
on different choices of discrete background parameters.}
\end{figure}

The next two figures 4 and 5, again from
\cite{Gmeiner:2008xq},
 show numbers of solutions of models on $T^6/\mathbb{Z}_6'$ with SM gauge group and three
generations of quarks and leptons. Specifically, in the next figure the total number of three
generation models including chiral exotics is shown. In total $\mathcal{O}(10^{19})$ models with three generations have been found.
\begin{figure}
\begin{center}
  \includegraphics[width=0.5\textwidth]{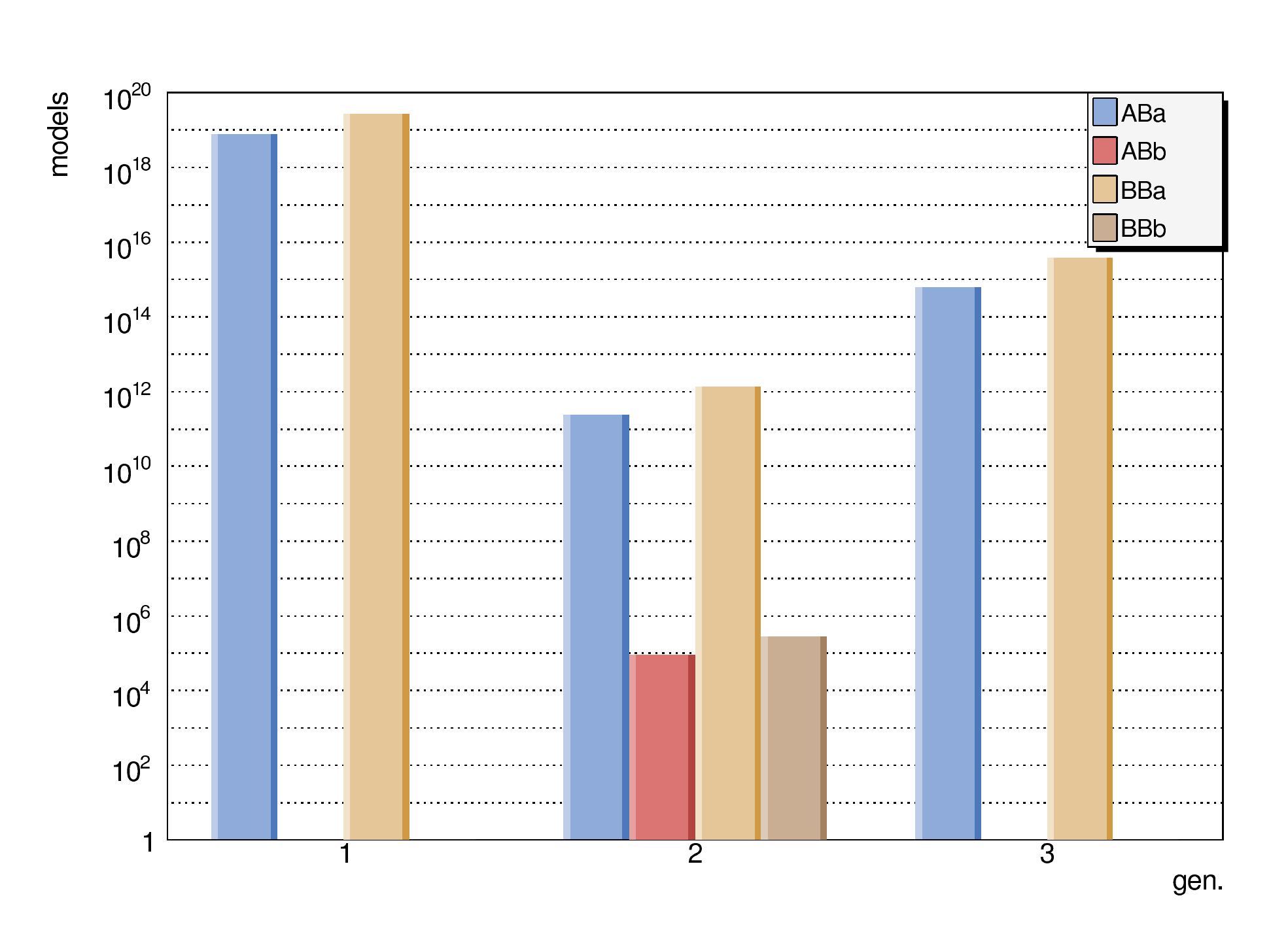}
\end{center}
\caption{The total number of consistent D6-brane embeddings on the $Z_6'$ orientifold leading to the
three generation SM including chiral exotics}
\end{figure}
Finally the total amount of exotic matter in models with three generations is depicted.
\begin{figure}
\begin{center}
  \includegraphics[width=0.5\textwidth]{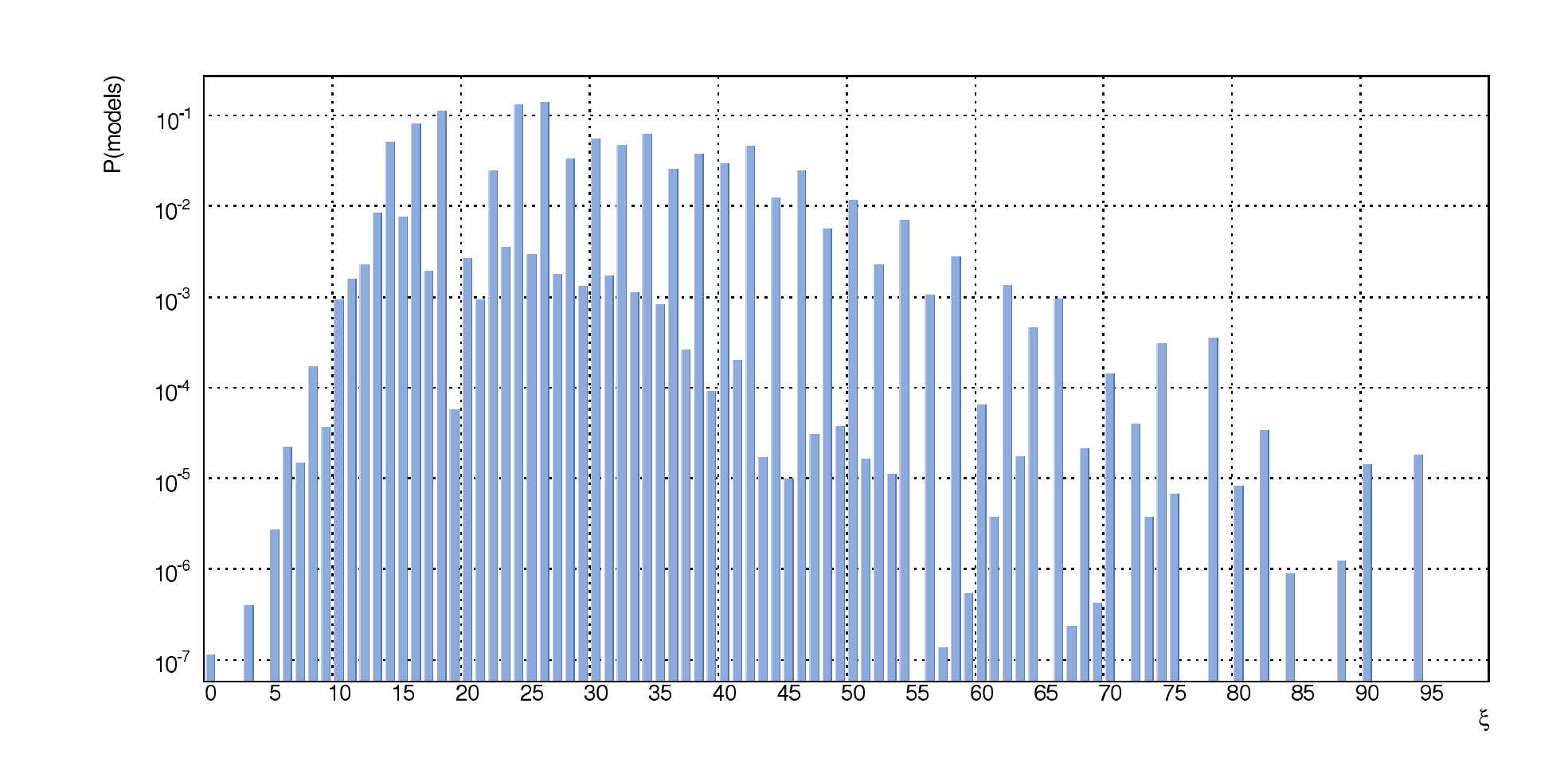}
\end{center}
\caption{The total number of chiral exotics in $Z_6'$ orientifolds with SM
spectra.}
\end{figure}
Asking for zero number of chiral exotics gets an additional suppression factor of 
$\mathcal{O}(10^{-7})$. Hence, models with only three generations of quarks and leptons
and no chiral exotics occur roughly with likelihood of $\mathcal{O}(10^{-11})$, compared
to the total number of solutions. Note that most of these models contain a relatively
large number of pairs of Higgs doublets, which  are not to be confused with chiral exotics,
since they are in vector-like representations of the SM gauge group. In addition
to the Higgs fields, there are other vector-like exotic states, which will get masses
due to some deformations (geometric deformations like blowing up orbifold singularities or
also deformations of brane positions) of the background parameters.

\section{Possible low energy (LHC) signatures of intersecting D-brane models}

\subsection{Low string scale in  intersecting brane compactifications}

There is some good reason to believe that the resolution of the
hierarchy problem lies in new
physics around the TeV mass scale. The LHC collider at CERN is designed to
discover new physics
precisely in this energy range, hopefully giving important clues about the
nature of dark matter and perhaps at the same time about the solution of the
hierarchy problem. In fact, there are at least three, not necessarily mutually
exclusive scenarios, offered as solutions of the hierarchy problem:
\vskip0.3cm

\noindent \hskip1cm $\bullet$\  Low energy supersymmetry at around 1 TeV.

\vskip0.2cm

\noindent \hskip1cm $\bullet$\
New strong dynamics at around 1 TeV (technicolor, little Higgs models, etc).

\vskip0.2cm

\noindent \hskip1cm $\bullet$\ Large extra dimensions and a low scale
for (quantum) gravity  at around 1 TeV.

\vskip0.3cm\noindent
Here we discuss some universal features
of the large extra dimensions scenario \cite{ArkaniHamed:1998rs, Antoniadis:1998ig}
relevant for its possible discovery at the LHC. In this scenario,
the gravitational and gauge interactions are unified at around 1 TeV,
and the observed weakness of gravity at lower energies is due to the existence
of large extra dimensions. Gravitons may scatter into the extra space and
by this the gravitational coupling constant is decreased to its observed value.
Extra dimensions arise naturally in string theory. Hence, one obvious question is how
to embed the above scenario into string theory. Then the next important question is what are the possible signatures  of large
extra dimensions and
low gravity in string theory, and how  to detect them at the LHC.
Large extra dimensions can appear in string theory in case that the intrinsic
scale of the string excitations, called the string mass $M_{\rm string}$ is
very low, namely at the
order of TeV.  In this case a whole tower of infinite  string excitations
will open up at around
$1\ {\rm TeV}$, where the new particles essentially follow the well known Regge
trajectories of vibrating strings,
\begin{equation}
j=j_0+\alpha' M^2\ ,
\end{equation}
with the spin $j$ and $\alpha'$ the Regge slope parameter that determines the fundamental
string  mass scale $M_{\rm string}^2=\alpha'^{-1}$.

Let us list what kind of string signatures from a low string scale and from
large
extra dimensions can be
possibly expected  at the LHC:

\vskip0.3cm

\noindent $\bullet$ The discovery of new exotic particles around $M_{\rm
string}$. For
example, many string models predict the existence of new, massive $Z'$ gauge
bosons
from additional $U(1)$ gauge symmetries.

\vskip0.2cm

\noindent $\bullet$ The discovery of (non-perturbative) quantum gravity effects in the form of
mini black
holes.

\vskip0.2cm

\noindent $\bullet$ The discovery of string Regge excitations with masses of
order $M_{\rm string}$.

Before we discuss the above mentioned stringy signatures, we like to describe
how large extra dimensions can be realized in Calabi-Yau orientifolds
and
how the local, SM
D-brane system has to be embedded into a large volume Calabi-Yau space.

First we discuss the gravitational and gauge couplings
in orientifold compactifications.
In the following we consider the type II superstring compactified on
a six--dimensional compactification manifold.
In addition, we consider a D$p$--brane wrapped on a $p-3$--cycle with the
remaining four dimensions extended into the uncompactified space--time.
We have $d_\parallel=p-3$ internal directions
parallel to the D$p$--brane world volume
and $d_\perp=9-p$ internal directions transverse to the D$p$--brane world volume.
Let us denote the radii (in the string frame) of the parallel directions by
$R^\parallel_i\ ,i=1,\ldots,d_\parallel$ and the
radii of the transverse directions by $R^\perp_j,\ j=1,\ldots,d_\perp$.
While the gauge interactions are localized
on the D--brane world volume the gravitational interactions are also spread into the
transverse space. This gives qualitatively different quantities for their
couplings.
In $D=4$ we obtain for the Planck mass
($\alpha'=M_{\rm string}^{-2}$) 
\begin{equation}\label{planckmass}
M_{\rm Planck}^2=8\ e^{-2\phi_{10}}\ M_{\rm string}^8\ \frac{V_6}{(2\pi)^{6}}\ ,
\end{equation}
where the internal six-dimensional (string frame) volume $V_6$ is expressed
in terms of the parallel and transversal
radii as
\begin{equation}\label{volumes}
V_6=(2\pi)^6\ \prod_{i=1}^{d_\parallel} R^\parallel_i\ \prod_{j=1}^{d_\perp} R^\perp_j\ .
\end{equation}
The dilaton field $\phi_{10}$ is related to the $D=10$
type II string coupling constant through $g_{\rm string}=e^{\phi_{10}}$.
The gravitational coupling constant follows from eq.(\ref{planckmass}) through the
relation $\kappa_4^{-2}=\frac{1}{8\pi}\ M_{\rm Planck}^2$.
On the other hand, in type II superstring theory
the gauge theory on the D--brane world--volume has the gauge coupling:
\begin{equation}\label{Dpgaugecoupling}
g_{Dp}^{-2}=(2\pi)^{-1}\ {\alpha'}^{\frac{3-p}{2}}\ e^{-\phi_{10}}\
\prod_{i=1}^{d_\parallel}R_i^\parallel\ .
\end{equation}
Here each factor $i$ accounts for an $1$--cycle wrapped along the $i$--th
coordinate segment.
While the size of the gauge couplings is determined by the size of the parallel
dimensions, the strength of gravity is influenced by all directions.

From (\ref{planckmass}) and the gauge coupling (\ref{Dpgaugecoupling}) we
may deduce a relation between the Planck mass $M_{\rm Planck}$,
the string mass $M_{\rm string}$ and the sizes $R_j$ of the compactified
internal directions. For type II we obtain:
\begin{equation}
g_{Dp}^2\ M_{\rm Planck}=2^{5/2}\pi\ M_{\rm string}^{7-p} \
\Biggl(\prod\limits_{j=1}^{d_\perp} R_j^\perp\Biggr)^{1\over 2}\
\Biggl(\prod\limits_{i=1}^{d_\parallel} R_i^\parallel\Biggr)^{-1/2}\ .
\end{equation}
Hence, by enlarging some of the transverse compactification
radii $R_j^\perp$ the string scale has to become lower in order to achieve the correct Planck
mass ($p<7$).
This is to be contrasted with a theory of closed (heterotic) strings only. In that case
the relation between the Planck mass and the string scale does not depend on
the volume. It is given by the relation
$M_{\rm string}=g_{\rm string}\ M_{\rm Planck}$, which requires
a high string scale $M_{\rm string} \sim 10^{17} {\rm GeV}$ for the correct Planck mass.

A priori,
there are no compelling reasons why the string mass scale should be much lower than the
Planck mass. In the large volume compactifications of 
\cite{Balasubramanian:2005zx,Conlon:2005ki,Conlon:2007xv}
it was shown that that one can indeed stabilize
moduli in such a way that the string scale $M_{\rm string}$
is at intermediate energies of about $10^{11-12}~{\rm GeV}$. Then
the internal CY volume $V$ is of order $VM_{\rm string}^6={\cal O}(10^{16})$. The motivation for
this scenario is to obtain a supersymmetry breaking scale around $1~{\rm TeV}$, since
one derives the following relation for the gravitino mass:
\begin{equation}
m_{3/2}\sim {M_{\rm string}^2\over M_{\rm Planck}}\, .
\end{equation}
However, giving up the requirement of supersymmetry at the TeV scale, one is free to
consider
CY manifolds with much larger volume.
In fact, if it happens for $M_{\rm string}$ to be within the range of LHC energies,
not too far beyond
1 TeV, string theory can be tested.
In this case the Calabi-Yau volume is as large as $VM_{\rm string}^6
={\cal O}(10^{32})$. Of course one has
to find
scalar potentials with minima that lead to such big internal volumes.

Let us now discuss the possible sizes of large  extra dimensions subject to the
experimental facts.
Cavendish type experiments test Newton's law up to a scale of millimeters. This provides
an upper bound on the large extra dimensions $R_j^\perp$ to be in the millimeter range.
On the other hand, QCD and electroweak scattering experiments give an upper bound
on the small extra dimensions $R_i^\parallel$ in the range of the electroweak scale
$M_{EW}^{-1}$.

A first look at the relations (\ref{planckmass}) and (\ref{volumes}) gives an estimate on the string
scale
$M_{\rm string}$ and the size of $d_\perp$ extra dimensions $R_j^\perp$.
For the $d_\parallel$ small directions to be of the order of the string scale
$M_{\rm string}$ and $d_\perp$ extra dimensions of size $R^\perp$ we
obtain the values shown in this table.\footnote{The above values are computed for $g_{\rm string}\simeq g^2=\frac{1}{25}$,
i.e. $\alpha=\frac{g^2}{4\pi}=0.003$. Furthermore, $E_R=\frac{hc}{R^\perp}$ and $1\
GeV^{-1}\sim 10^{-15}m$.}

\vskip0.1cm\def\ss#1{{\scriptstyle{#1}}}
{\vbox{
{$$
\vbox{\offinterlineskip\tabskip=0pt
\halign{\strut\vrule#
&~$#$~\hfil
&\vrule$#$
&~$#$~\hfil
&\vrule$#$
&~$#$~\hfil
&\vrule$#$
&~$#$~\hfil
&\vrule$#$
&~$#$~\hfil
&\vrule$#$
&~$#$~\hfil
&\vrule$#$
&~$#$~\hfil
&\vrule$#$\cr
\noalign{\hrule}
&  &&d_\perp=1
&&d_\perp=2 && d_\perp=3  && d_\perp=4&& d_\perp=5&& d_\perp=6 &\cr
\noalign{\hrule}
&R^\perp\ [GeV^{-1}]&&1.6\cdot10^{26}&&4\cdot10^{11} && 5.4\cdot10^{6}&&2\cdot 10^4&&
693 &&74  &\cr
&R^\perp\ [m]&&1.6\cdot10^{11}&&4\cdot10^{-4} && 5.4\cdot10^{-9}&&2\cdot10^{-11}&&
7\cdot 10^{-13}&&7\cdot 10^{-14} &\cr
&E_R\ [MeV]&&7.7\cdot 10^{-24}   &&3\cdot 10^{-9}  &&2\cdot 10^{-4}&&0.06&&1&&16&\cr
\noalign{\hrule}}}$$
\vskip-6pt
\centerline{\noindent{}
{\it Size of $d_\perp$ large extra dimensions for a string scale of $M_{\rm
string}=1\ TeV$.}}
\vskip10pt}}}
\noindent
So, the case $d_\perp=1$ is ruled out experimentally.

In fact, it is not
completely straightforward to construct SM-like D-brane models on CY spaces with large transverse dimensions.
In order to combine D-branes with SM particle content with the scenario of large extra dimensions, one
has to consider specific types of Calabi-Yau compactifications.
The three or four stacks of intersecting D-branes that give rise to the spectrum of the SM are
just local modules that have to be embedded into a global large volume CY-manifold in order to obtain a consistent
string compactification. For internal consistency several tadpole and stability conditions have to be
satisfied that depend on the details of the compactification, such as background fluxes etc. In this
work we will not aim to provide fully consistent orientifold compactifications with all tadpoles cancelled, since it is enough for us to
know the properties of the local SM D-brane modules for the
computation of the  scattering amplitudes among the SM open strings. However it is important to
emphasize that in order to allow for large volume compactification, the D-branes eventually cannot be wrapped
around  untwisted 3- or 4-cycles of a compact
torus or of orbifolds, but one has to consider twisted, blowing-up cycles of an orbifold or more general Calabi-Yau spaces
with blowing-up cycles.
The reason for this is that wrapping the three or four
stacks of D-branes around internal cycles of a six-torus or untwisted orbifold cycles, the volumes of these
cycles involve the toroidal radii. Therefore these volumes cannot be kept small
while making the overall volume of the six-torus very big. Hence,
the SM D-branes  must be wrapped around small cycles inside a blown up orbifold
or a CY manifold. Other
cycles have to become large, in order to get a CY space with large volume and a low string scale
$M_{\rm string}$.

Let us here give some
short discussion on the volume  dependence of the gauge couplings in type IIA orientifolds..
The corresponding D6-brane gauge coupling constants
are proportional to the volumes of the wrapped 3-cycles, i.e.:
\begin{equation}
g^{-2}_{D6_a}=(2\pi)^{-1}\ \alpha'^{-2}\ {\rm Vol}(\Pi_a)\ .\end{equation}
The volume of the cycle $\Pi_a$ is given in terms of the associated complex 
structure moduli $U_a$ of a Calabi-Yau manifold $X$.
To accommodate type IIA orientifolds with low string scale and large overall volume,
the corresponding complex structure moduli $U^s_\beta$, around which the SM D6-branes are
wrapped, must be small
compared to the volume of $X$
to achieve finite values for the corresponding gauge coupling constants.
For this, the Calabi-Yau spaces $X$ must satisfy certain
restrictions for large volume
compactifications to be possible. In principle the structure of the allowed
IIA Calabi-Yau spaces can be
inferred from type IIB via mirror symmetry. E.g. one can wrap the D6-branes
around certain rigid
(twisted) 3-cycles of orbifold compactifications (see e.g. \cite{Blumenhagen:2005tn}), 
which can be kept small, whereas
the overall volume is made very large.

To perform the computation of the matter field scattering amplitudes, 
as in type IIB  we assume
that the 3-cycles, which are are wrapped by the SM D6-branes, 
are flat and have a kind of toroidal like intersection pattern. 
Specifically, we assume that the SM sector is wrapped around 3-cycles inside
a local $T^2\times T^2\times T^2$,
and the D6-brane wrappings around the tree 2-tori are described by wrapping numbers
$(n^i_a,m^i_a)$ $(i=1,2,3)$, where the lengths $L^i_a$ of the wrapped 
1-cycles in each $T^2$ is given by the following
equation:
\begin{equation}
L^i_a=\sqrt{(n^i_a)^2\ (R_i)^2+(m^i_a)^2\ (R_{i+1})^2}\ .\end{equation}
Then the gauge coupling on  a D$6$--brane which is wrapped around a
$3$--cycle, is:
\begin{equation}
g_{D6_{a}}^{-2}=(2\pi)^{-1}\ {\alpha'}^{-3/2}\ e^{-\phi_{10}}\
\prod_{i=1}^{3} L^i_a\ .\end{equation}
Here, the $3$--cycle $\Pi_a$ is assumed to be a direct product of
three $1$--cycles with wrapping numbers $(n^i,m^i)$  w.r.t. a pair of two internal
directions\footnote{In type IIB orientifolds, the gauge coupling of a D7-brane, 
wrapped around the 4-cycle $T^{2,j}\times T^{2,k}$ with wrapping numbers
$m^j$, $m^k$ and magnetic fluxes
$f^j$, $f^k$ is 
\begin{equation}
g^{-2}_{D7_i}=(2\pi)^{-1}\ \alpha'^{-2}\ |m^jm^k|\ {\rm Re}(T_j-f^jf^kS)\ .
\end{equation}
}
In terms of the corresponding three complex structure moduli $U_i$ of the $T^2$'s
this equation becomes
\begin{equation}
g_{D6_{a}}^{-2}=(2\pi)^{-1}e^{-\phi_4}\prod_{i=1}^3\ 
{|n^i_a-m^i_aU_i|\over\sqrt{{\rm Im}(U_i)}}\ .\end{equation}
Finally, the intersection angles of the D6-branes with the O6-planes along the
three $y_i$ directions can be expressed as
\begin{equation}
\tan(\theta^i_a)={m^i_aR_{i+1}\over n^i_aR_i}\ ,
\end{equation}
and the D6-brane intersection angles are simply given as $\theta_{ab}^i=
\theta_b^i-\theta_a^i$.
More details about the effective gauge couplings, 
and also about matter field metrics of these 
kind of intersecting D-brane models can be found in \cite{Lust:2004cx}.

\subsection{Production of mini black holes at the LHC}

One of the most exciting possibilities for the LHC is the discovery of small
higher-dimensional black holes that can be
formed when two sufficiently energetic particles collide 
\cite{Giddings:2001bu,Dimopoulos:2001hw,Meade:2007sz}.
This means that effects of higher dimensional quantum gravity can get  strong if the string scale
is low around the TeV scale, and if the volume of the extra dimensions is large.
The geometrical cross section for the production of mini black holes is of the order
\begin{equation}
\sigma(E)\sim{1\over M_{b.h.}^2}\Biggl({E\over M_{b.h.}}\Biggr)^\alpha\, ,
\end{equation}
where $M_{b.h.}$ is the black hole mass, i.e. the effective scale of quantum gravity, and $\alpha\leq 1$
for higher dimensional black holes. Since the production of mini black holes is basically
a non-perturbative effect, the black hole mass is suppressed by the string
coupling constant compared to the string scale:
\begin{equation}
M_{b.h.}\sim{M_{\rm string}\over g_{\rm string}}\,.
\end{equation}
Therefore, for weak string coupling, the onset for non-perturbative black hole production is higher than
for the production of perturbative Regge excitations, 
the threshold for an increase in the $2\rightarrow 2$ scattering cross section is almost inevitably lower
than the threshold for black hole production (see chapter 3.4).
 
\subsection{Production of (heavy) $Z'$ gauge bosons and mini-charged particles}

Another very interesting signal for new stringy physics at the LHC is the production
of heavy neutral $Z'$ gauge bosons (see e.g. \cite{Kiritsis:2002aj,Ghilencea:2002da}). These particles are quite generic in any string compactification,
and they receive their mass via a Green-Schwarz mixing with axionic scalar fields. E.g.
in the four stack D6-brane model with gauge group $U(3)\times U(2)\times U(1)\times U(1)$
three $U(1)$ gauge bosons will get a mass by the Green-Schwarz effect, and only the
hyper charge gauge field related to $U(1)_Y$ stays massless.

To understand the basis of the mechanism 
giving masses to the $U(1)$'s 
let us consider the following
Lagrangian coupling an Abelian gauge field $A_\mu$ to an antisymmetric
tensor $B_{\mu\nu}$:

\begin{equation}
\label{dualuno} 
{\cal L}\ =\ -\frac{1}{12} H^{\mu\nu\rho}
H_{\mu\nu\rho}-\frac{1}{4g^2} F^{\mu\nu} F_{\mu\nu} 
+ \frac{c}{4}\ \epsilon^{\mu\nu\rho\sigma} B_{\mu\nu}\ F_{\rho\sigma},
\end{equation}
where 
\begin{equation}
\label{definicion}
H_{\mu\nu\rho}=\partial_\mu B_{\nu\rho}+\partial_\rho B_{\mu\nu}
+\partial_\nu B_{\rho\mu}, \qquad  F_{\mu\nu}=\partial_\mu
A_\nu-\partial_\nu A_\mu
\end{equation}
and $g, c$
are arbitrary  constants. This corresponds to
the kinetic term for the fields $B_{\mu\nu}$ and $A_{\mu}$ together
with the $B\wedge F$ term. We will now proceed to dualize this
Lagrangian in two equivalent ways. First we can re-write it in terms of
the (arbitrary) field $H_{\mu\nu\rho}$ imposing the constraint $H=dB$ by the
standard introduction of a Lagrange multiplier field $\eta$ in the
following way:
\begin{equation}
\label{dualdos}
{\cal L}_0=\ -\frac{1}{12} H^{\mu\nu\rho}\
H_{\mu\nu\rho}-\frac{1}{4g^2} F^{\mu\nu}\ F_{\mu\nu} 
- \frac{c}{6}\ \epsilon^{\mu\nu\rho\sigma} H_{\mu\nu\rho}\ A_{\sigma} 
-\frac{c}{6}\eta \epsilon^{\mu\nu\rho\sigma} \partial_\mu H_{\nu\rho\sigma}.
\end{equation}
 Notice that integrating out $\eta$ implies $d^*H=0$ which in turn
implies that (locally) $H=dB$ and then we recover (\ref{dualuno}).  
Alternatively, integrating by parts the last term in (\ref{dualdos}) we
are left with a quadratic action for $H$ which we can solve
immediately to find
\begin{equation}
H^{\mu\nu\rho}= - {c}\ \epsilon^{\mu\nu\rho\sigma}
\left(A_\sigma+\partial_\sigma \eta\right).
\end{equation}
Inserting  this back into (\ref{dualdos}) we find:
\begin{equation}
{\cal L}_{A}\ =\ -\frac{1}{4g^2}\ F^{\mu\nu}\ F_{\mu\nu} -
\frac{c^2}{2} \left(A_\sigma+\partial_\sigma \eta\right)^2
\end{equation}
which is just a mass term for the gauge field $A_\mu$ after ``eating'' 
the scalar $\eta$ to acquire a mass $m^2=g^2 c^2$.
 Notice that this is similar to the St\"uckelberg mechanism
where we do not need a scalar field with a  vacuum expectation value
to give a mass to the gauge boson, nor do we have a massive Higgs-like
field at the end.

In intersecting D6-brane models with four stacks of D-branes, 
there are four 
 RR two-form  fields  $B_i$ with couplings
to  the $U(1)_{\alpha }$ field strengths:
\beq
\sum_i c_i^{\alpha}  \ B_i \wedge  tr (F^{\alpha }),\quad  \ i=1,2,3,4;\quad 
\alpha = a,b,c,d \ 
\label{gsuno}
\eeq
and in addition
there are  couplings of the Poincar\'e  dual scalars 
(representing the same degrees of freedom) $\eta_i$ 
of the $B_i$ fields:
\beq
\sum_i d_i^{\beta } \eta_i tr(F^{\beta }\wedge F^{\beta }),
\label{gsdos}   
\eeq
 where $F^{\beta }$ are the field strengths of any of the gauge groups.
The combination of both couplings,
by tree-level exchange of the RR-fields,  
 cancels the
mixed $U(1)_\alpha $ anomalies  $A_{\alpha \beta }$ 
with any other group $G_\beta $
as: 
\beq
A_{\alpha \beta }\ +\ \sum_i c_i^{\alpha} d_i^{\beta } \ =\ 0   \ .  
\label{gstres}
\eeq
The coefficient  $c_i^{\alpha }$ and $d_i^{\alpha}$ may be
computed  explicitly for each given D-brane configuration. 
Now,
after a duality transformation the $B\wedge F$ couplings turn into   
explicit mass terms for the Abelian gauge bosons given by
the expression:
\begin{equation} 
(M^2)_{\alpha\beta}= g_\alpha g_\beta M_{\rm string}^2 \sum_{i=1}^{3} c_i^\alpha
c_i^\beta,\qquad \alpha,\beta=a,b,c,d.
\label{masones}
\end{equation}
where the sum runs over the massive RR-fields present in the models
and where $g_\alpha$ is the coupling of $U(1)_\alpha$.
Here we have normalized to unity the gauge boson kinetic functions. 
We see that at weak coupling the masses of the $Z'$ gauge bosons are possibly even
lower that the string scale, such that they could be produced at the LHC. Another
effect of heavy $Z'$ gauge field is the contribution to the SM $\rho$-parameter.
Finally, anomalous $U1)'$ gauge bosons can also contribute the anomalous magnetic moment
of the muon.

Another interesting effect is the mixing of massless or very light
$Z'$ gauge bosons in the hidden sector with the standard
photon (resp. with the $U(1)_Y$ gauge bosons) by
their
kinetic energies at one loop string perturbation theory (see e.g. \cite{Abel:2008ai}).
This effect can by described by a mixing term
in the effective low energy action of the form
\begin{equation}
{\cal L_{\rm mix}}={\chi\over g_ag_b}F_{\mu\nu}^{(a)}F^{(b)\mu\nu}\, .
\end{equation}
This is nothing else than an off-diagonal 1-loop string threshold effect due to massive string excitations
which carry both electric and also $U(1)'$ gauge charges. 
If in addition, the hidden sector contains light hidden sector matter particles, which are charged under
$U(1)'$, then these particles also acquire a tiny electric charge of the order
\begin{equation}
Q_e^{(a)}=\chi g_b\, .
\end{equation}
Hence in a wide class of models one can experimentally look for
signatures of electrically
minicharged particles (MCPs) in high precision experiments.
This kind of non-accelerator experiments  could provide a very powerful test of the hidden sector in string compactifications.

\subsection{Four-point string scattering amplitudes -- production of heavy string Regge excitations and KK/winding states}

The production of string Regge excitations  will lead to new contributions to standard model
 scattering 
processes, like QCD jets or scattering of quarks
into leptons or gauge bosons, which can be measurable at LHC in case the
string scale is low \cite{AAB,Dudas,Cullen:2000ef,ChialvaGT,Anchordoqui:2008ac,Lust:2008qc,Anchordoqui:2008di,Anchordoqui:2009mm}.\footnote{For a recent
study on the effect of string Regge excitations at the LHC in warped compactifications see \cite{Hassanain:2009at}.}
Second there are the KK and winding excitations along the small internal
dimensions, i.e.
KK and winding excitations of the
SM fields. Their masses depend
on the internal volumes, 
and they should be also near the string scale $M_{\rm string}$.

For those amplitudes
involving four gauge bosons or two gauge bosons and two matter fermions, the 
amplitudes do not depend on the geometry of the underlying Calabi-Yau spaces \cite{Lust:2008qc}. This model
independence still also holds for the four--fermion matter amplitudes, but only
w.r.t. their dependence on the four-dimensional kinematical variables
$s,t,u$. On the other hand, the four--fermion amplitudes
do depend on the internal Calabi-Yau geometry and topology. Concretely, the
four--fermion amplitudes in general
depend on the Calabi-Yau intersection numbers, and also on the rational instanton numbers
of the Calabi-Yau space.
However, to perform the open string CFT computations for the scattering amplitudes 
of matter fields we shall assume that the SM D--branes are wrapped around
flat, toroidal like cycles. Therefore the four--fermion
amplitudes are functions of toroidal wrapping numbers. This sounds
in contradiction to what we have stated before about the large volume
compactifications. Hence, eventually switching from our toroidal-like 
results to more general Calabi-Yau
expressions, some of the factors, which depend on the toroidal geometry, have to be replaced
by geometrical or topological Calabi-Yau parameters.
However, the kinematical structure of the matter 
field amplitudes is universal and not affected
by the underlying Calabi-Yau geometry. At any rate, as we shall argue later, for the case that the longitudinal brane directions are somewhat
greater than the string
scale $M_{\rm string}$ the four--fermion couplings depend only on the local
structure of the brane intersections, but not on the global CY geometry.

The general structure of a four point amplitude of four open string states is as follows.
Let $\Phi^i$, $i=1,2,3,4$, represent gauge bosons, quarks of leptons of the standard model realized on three or more stacks of intersecting D-branes. The corresponding string vertex operators $V_{\Phi^i}$ are constructed from the fields of the underlying superconformal field theory (SCFT) and contain explicit (group-theoretical) Chan-Paton factors. In order to obtain the scattering amplitudes, the vertices are inserted at the boundary of a disk world-sheet, and the following SCFT correlation function is evaluated:
\begin{equation}\label{diskcor}
{\cal A}(\Phi^{1},\Phi^{2},\Phi^{3},\Phi^{4})=\sum_{\pi\in S_4/Z_2}
V_{CKG}^{-1}\int\limits_{{\cal I}_\pi}
\Biggl(\prod_{k=1}^4 dz_k\Biggr)\ \langle{V_{\Phi^{1}}(z_1)\ V_{\Phi^{2}}(z_2)\
V_{\Phi^{3}}(z_3)\ V_{\Phi^{4}}(z_4)}\rangle\ .
\end{equation}
Here, the sum runs over all six cyclic inequivalent orderings $\pi$ of the four vertex
operators along the boundary of the disk. Each permutation $\pi$ gives rise to
an integration region
${\cal I}_\pi=\{z\in R\ | z_{\pi(1)}<z_{\pi(2)}<z_{\pi(3)}<z_{\pi(4)}\}$. The group-theoretical factor is determined by the trace of the product of individual Chan-Paton factors,
ordered in the same way as the vertex positions. The disk boundary contains four segments which may be associated to as many as four different stacks of D-branes, since each vertex of a field originating from a D-brane intersection connects two stacks. Thus the Chan-Paton factor may actually contain as many a four traces, all in the fundamental representations of gauge groups associated to the respective stacks. However, purely partonic amplitudes for the scattering of quarks and gluons involve no more than three stacks.

In order to cancel the total background ghost charge of $-2$ on the disk,
the vertices in the
correlator (\ref{diskcor}) have to be chosen in the appropriate ghost picture and the picture ``numbers'' must add to $-2$.
Furthermore, in Eq.(\ref{diskcor}), the factor $V_{CKG}$
accounts for the volume of the
conformal Killing group of the disk after choosing the conformal gauge.
It will be canceled by fixing three vertex positions and introducing the
respective $c$--ghost correlator.
Because of the $PSL(2,R)$ invariance on the disk, we can fix three positions
of the vertex operators. Depending on the ordering ${\cal I}_\pi$
of the vertex operator positions we obtain six partial amplitudes.
The first set of three partial amplitudes may be obtained by the choice
\begin{equation}
z_1=0\ \ \ ,\ \ \ z_3=1\ \ \ ,\ \ \ z_4=\infty\ ,\end{equation}
while for the second set we choose:
\begin{equation}
z_1=1\ \ \ ,\ \ \ z_3=0\ \ \ ,\ \ \ z_4=\infty\ .\end{equation}
The two choices imply the ghost factor $\langle{c(z_1)c(z_2)c(z_3)}=z_{13}z_{14}z_{34}\rangle$.
The remaining vertex position $z_2$ takes arbitrary values along the
boundary of the disk. After performing all Wick contractions in eq.(\ref{diskcor}) the
correlators become basic, and 
generically for each partial amplitude the integral 
may be reduced to the Euler Beta function:
\begin{equation}
B(s,u)=\int_0^1 x^{s-1}\ (1-x)^{u-1}=\frac{\Gamma(s)\
\Gamma(u)}{\Gamma(s+u)}=\frac{1}{s}+\frac{1}{u}-{\pi^2\over 6}\ (s+u)+{\cal O}({\alpha'}^2)\ .
\end{equation}

Due to the extended nature of strings, the world--sheet string amplitudes
are generically non--trivial functions in $\ap$ in addition to the
usual dependence on the kinematic invariants and degrees of freedom of the
external states.
In the effective field theory description this $\ap$--dependence gives rise to a
series of infinite many resonance channels due to Regge excitations and/or
new contact interactions.
Generically, as we already saw, tree--level string amplitudes involving
four gluons or amplitudes with two gluons and two fermions
are described by the Euler Beta function depending on the kinematic
invariants $s=(k_1+k_2)^2,\ t=(k_1-k_3)^2,\ u=(k_1-k_4)^2$, with $s+t+u=0$ and
$k_i$ the four external momenta.
The whole amplitudes $A(k_1,k_2,k_3,k_4;\ap)$ may be understood as an
infinite sum over $s$--channel poles with intermediate string states
$|k;n\rangle$ exchanged, as it can be seen in the figure 6.
\begin{figure}
\begin{center}
  \includegraphics[width=0.5\textwidth]{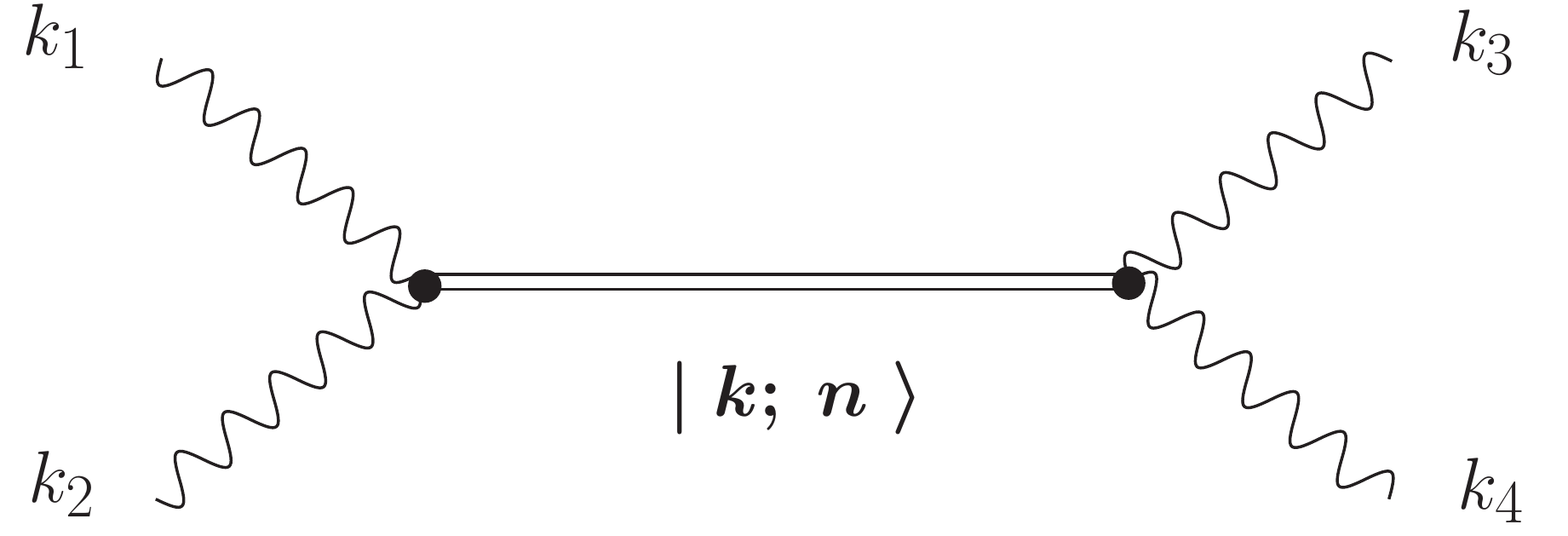}
\end{center}
\caption{Exchange of an infinite tower of Regge excitations in open string scattering processes.}
\end{figure}
After neglecting kinematical factors the string amplitude $A(k_1,k_2,k_3,k_4;\ap)$
assumes the form
\begin{equation}
\label{reggeexchange}
A(k_1,k_2,k_3,k_4;\ap)
\sim-\fc{\Gamma(-\ap s)\ \Gamma(1-\ap u)}{\Gamma(-\ap s-\ap u)}
=\sum_{n=0}^\infty\ \ \fc{\gamma(n)}{s-M_n^2}\end{equation}
as an infinite sum over $s$--channel poles at the masses
\begin{equation}
M_n^2=M_{\rm string}^2\ n
\end{equation}
of the string Regge excitations.
In eq.(\ref{reggeexchange}) the residues $\gamma(n)$  are determined by the three--point coupling of the intermediate
states $|k;n\rangle$ to the external particles and given by
\begin{equation}
\gamma(n)={t\over n!}\fc{\Gamma(-u\ap+ n)}{\Gamma(-u\ap)}=\fc{t}{n!}\ \prod\limits_{j=1}^n
[-u\alpha'-1+j]\sim (-\ap\ u)^n\ \ \ ,
\end{equation}
with $n+1$ being the highest possible
spin of the state $|k;n\rangle$.

Another way of looking at the expression (\ref{reggeexchange}) appears, when we express each
term in the sum as a power series expansion in $\ap$:
\begin{eqnarray}\label{newexp}
A(k_1,k_2,k_3,k_4;\ap)\ &\sim&\ \fc{t}{s}\ -\ {\pi^2\over 6}\ tu\ \ap^2+\ldots\ .\nonumber \\
& {~}&\hskip0.3cm\underbrace{\hskip0.5cm}_{n=0}\hskip0.4cm
\underbrace{\hskip2.5cm}_{n\neq 0}
\end{eqnarray}
In this form, the massless state $n=0$ gives rise to a field--theory
contribution ($\ap=0$), while at the order $\ap^2$
all massive states $n\neq 0$ sum up to a finite term.
The $n=0$ term in   (\ref{newexp}) describes the field--theory contribution to
the scattering diagram, e.g. the exchange of a massless gluon.
On the other hand, the term at the order $\ap^2$ describes a new string contact
interaction as a result of summing up all heavy string states.
E.g. for a four gluon superstring amplitude the first string contact interaction is
given by  ${\alpha'}^2\ g_{Dp}^{-2}\ {\rm t}r F^4$, which represent a correction to YM theory, as shown
in figure 7.
\begin{figure}
\begin{center}
  \includegraphics[width=0.5\textwidth]{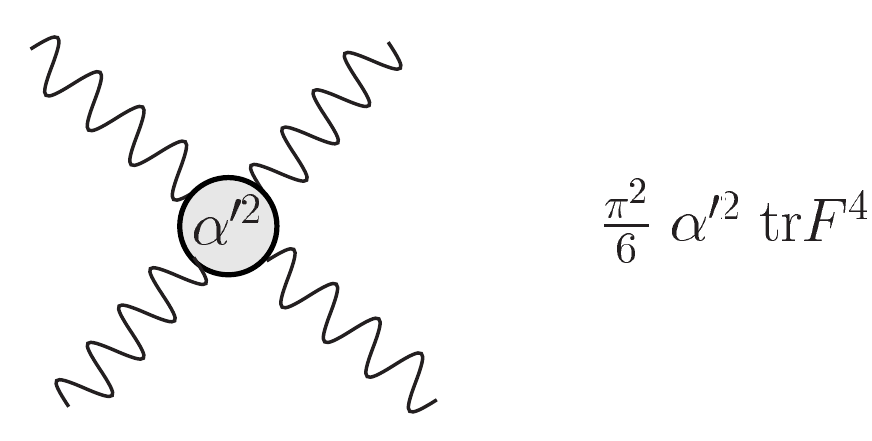}
\end{center}
\caption{Order $(\alpha')^2$ contact interaction in the scattering of four open string gluons.}
\end{figure}

\subsubsection{Four gluon scattering amplitude}

Let us start with the open string tree level scattering of four gauge bosons on the disk.
The gauge bosons are open strings with ends on same brane, for gluons say the QCD stack $a$.
The gauge boson vertex operator in the $(-1)$-ghost picture reads
\begin{equation}\hskip -3.7cm
V_{A^a}^{(-1)}(z,\xi,k) ~=~ g_{A} [T^a]^{\alpha_1}_{\alpha_2}\ e^{-\phi(z)}\ \xi^\mu\ \psi_\mu(z)\
e^{ik_\rho X^\rho(z)}\ ,
\end{equation}
while in the zero--ghost picture we have:
\begin{equation}
V_{A^a}^{(0)}(z,\xi,k)=\fc{g_{A}}{(2\ap)^{1/2}} [T^a]^{\alpha_1}_{\alpha_2} \xi_\mu\
[\ i\partial X^\mu(z)+2\ap\ (k\psi)\ \psi^\mu(z)\ ]\ e^{ik_\rho X^\rho(z)}\ .\end{equation}
where $\xi^\mu$ is the polarization vector. The vertex must be inserted on the segment of disk boundary on stack $a$, with the indices $\alpha_1$ and $\alpha_2$ describing the two string ends.

Four-gluon amplitudes have been known for many years. 
The corresponding string disk  diagram is shown in the figure 8.
\begin{figure}
\begin{center}
  \includegraphics[width=0.4\textwidth]{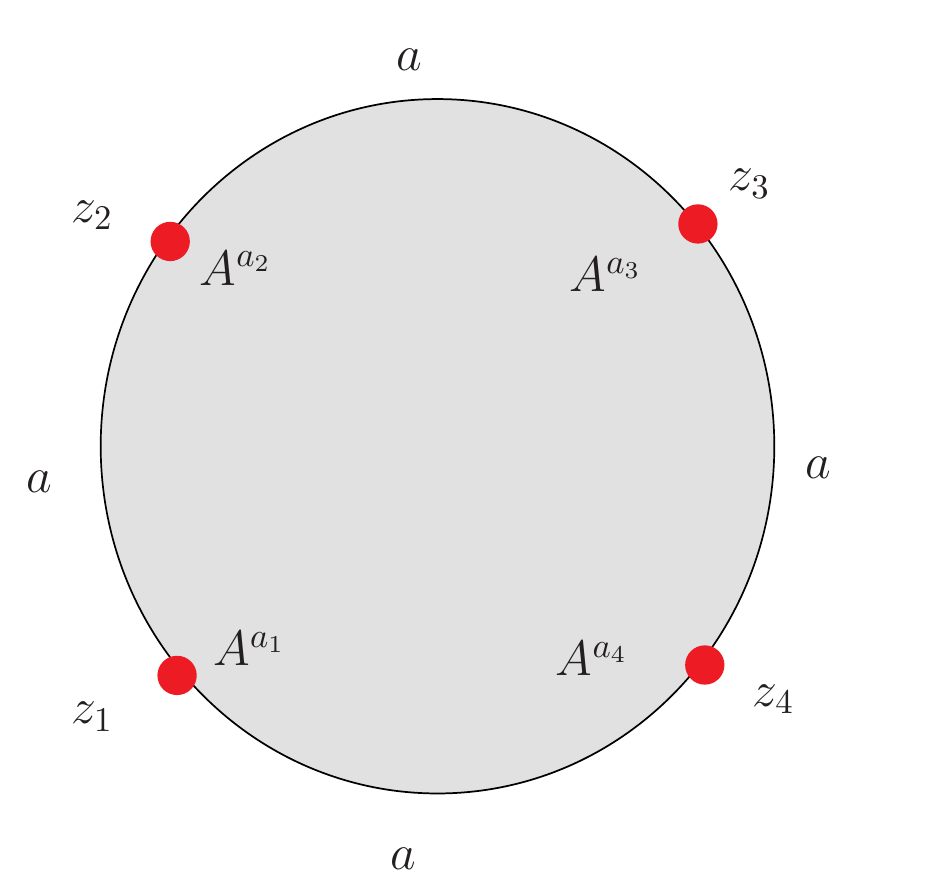}
\end{center}
\caption{The four gauge boson disk diagram.}
\end{figure}
The complete amplitude can be generated from the maximally helicity violating MHV amplitudes 
\cite{Stieberger:2006bh,Stieberger:2006te}.
Averaging over helicities and colors of the incident partons and summed over helicities and colors of the outgoing particles, we obtain for gluon scattering SU(3)
\begin{equation}
|{\cal M}(gg\to
 gg)|^2~=\Big({1\over s^2}+{1\over t^2}+{1\over u^2}\Big)
\bigg[\,{9\over 4}\!\left(\, s^2V^2_s+ t^2V^2_t
+u^2V^2_u\,\right) - {1\over 3}\!\left(\, sV_s
+ tV_t+uV_u\,\right)^{2}\bigg]   
 \end{equation}
 
 In the D-brane models under consideration,
the ordinary $SU(3)$ color gauge symmetry is extended to $U(3)$, so
that the open strings terminating on the stack of ``color'' branes
contain an additional $U(1)$ gauge boson $C$.
  Replacing one gluon by the $U(1)$  color singlet gauge boson component  $C$ in the QCD stack $a$,
  there is also a non-vanishing string amplitude with one photon or one $Z$-boson ($C=\gamma,Z$),
  since the photon or the Z-boson always has an admixture of this $U(1)$ gauge group:
 
 \begin{equation}
 |{\cal M}(gg\to
 gC)|^2~=
 {5\over 6}Q_C^2\Big({1\over s^2}+{1\over t^2}+{1\over
u^2}\Big)\left(\, sV_s+ tV_t+uV_u\,\right)^{2}  
\end{equation}
 
 In the zero-slop field theory limit $\alpha'\rightarrow 0$ the functions $V_s,V_t,V_u\rightarrow 1$,
 and the four gauge boson amplitudes get contributions only from the exchange of SM
 fields. In this limit the string amplitudes approach the known results true in the SM
 Note that the ${\cal M}(gg\to
 gC)\rightarrow 0$, as required in the tree level SM.
 Note that the four gauge boson amplitude is completely model independent,
 there are no KK-particles being exchanged in the s-channel.

\subsubsection{Two gluon, two quark scattering amplitudes}

We now consider the following correlation function between two gauge bosons and two matter fermions:
\begin{equation}
\langle{V_{A^{x}}^{(0)}(z_1,\xi_1,k_1)\ V_{A^{y}}^{(-1)}(z_2,\xi_2,k_2)\
V^{(-1/2)}_{\psi^{\alpha_3}_{\beta_3}}(z_3,u_3,k_3)
\ V^{(-1/2)}_{\bar\psi^{\beta_4}_{\alpha_4} }(z_4,\bar u_4,k_4)}\rangle\ .
\end{equation}
The fermion vertex operators are boundary changing operators, being inserted at the intersection of brane stack $a$ and $b$.
Specifically, 
the chiral fermion vertex operators of the quarks and leptons are:
\begin{eqnarray}
V^{(-1/2)}_{\psi^{\alpha}_{\beta}}(z,u,k)&=&g_\psi [T^{\alpha}_{\beta}]_{\alpha_1}^{\beta_1}e^{-\phi(z)/2}\ u^{\lambda}
S_{\lambda}(z)\ \Xi^{\scriptscriptstyle a\cap b}(z)\ e^{ik_\rho X^\rho(z)}\ ,\nonumber \\
V^{(-1/2)}_{\bar\psi^{\beta}_{\alpha}}(z,\bar u,k)&=&g_\psi [T_{\alpha}^{\beta}]^{\alpha_1}_{\beta_1}
e^{-\phi(z)/2}\ \bar u_{\dot\lambda}
S^{\dot\lambda}(z)\ \overline\Xi^{\scriptscriptstyle a\cap b}(z)\ e^{ik_\rho X^\rho(z)}\ .
\end{eqnarray}
These vertices connect two segments of disk boundary, associated to stacks $a$ and $b$, with the indices $\alpha_1$ and $\beta_1$ representing the string ends on the respective stacks.
The internal field $\Xi^{\scriptscriptstyle a\cap b}$ of conformal dimension $3/8$
is the fermionic boundary
changing operator.
In the intersecting D-brane models, the intersections are characterized by angles $\theta_{ba}$. Then
$\Xi^{\scriptscriptstyle a\cap b}$ can be expressed
in terms of bosonic and fermionic twist fields $\sigma$ and $s$:
\begin{equation}
\Xi^{\scriptscriptstyle a\cap b}=\prod_{j=1}^3\sigma_{\theta^j_{ba}}\ s_{\theta^j_{ba}}\ \ \ ,\ \ \
\overline\Xi^{\scriptscriptstyle a\cap b}=\prod_{j=1}^3\sigma_{-\theta_{ba}^j}\ s_{-\theta_{ba}^j}\ .
\end{equation}
The spin fields
\begin{equation}
s_{\theta^j}=e^{i(\theta^j-\h)H^j}\ \ \ ,\ \ \
s_{-\theta^j}=e^{-i(\theta^j-\h)H^j}
\end{equation}
have conformal dimension $h_s=\h(\theta^j-\h)^2$ and twist the internal part
of the Ramond ground state spinor.
The field $\sigma_{\theta}$ has conformal dimension
$h_\sigma=\h\theta^j(1-\theta^j)$
and produces discontinuities in the boundary conditions of the
internal complex bosonic Neveu--Schwarz coordinates $Z^j$.

The fact that fermions originate from the same pair of stacks, say $a$ and $b$
is forced
upon us by the conservation of twist charges, in a similar way as their
opposite helicities are
forced  by the internal charge conservation.
It follows that both gauge bosons must be associated either to one of these stacks, say
$(x,y)=(a_1,a_2)$, or one of them is associated to $a$ while the other to $b$,
say $(x,y)=(a,b)$. The corresponding disk diagrams are shown in the figure 9.
\begin{figure}
\begin{center}
  \includegraphics[width=0.6\textwidth]{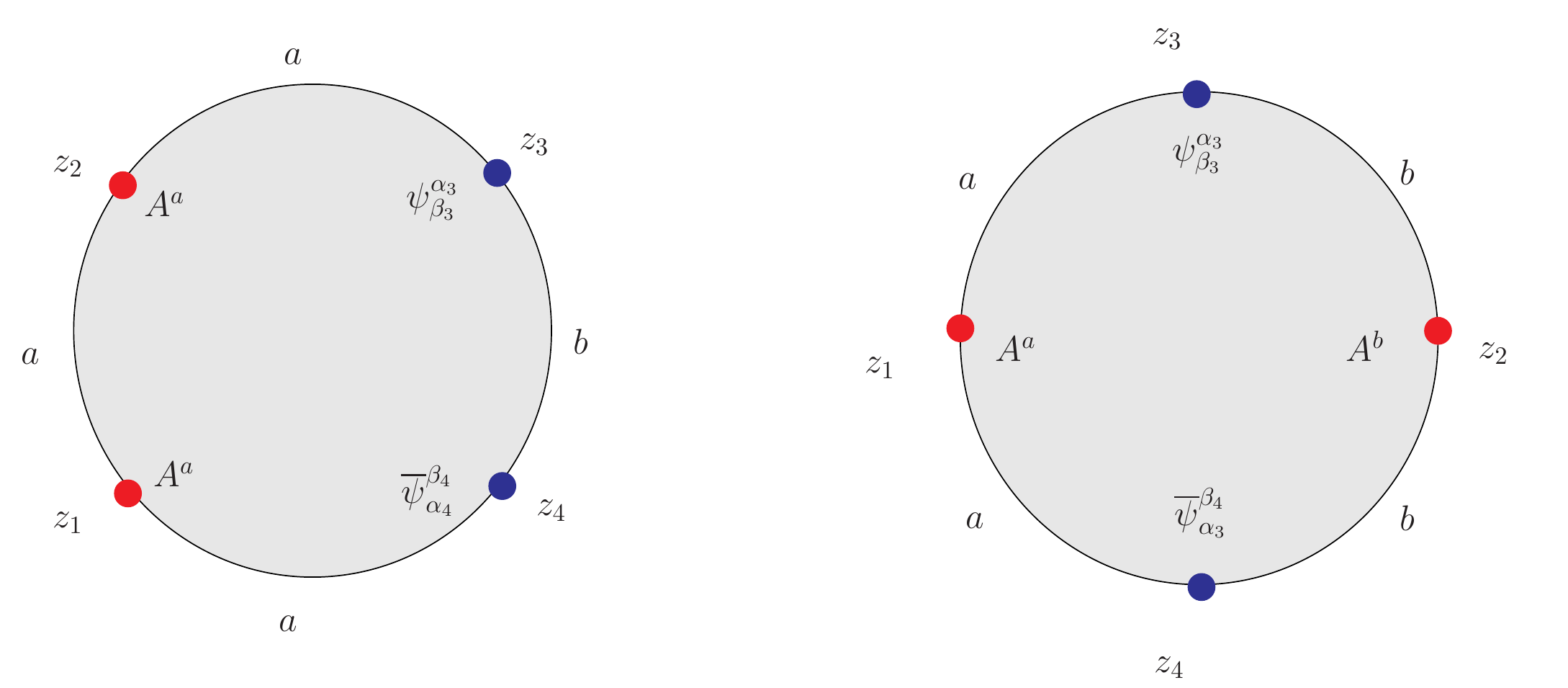}
\end{center}
\caption{The two gauge boson - two fermion disk diagrams.}
\end{figure}
Using these informations one obtains the following tree level (squared) amplitudes for
two gauge boson, two fermion scattering processes \cite{Lust:2008qc},
\begin{equation}
|{\cal M}(gg\to
 q\bar q)|^2~=
{t^2+u^2\over s^2}\bigg[\, {1\over 6}{1\over ut}(tV_t+uV_u)^2
-{3\over 8}V_tV_u\bigg] \end{equation}
and
\begin{equation}
|{\cal M}(gq\to
 gq)|^2~=
{s^2+u^2\over t^2}\bigg[ V_sV_u-{4\over 9}{1\over su}(sV_s+uV_u)^2
\bigg]  
\end{equation}
Again, in the s-channel there can be only the exchange of heavy Regge states and no KK-states.
Hence also these two amplitudes are completely independent from the internal geometry.

\subsubsection{Four quark scattering amplitudes}

In the 
most general case, all fermions are at different intersections, as seen in figure 10.
\begin{figure}
\begin{center}
  \includegraphics[width=0.6\textwidth]{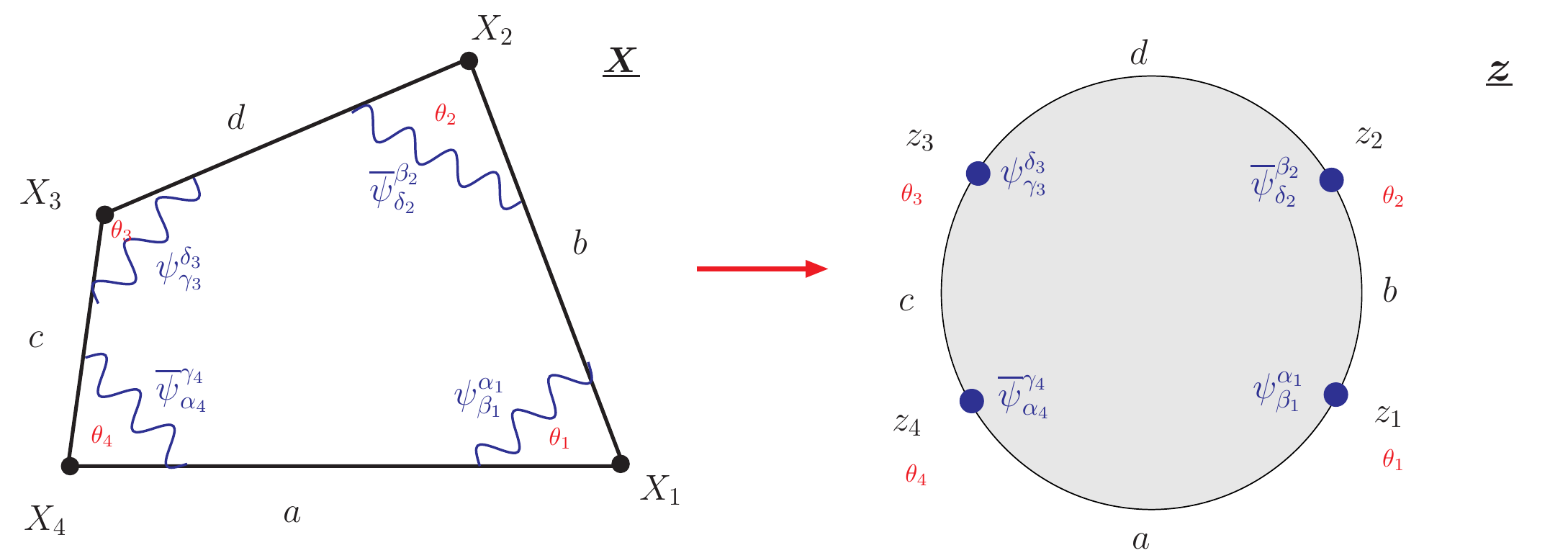}
\end{center}
\caption{The SM by four stacks of intersecting branes and the corresponding  four fermion disk diagram.}
\end{figure}
Then, without going into further details the corresponding tree level four-fermion scattering amplitudes
take the following form \cite{Lust:2008qc}\footnote{Four fermion amplitudes in intersecting
brane models in the context of proton decay, FCNC currents and Yukawa couolings were
also computed in     
\cite{Cvetic:2003ch,Klebanov:2003my,Abel:2003yx,Abel:2003yh,Cvetic:2006iz}.}:
\begin{eqnarray}
|{\cal M}(qq\to
 qq)|^2~&=&
{2\over 9}{1\over t^2}\Big[\big(sF^{bb}_{tu}\big)^2  +\big(sF^{cc}_{tu}\big)^2 +\big(uG^{bc}_{ts}\big)^2  +\big(uG^{cb}_{ts}\big)^2     \Big]  \nonumber \\&+&
{2\over 9}{1\over u^2}\Big[\big(sF^{bb}_{ut}\big)^2  +\big(sF^{cc}_{ut}\big)^2 +\big(tG^{bc}_{us}\big)^2  +\big(tG^{cb}_{us}\big)^2\Big]\nonumber \\
  &-&{4\over 27} {s^2\over tu}\big(
F^{bb}_{tu} F^{bb}_{ut}+F^{cc}_{tu} F^{cc}_{ut}\big) 
\end{eqnarray}
Here the functions $F$ and $G$ depend on $\alpha$ and now also on the masses
of the internal KK (and winding) states. Therefore these amplitudes depend on the
internal geometry and are not anymore model independent. Furthermore, due
to the quantum numbers of the fermions there are no s-channel poles, 
but in the t,u-channels there is the exchange of
Regge, KK and winding modes, as it can be seen by appropriate expansions
of the functions $F$ and $G$.  

\subsubsection{Dijet signals for lowest mass strings at the LHC}

In this section we will determine the contribution from the exchange of excited, heavy
Regge states to dijet processes at the LHC \cite{Anchordoqui:2008di}.\footnote{A recent update and possible signatures from
Kaluza-Klein particles was presented in \cite{Anchordoqui:2009mm}.}
The first Regge
excitations of the gluon $(g)$ and quarks $(q)$ will be denoted by
$g^*,\ q^*$, respectively. The first excitation of the
$C$ will be denoted by $C^*$.
In the following we isolate the contribution to the partonic cross
section from the first resonant state. Note that far below the
string threshold, at partonic center of mass energies $\sqrt{s}\ll
M_s$, the form factor $V(s,t,u)\approx 1-\frac{\pi^2}{6}{su}/M^4_s$
and therefore the contributions of Regge
excitations are strongly suppressed. The $s$-channel pole terms of
the average square amplitudes contributing to dijet production at
the LHC can be obtained from the general formulae given in
in the previous subsection. However, for
phenomenological purposes, the poles need to be softened to a
Breit-Wigner form by obtaining and utilizing the correct {\em
  total} widths of the resonances~\cite{Anchordoqui:2008hi}. After
this is done, the contributions of the various channels are as
follows:
\begin{eqnarray}
|{\cal M} (gg \to gg)| ^2 & = & \frac{19}{12} \
\frac{g^4}{M_s^4} \left\{ W_{g^*}^{gg \to gg} \, \left[\frac{M_s^8}{(  s-M_s^2)^2
+ (\Gamma_{g^*}^{J=0}\ M_s)^2} \right. \right.
\left. +\frac{  t^4+   u^4}{(  s-M_s^2)^2 + (\Gamma_{g^*}^{J=2}\ M_s)^2}\right] \nonumber \\
   & + &
W_{C^*}^{gg \to gg} \, \left. \left[\frac{M_s^8}{(  s-M_s^2)^2 + (\Gamma_{C^*}^{J=0}\ M_s)^2} \right.
\left. +\frac{  t^4+  u^4}{(  s-M_s^2)^2 + (\Gamma_{C^*}^{J=2}\ M_s)^2}\right] \right\},
\label{gggg2}
\end{eqnarray}
\begin{eqnarray}
|{\cal M} (gg \to q \bar q)|^2 & = & \frac{7}{24} \frac{g^4}{M_s^4}\ N_f\
\left [W_{g^*}^{gg \to q \bar q}\, \frac{  u   t(   u^2+   t^2)}{(  s-M_s^2)^2 + (\Gamma_{g^*}^{J=2}\ M_s)^2} \right. \nonumber \\
 & + &  \left. W_{C^*}^{gg \to q \bar q}\, \frac{  u   t (   u^2+   t^2)}{(  s-M_s^2)^2 +
(\Gamma_{C^*}^{J=2}\ M_s)^2} \right]
\end{eqnarray}
\begin{eqnarray}
|{\cal M} (q \bar q \to gg)|^2  & = &  \frac{56}{27} \frac{g^4}{M_s^4}\
\left[ W_{g^*}^{q\bar q \to gg} \,  \frac{  u   t(   u^2+   t^2)}{(  s-M_s^2)^2 + (\Gamma_{g^*}^{J=2}\ M_s)^2} \right. \nonumber \\
 & + & \left.  W_{C^*}^{q\bar q \to gg} \, \frac{  u   t(   u^2+   t^2)}{(  s-M_s^2)^2 + (\Gamma_{C^*}^{J=2}\ M_s)^2} \right] \,\,,
\end{eqnarray}
\begin{equation}
|{\cal M}(qg \to qg)|^2  =  - \frac{4}{9} \frac{g^4}{M_s^2}\
\left[ \frac{M_s^4   u}{(  s-M_s^2)^2 + (\Gamma_{q^*}^{J=1/2}\ M_s)^2} + \frac{u^3}{(s-M_s^2)^2 + (\Gamma_{q^*}^{J=3/2}\ M_s)^2}\right],
\label{qgqg2}
\end{equation}
where $g$ is the QCD coupling constant $(\alpha_{\rm QCD}=\frac{g^2}{4\pi}\approx 0.1)$
 and $\Gamma_{g^*}^{J=0} = 75\, (M_s/{\rm TeV})~{\rm GeV}$,
$\Gamma_{C^*}^{J=0} = 150 \, (M_s/{\rm TeV})~{\rm GeV}$,
$\Gamma_{g^*}^{J=2} = 45 \, (M_s/{\rm TeV})~{\rm GeV}$,
$\Gamma_{C^*}^{J=2} = 75 \, (M_s/{\rm TeV})~{\rm GeV}$,
$\Gamma_{q^*}^{J=1/2} = \Gamma_{q^*}^{J=3/2} = 37\, (M_s/{\rm
  TeV})~{\rm GeV}$ are the total decay widths for intermediate states
$g^*$, $C^*$, and $q^*$ (with angular momentum
$J$)~\cite{Anchordoqui:2008hi}. The associated weights of these
intermediate states are given in terms of the probabilities for the
various entrance and exit channels
\begin{equation}
W_{g^*}^{gg \to gg} = \frac{(\Gamma_{g^* \to gg})^2}{(\Gamma_{g^* \to gg})^2 +
(\Gamma_{C^* \to gg})^2} = 0.09 \,,
\label{w1}
\end{equation}
\begin{equation}
W_{C^*}^{gg \to gg} = \frac{(\Gamma_{C^*
  \to gg})^2}{(\Gamma_{g^* \to gg})^2 + (\Gamma_{C^* \to gg})^2} =
0.91 \, ,
\label{w2}
\end{equation}
\begin{equation}
W_{g^*}^{gg \to q \bar q}  = W_{g^*}^{q \bar q \to gg} =
\frac{\Gamma_{g^* \to gg} \,
\Gamma_{g^* \to q \bar q}} {\Gamma_{g^* \to gg} \,
\Gamma_{g^* \to q \bar q} + \Gamma_{C^* \to gg} \,
\Gamma_{C^* \to q \bar q}} = 0.24 \, ,
\label{w3}
\end{equation}
\begin{equation}
W_{C^*}^{gg \to q \bar q} = W_{C^*}^{q \bar q \to gg}  =
\frac{\Gamma_{C^* \to gg} \,
\Gamma_{C^* \to q \bar q}}{\Gamma_{g^* \to gg} \,
\Gamma_{g^* \to q \bar q} + \Gamma_{C^* \to gg} \,
\Gamma_{C^* \to q \bar q}} = 0.76 \, .
\label{w4}
\end{equation}
Superscripts $J=2$ are understood to be inserted on all the $\Gamma$'s in
Eqs.(\ref{w1}), (\ref{w2}), (\ref{w3}), (\ref{w4}).
Equation~(\ref{gggg2}) reflects the fact that weights for $J=0$ and
$J=2$ are the same~\cite{Anchordoqui:2008hi}. In what follows we set
the number of flavors $N_f =6$.

\begin{figure}
\begin{center}
  \includegraphics[width=0.55\textwidth]{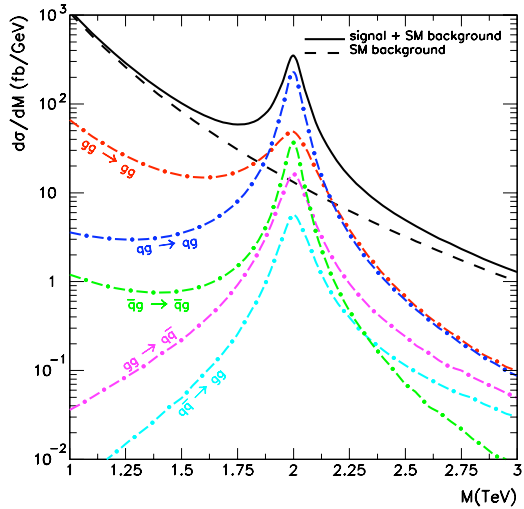}
\end{center}
\caption{Dijet cross sections.}
\end{figure}

\begin{figure}
\begin{center}
  \includegraphics[width=0.5\textwidth]{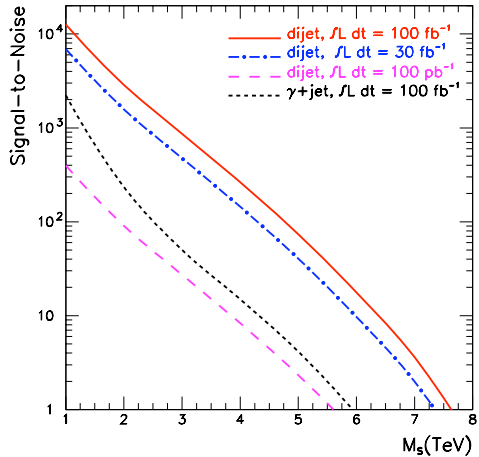}
\end{center}
\caption{Signal to noise ratio.}
\end{figure}

In figure 11 we show a
representative plot of the invariant mass spectrum, for $M_s =2$~TeV,
detailing the contribution of each subprocess.  The QCD background has
been calculated at the partonic level from the same processes as
designated for the signal, with the addition of $qq\to qq$ and $q \bar
q \to q \bar q$.  Our calculation, making use of the CTEQ6D parton
distribution functions~\cite{Pumplin:2002vw} agrees with that
presented in~\cite{Bhatti:2008hz}.
Finally we estimate (at the parton level) the LHC discovery
reach, namely one may calculate a signal-to-noise ratio, with the signal
rate estimated in the invariant mass window $[M_s - 2 \Gamma, \, M_s +
2 \Gamma]$. The noise is defined as the square root of the number of
background events in the same dijet mass interval for the same
integrated luminosity.
The top two and bottom curves in figure 12 show the behavior
of the signal-to-noise (S/N) ratio as a function of the string scale
for three integrated luminosities (100~fb$^{-1},$ 30~fb$^{-1}$ and
100~pb$^{-1}$) at the LHC. It is remarkable that within 1-2 years of
data collection, {\it string scales as large as 6.8 TeV are open to
 discovery at the $\geq 5\sigma$ level.} For 30~fb$^{-1},$ the
presence of a resonant state with mass as large as 5.7 TeV can provide a
signal of convincing significance $(S/N\geq13)$. The bottom curve in figure 12,
corresponding to data collected in a very early run of 100~pb$^{-1},$
shows that a resonant mass as large as 4.0~TeV can be observed with
$10\sigma$ significance! Once more, we stress that these results
contain no unknown parameters. They depend only on the D-brane
construct for the standard model, and {\it are independent of
  compactification details.}

\section{Flux compactifications, moduli stabilization and the cosmological constant}

In this section we  discuss a few aspects about the moduli 
stabilization process due to background fluxes  (see ref. \cite{Grana:2005jc,Denef:2007pq} for reviews on flux compactifications) and
non-perturbative effects.
The number of flux vacua
on a given CY background space is very huge 
\cite{Bousso:2000xa,Douglas:2003um,Ashok:2003gk,Denef:2004ze}:
${\cal N}_{vac}\sim 10^{500}$.
This number arises by counting all possible flux combinations that
are constrained by satisfying similar tadpole conditions as the D-branes discussed before.
Again one can try to make some interesting statistical predictions within the
flux landscape, like the question what is the likelihood for obtaining a tiny cosmological
constant, or if supersymmetry is broken at high or low energy scales 
\cite{Dine:2004is,ArkaniHamed:2004fb,Denef:2004cf,Arkani-Hamed:2005yv}.
Of course, in order to make more concrete predictions in the string landscape, the flux vacua statistics
must be eventually combined  with the D-brane
statistics, described before. Since the D-branes of SM contribute to the tadpole by a certain amount,
the possibilities for introducing backgrounds are limited in the presence of D-branes.
Therefore, if we assume that the SM is present, the number of fluxes is most likely much lower than
the ${\cal N}_{vac}\sim 10^{500}$,  quoted above. This reduction in flux possibilities should
be taken into account when making statistic statements about flux vacua, in particular
in connection about the likelihood to obtain a small cosmological constant.

Flux vacua constitute a very  interesting region in the string landscape by the following
arguments:

\begin{itemize}
\item Background fluxes can stabilize moduli. In the description of an effective
action, background fluxes generically create a potential for the moduli fields, which leads
to a set of discrete vacua. This discretuum of vacua is often called the string landscape,
although we have introduced the string landscape os the space of all
consistent string solutions. In several cases, the background fluxes do not stabilize
all moduli, but additional non-perturbative effects are needed to obtain a discrete
landscape of solutions.
\item
Since many low energy couplings of string compactifications, like gauge couplings or
Yukawa couplings are moduli dependent functions, moduli stabilization by fluxes opens
at least in principle the possibility to compute these couplings in the string landscape.
This is relevant for eventually making contact between the string landscape and
the parameters in particle physics, e.g. in the SM.

\item
The discrete flux vacua (plus possible non-perturbative effects) are a good starting point
for string cosmology. Often one needs additional ingredients, like uplift from discrete anti-de Sitter
vacua with a negative cosmological constant to de Sitter vacua with a positive cosmological constant.
This is also necessary for string inflationary models, where one of the scalar fields plays
the role of the inflaton field after moduli stabilization and must have a rather flat, positive
effective potential (see section 5).
\end{itemize}

In this section we will describe some aspects about moduli stabilization in string theory.
Massless moduli occur typically in many geometric string compactifications as the parameters,
which describe the size and the shape of the internal geometry, as well as the positions of
the D-branes in the compact space. In string theoretical language these parameters
correspond to marginal conformal fields of conformal dimension $(h,\bar h)=1,1$. Turning on discrete
fluxes the moduli fields become massive and disappear as deformation parameters of the
underlying conformal field theory. This effect can be described in the effective, 4-dimensional supergravity
description by an effective potential, which fixes the vacuum expectation values of the moduli
fields to discrete values and giving them at the same time  non-vanishing masses. 
The associated vacuum energy can be negative ($AdS_4$ vacua), or positive
($dS_4$ vacua) or also zero (Minkowski $R^{3,1}$ vacua). In the following we will discuss
how moduli stabilization occurs in the effective field theory. We will use
the effective superpotential approach (F-terms), neglecting contributions to
the scalar potential from D-terms.

After discussing some general and also more mathematical aspects of flux
compactifications (see e.g.\cite{Grana:2005jc} for a comprehensive review),
 we will describe type IIB vacua
with fluxes and also possible non-perturbative superpotentials. Then we will focus on
$AdS_4$ vacua in type IIA orientifold compactifications.

\subsection{General and mathematical aspects of flux compactifications}

The bosonic content of type II supergravity consists of a metric $g$, a dilaton $\Phi$, an NSNS 3-form flux field $H$ and RR n-form flux fields
$F_{n}$. In the democratic formalism, where the number
 of  RR-fields is doubled,  $n$ runs over $0,2,4,6,8,10$ in IIA and over $1,3,5,7,9$ in type IIB.
We write $n$ to denote
the dimension of the RR-fields; for example $(-1)^n$ stands for $+1$ in type IIA and $-1$ in type IIB.
After deriving the equations of motion from the action, the redundant RR-fields are to be removed by hand
by means of the duality condition:
\begin{equation}{
\label{Fduality}
F_{n} = (-1)^{\frac{(n-1)(n-2)}{2}} e^{\frac{n-5}{2}\Phi}\star_{10} F_{(10-n)} ~,
}
\end{equation}
given here in the Einstein frame.
We will collectively denote the RR-fields, and the corresponding potentials, with polyforms
$F=\sum_n F_{n}$ and $C=\sum_n C_{(n-1)}$, so that: $F=\d_H C$.

In the Einstein frame, the bosonic part of the bulk action  reads:
\begin{equation}{
S_{\text{bulk}} = \frac{1}{2 \kappa_{10}^2} \int \d^{10} x \sqrt{-g} \left[ R -\frac{1}{2} (\partial \Phi)^2 - \frac{1}{2}
e^{-\Phi}H^2  -\frac{1}{4} \sum_n e^{\frac{5-n}{2}\Phi} F_{n}^2 \right] \, ,
}
\end{equation}
where for an $l$-form $A$ we define
\begin{equation}{
A^2 = A \cdot A = \frac{1}{l!} \, A_{M_1 \ldots M_l} A_{N_1 \ldots N_l} g^{M_1N_1} \cdots g^{M_lN_l} \, .
}
\end{equation}
Since (\ref{Fduality}) needs to be imposed by hand this is strictly-speaking only a pseudo-action.
Note that the doubling of the RR-fields leads to factors of $1/4$ in their kinetic terms.

The contribution from the calibrated (supersymmetric) brane sources \cite{Koerber:2005qi,Martucci:2005ht}
can be written as:
\begin{equation}{\label{dbia}
S_{\text{source}} = \int \langle C, j \rangle
-\sum_n e^{\frac{n}{4}\Phi} \int\langle  \Psi_n,j\rangle \, ,
}
\end{equation}
with
\begin{equation}{\label{dbib}
\Psi_n = e^A \d t \wedge \frac{e^{-\Phi}}{(n-1)! \hat{\epsilon}_1{}^T \epsilon_1} \hat\epsilon_1{}^T \gamma_{M_1 \ldots M_{n-1}} \hat\epsilon_2 \, \d X^{M_1} \wedge \ldots \wedge \d X^{M_{n-1}} \, ,
}
\end{equation}
with $\hat{\epsilon}_{1,2}$ nine-dimensional internal supersymmetry generators,
and with the Mukai pairing $\langle \cdot, \cdot \rangle$  given by
\begin{equation}{
\label{mukai}
\langle \phi_1, \phi_2 \rangle = \phi_1 \wedge \alpha(\phi_2)|_{\text{top}} \, .
}
\end{equation}

The dilaton equation of motion and the Einstein equation read
\begin{eqnarray}
\label{dilaton}
0&=& \nabla^2 \Phi +\frac{1}{2}e^{-\Phi} H^2 - \frac{1}{8} \sum_n (5-n) e^{\frac{5-n}{2}\Phi}F_{n}^2
+\frac{\kappa_{10}^2}{2} \sum_n (n-4)e^{\frac{n}{4}\Phi}  \star \! \langle \Psi_n,j\rangle
\, , \\
\label{einstein}
0&= &R_{MN} + g_{MN} \left( \frac{1}{8}e^{-\Phi} H^2 + \frac{1}{32} \sum_n (n-1)e^{\frac{5-n}{2}\Phi}F_{n}^2\right)\\
 & &~~~~~~~~~~~-\frac{1}{2}\partial_M \Phi \partial_N \Phi - \frac{1}{2}e^{-\Phi} H_{M} \cdot H_{N}
-\frac{1}{4} \sum_n e^{\frac{5-n}{2}\Phi}F_{n\,M} \cdot F_{n\, N} \nonumber\\
 & &~~~~~~~~~~~- 2 \kappa_{10}^2 \sum_n e^{\frac{n}{4}\Phi}\star \! \langle  \left(-\frac{1}{16} n g_{MN} + \frac{1}{2} g_{P(M} dx^P \otimes \iota_{N)}\right)\Psi_n,j\rangle\nonumber \, ,
 \, ,
\end{eqnarray}
where we defined for an $l$-form $A$
\begin{equation}{
A_{M} \cdot A_{N} = \frac{1}{(l-1)!} A_{MM_2 \ldots M_l} A_{NN_2 \ldots N_l} g^{M_2N_2} \cdots g^{M_lN_l} \, .
}
\end{equation}
The Bianchi identities and the equations of motion
for the RR-fields, including  the contribution from the `Chern-Simons' terms of the sources,
take the form
\begin{eqnarray}
\label{eomBian}
\label{Bianchis}
0&= &\d F+H\wedge F  + 2 \kappa_{10}^2 \,  j\, , \\
\label{eomF}
0&= &\d\left( e^{\frac{5-n}{2}\Phi}\star F_{n} \right) -e^{\frac{3-n}{2}\Phi}H\wedge  \star F_{(n+2)}
-2 \kappa_{10}^2 \, \alpha(j)
\, .
\end{eqnarray}
Finally, for the equation of motion for $H$ we have:
\begin{equation}{
\label{eomH}
0= \d (e^{-\Phi} \star \! H)
- \frac{1}{2} \sum_n e^{\frac{5-n}{2}\Phi}\star F_{n} \wedge F_{(n-2)}
+ \left. 2 \kappa_{10}^2 \sum_n e^{\frac{n}{4}\Phi}\Psi_n \wedge \alpha(j)\right|_8\, .
}
\end{equation}

For the ten-dimensional metric one  uses a general warped ansatz of the form 
\begin{equation}
ds^2 = g^0_{MN} \, dx^M \otimes dx^N = 
{\rm e}^{2\Delta(y)} \left( dx^\mu \otimes dx^\nu \, \hat g_{\mu\nu}(x)
+ dy^m \otimes dy^n \, \hat g_{mn}(y)\right)\,.
\label{metric}
\end{equation}
The four-dimensional metric $g_{\mu\nu}$ describes either a
Minkowski, de Sitter ($dS_4$), or anti-de Sitter ($AdS_4$) space.
In general other bosonic fields are also allowed to acquire
non-trivial profiles and vacuum expectation values in the
background, but all fermion fields vanish.

In ten space-time dimensions the type II supersymmetry variations
for the two gravitinos $\psi_M^A$ ($A=1,2)$ and the two dilatinos
take the following form (in string frame)
\begin{eqnarray}\label{susycond}
\delta \psi_M &=& \nabla_M \epsilon
+ \frac{1}{4}\slash\!\!\!\!{H_M} \mathcal{P}\epsilon +
\frac{1}{16}e^{\Phi}
\sum_n \slash\!\!\! \!{F_{n}}\Gamma_M \mathcal{P}_n \epsilon\ , \nonumber\\
\delta \lambda &=&
\slash\!\!\!{\partial}\Phi\epsilon
+ \frac{1}{2}\slash\!\!\!\!{H} \mathcal{P}\epsilon +
\frac{1}{8}e^{\Phi}
\sum_n (-1)^n(5-n)
\slash\!\!\! \!{F_{n}}\mathcal{P}_n \epsilon\ .
\end{eqnarray}
Here the spinors $\psi_M$, $\lambda$ and $\epsilon$ always combine two spinors
but we suppress the label $A$.
The $\mathcal{P}$ and $\mathcal{P}_n$ are $2\times 2$ projection matrices,
whose form we do not need explicitly. The vanishing of these variations is required for supersymmetry. The
number of Killing spinors $\epsilon$ determines
the number of supercharges that are preserved.
It is now evident that without fluxes and with constant dilaton the solutions are just the
covariantly constant spinors of the Calabi-Yau, while
fluxes and dilaton profiles turn the conditions into much more
complicated looking differential equations.

Now we assume the following $\mathcal{N}=1$ compactification ansatz for the ten-dimensional supersymmetry generators  into four- and six-dimensional spinors:
\begin{eqnarray}\label{spinansatz}
\epsilon_1 & =&\zeta_+\otimes \eta^{(1)}_+ \, + \, \zeta_-\otimes \eta^{(1)}_- \ , \\
\epsilon_2 & =&\zeta_+\otimes \eta^{(2)}_\mp \, + \, \zeta_-\otimes \eta^{(2)}_\pm \ ,
\end{eqnarray}
for IIA/IIB, where $\zeta_\pm$ are four-dimensional and $\eta^{(1,2)}_\pm$ six-dimensional Weyl spinors. The Majorana conditions for $\epsilon_{1,2}$ imply the four- and six-dimensional reality conditions $(\zeta_+)^*=\zeta_-$ and $(\eta^{(1,2)}_+)^*=\eta^{(1,2)}_-$. This reduces the structure of the {\em generalized} tangent bundle to SU(3)$\times$SU(3). The structure of the tangent bundle itself on the other hand is a subgroup of SU(3) since there is at least one invariant internal spinor. What subgroup
exactly depends on the relation between $\eta^{(1)}$ and $\eta^{(2)}$. Following the terminology of 
\cite{Andriot:2008va,Caviezel:2008ik} the following classification can be made:
\begin{itemize}
\item strict SU(3)-structure: $\eta^{(1)}$ and $\eta^{(2)}$ are parallel everywhere;
\item static SU(2)-structure: $\eta^{(1)}$ and $\eta^{(2)}$ are orthogonal everywhere;
\item intermediate SU(2)-structure: $\eta^{(1)}$ and $\eta^{(2)}$ at a fixed angle, but neither a zero angle nor a right angle;
\item dynamic SU(3)$\times$SU(3)-structure: the angle between $\eta^{(1)}$ and $\eta^{(2)}$ varies, possibly becoming a zero angle or a right angle at a special locus.
\end{itemize}
Since for static and intermediate SU(2)-structure there are two independent internal spinors the structure of the tangent bundle reduces to SU(2), while
for dynamic SU(3)$\times$SU(3)-structure no extra constraints beyond SU(3) are imposed on the topology of the tangent bundle since the two internal spinors $\eta^{(1)}$ and $\eta^{(2)}$ might not be everywhere independent. 

\vskip0.2cm
\noindent
{\it SU(3) structure:}

\vskip0.2cm
\noindent
Let us consider the $SU(3)$ structure case in more detail 
(see e.g. \cite{Chiossi:2002tw,Gurrieri:2002wz,Lopes Cardoso:2002hd,Grana:2004bg,Grana:2005tf}
A  globally well defined non-vanishing
spinor exists only on manifolds that
have {\it reduced structure}. The structure group 
of a manifold is the 
group of transformations required to patch the orthonormal frame bundle. 
A Riemannian manifold of dimension $d$ has automatically structure group 
$SO(d)$. All vector, tensor and spinor representations can therefore be decomposed in 
representations of $SO(d)$. If the manifold has reduced structure group $G$, then
every representation can be further decomposed in representations of 
$G$. 
For $d=6$, supersymmetry in the absence of fluxes thus leads to the constraint
(\ref{susycond}) implying reduced holonomy for the internal space,
from $SO(6)\simeq SU(4)$ to $SU(3)$, so that ${\cal X}$ is a Calabi-Yau manifold.
The two covariantly constant spinors of type II theories
of course lead to ${\cal N}=2$ supersymmetry in four dimensions.
With non-vanishing fluxes $\eta_\pm$ can be viewed as covariantly constant with respect
to a new connection $\nabla'$ different from the Levi-Civita connection.
The internal manifold will no longer have $SU(3)$ holonomy (with
respect to the Levi-Civita connection).
Instead, the  requirement of having $SU(3)$ holonomy
with respect to the new connection means that the six-dimensional
internal manifold has a $SU(3)$ structure group, i.e.\ the
transition functions of the frame bundle take values in an
$SU(3)\subset SO(6)$ subgroup. These non-Calabi-Yau space nowadays are
characterized by so-called generalized geometry.

The $SU(3)$ group structure allows to decompose vectors, spinors and forms of the internal
six-dimensional manifold with respect to their transformation
properties under $SU(3)$. This is done by decomposing $SO(6)$ representations in terms 
of $SU(3)$ representations: ${\bf 4} \rightarrow {\bf 3}+ {\bf 1}$,
${\bf 6} \rightarrow {\bf 3}+ {\bf \bar 3}$, 
${\bf 15} \rightarrow {\bf 8}+ {\bf 3}+ {\bf \bar 3} + {\bf 1}$, 
${\bf 20} \rightarrow {\bf 6}+ {\bf \bar 6} + {\bf 3}+ {\bf \bar 3}+ 
{\bf 1}+ {\bf 1} $. The spinor representation is in the ${\bf 4}$ of $SO(6)$ which contains a singlet
under $SU(3)$. This means that there exists a globally well defined spinor on the manifold.
We furthermore can see that 
there are also singlets in the decomposition of 2-forms and 3-forms.
This means that there is also a non-vanishing globally well defined
real 2-form, and complex 3-form. These are called respectively $J$ and
$\Omega$. 
More precisely, a real non-degenerate two-form $J$ and a complex decomposable
three-form $\Omega$ completely specify an SU(3)-structure on the six-dimensional
 manifold $\mathcal{M}$ iff:
\begin{eqnarray}\label{OmegaJ}
\Omega\wedge J&=&0 \, , \\
\Omega\wedge\Omega^*&=&\frac{4i}{3}J^3\neq 0 ~,
\label{a1}
\end{eqnarray}
and the associated metric  is positive definite.
Up to a choice of orientation, the volume normalization can be taken such that
\begin{equation}
\frac{1}{6}J^3=-\frac{i}{8}\Omega\wedge\Omega^*=\mathrm{vol}_6~.
\label{voln}
\end{equation}

When the internal supersymmetry generators of (\ref{spinansatz}) are proportional,
\begin{equation}{\label{su3ansatz}
\eta^{(2)}_+ = (b/a) \eta^{(1)}_+ \, ,
}\end{equation}
with $|\eta^{(1)}|^2 = |a|^2, |\eta^{(2)}|^2=|b|^2$,
they define an SU(3)-structure as follows. First let us define a normalized
spinor $\eta_+$ such that $\eta^{(1)}_+= a \eta_+$ and $\eta^{(2)}_+ = b \eta_+$
and moreover we choose the phase of $\eta$ such that $a=b^*$. Note that in compactifications
to AdS$_4$ the supersymmetry imposes $|a|^2=|b|^2$ such that $b/a=e^{i\theta}$ is just a phase.
Now we can construct
$J$ and $\Omega$ as follows
\begin{equation}
\label{defJOm}
J_{mn}= i \eta^{\dagger}_+ \gamma_{mn} \eta_+ \, , \qquad \Omega_{mnp}= \eta^{\dagger}_- \gamma_{mnp} \eta_+ \, .
\end{equation}

The intrinsic torsion of  $\mathcal{M}$ decomposes into five modules (torsion classes)
${\cal W}_1,\dots,{\cal W}_5$. These also appear in the
SU(3) decomposition of the exterior derivative of $J$, $\Omega$. Intuitively,
 this is because the intrinsic torsion parameterizes the
failure of the manifold to be of special holonomy, which can also
be thought of as the deviation from closure of $J$, $\Omega$.
In fact, the
classification of the different classes of torsion under $SU(3)$
helps in understanding the properties of the underlying geometry.
In fact, on a manifold with $SU(3)$ group structure there is always a connection
$\nabla_m'$ with torsion that has $SU(3)$ holonomy, i.e. $\nabla_m'\eta=0$.
In case the connection is torsionless, the manifold is Calabi-Yau.
The torsion tensor can be decomposed in terms of $SU(3)$ representations as follows:
\begin{eqnarray} \label{torsion}
T^p_{mn}  &\in & ({\bf 3} \oplus {\bf \bar
3}) \otimes  ({\bf 1} \oplus {\bf 3} \oplus {\bf \bar 3}) \nonumber \\
&=& ({\bf 1} \oplus {\bf 
1}) \oplus  ({\bf 8} \oplus {\bf 8}) \oplus ({\bf 6} \oplus {\bf \bar
6}) \oplus  2 \, ({\bf 3} \oplus {\bf \bar 3} ) \nonumber\\
& & \quad {\cal W}_1 \quad  \quad\quad   {\cal W}_2 \quad  \quad\quad{\cal W}_3 \quad
\quad
 {\cal W}_4, {\cal W}_5  
\end{eqnarray}
${\cal W}_1, ... , {\cal W}_5$ are the five torsion classes that appear in the 
covariant derivatives of the spinor, of $J$ and of $\Omega$. 
${\cal W}_1$ is a complex scalar, ${\cal W}_2$ is  
a complex primitive (1,1) form, ${\cal W}_3$ is a real primitive $(2,1) + (1,2)$
form and ${\cal W}_4$ and ${\cal W}_5$ are real vectors.

The exterior derivative of $J$ and $\Omega$ can now be expressed
using these torsion classes:
\begin{eqnarray}
\d J&=&\frac{3}{2}\Im(\mathcal{W}_1\Omega^*)+\mathcal{W}_4\wedge J+\mathcal{W}_3 \, , \\
\d \Omega&= &\mathcal{W}_1 J\wedge J+\mathcal{W}_2 \wedge J+\mathcal{W}_5^*\wedge \Omega ~,
\label{torsionclassesv}
\end{eqnarray}
where $\mathcal{W}_1$ is a scalar, $\mathcal{W}_2$ is a primitive (1,1)-form, $\mathcal{W}_3$ is a real
primitive $(1,2)+(2,1)$-form, $\mathcal{W}_4$ is a real one-form and $\mathcal{W}_5$ a complex (1,0)-form.

A manifold with $SU(3)$ structure is  complex
if ${\cal W}_1={\cal W}_2=0$, i.e. $d\Omega$ is
a (3,1)-form. It is symplectic if ${\cal W}_1={\cal W}_3={\cal W}_4=0$, i.e. $J$ is closed. A
K\"ahler manifold is at the same time complex and symplectic, and
therefore the only non-zero torsion class can be ${\cal
W}_5$.\footnote{For a K\"ahler manifold the Levi-Civita connection has $U(3)$ holonomy.}
Finally for a Calabi-Yau manifold with $SU(3)$ holonomy all five
torsion classes are zero.

\vskip0.2cm
\noindent
{\it Geometric fluxes, T-duality and generalized geometry:}

\vskip0.2cm
\noindent
T-duality is a symmetry of string theory which related string theory on a circle
of radius $R$ to a string theory on a dual circle of radius $\alpha'/R$. Hence it is natural
how T-duality acts on flux backgrounds,
From the geometrical point of view T-dual backgrounds
look rather different, but from the string theory point if view they are equivalent.
Also from the low-energy supergravity point of view T-dual backgrounds are seemingly different,
in particular two T-dual backgrounds possess in general different $SU(3)$ group
structures and torsion classes. Later we will discuss a method to
provide a mathematically covariant  description of T-duality in terms of
generalized $SU(3)\times SU(3)$ group structures.

First consider the
Ramond fluxes of the type II superstring theories. T-duality (resp. an odd number of T-duality transformations) exchanges the type IIA superstring with
the type IIB superstring and vice versa. It follows that the even RR potentials $A^{(n})$ 
($n$ even) of the type IIB superstring are exchanged with the odd RR potentials
 $A^{(n})$ 
($n$ odd) of the type IIA superstring. So performing a T-duality transformation along the
coordinate $x$, we get the following T-duality rule for the Ramond fluxes:
\begin{equation}
F_{x\alpha_1 \cdots \alpha_p}^{(p+1)}
\stackrel{T_x}{\longleftrightarrow}
 F_{\alpha_1 \cdots \alpha_p}^{(p)} \,.
\label{eq:ft}
\end{equation}

Second, for the universal NS flux field strength $H^{(3)}$ we can use the Buscher rules 
\cite{Buscher:1987sk}
that were already derived in the world sheet approach. to recall, T-duality in the $x$ direction
provides the following new background:
\begin{eqnarray}
G^\prime_{xx} &=& \frac{1}{G_{xx}}, \qquad
G^\prime_{x \mu}  = -\frac{B_{x \mu}}{G_{xx}}, \qquad
B^\prime_{x \mu} = -\frac{G_{x \mu}}{G_{xx}} \\
G^\prime_{\mu \nu}  &=&  G_{\mu \nu} - \frac{G_{x \mu}G_{x \nu} - B_{x \mu} B_{x \nu}}{G_{xx} } \\
B^\prime_{\mu \nu}  &=&  B_{\mu \nu} - \frac{G_{x \mu}B_{x \nu} - B_{x \mu} G_{x \nu}}{G_{xx} } \\
e^{\phi^\prime} &=& \frac{e^{\phi}}{\sqrt{G_{xx}}}
\label{buscherrules}
\end{eqnarray}
This basically means that a flux background with non-vanishing $B_{x\mu}$ is T-dualized into
a purely geometric back ground with off-diagonal metric $G_{x\mu}$ and vice versa.
Switching from $B$ to $H$, a non-vanishing $H_{xyz}$ gets T-dualized along the $x$
direction in a metric background, which we call $G_{yz}=f^x_{yz}$
\begin{equation}
H_{xyz} \stackrel{T_x}{\longrightarrow} f^x_{yz}.
\end{equation}
Since these metric components arise from the NS H-flux after T-duality, one often calls
the $f^x_{yz}$'s NS metric, or also geometric fluxes. These constants often appear
in so-called twisted tori compactifications
\cite{Kachru:2002sk,Hull:2005hk,Grana:2006kf}, 
where they correspond to a certain, underlying algebraic
structure.

Let us demonstrate the T-duality transformation rules NS fluxes
by an explicit example
that of a $T^3$ with $H_3$ flux. Note that this background does not satisfy the supersymmetry conditions, since it is a flat background with nontrivial $H$-flux. This is not a problem, however, as we only use this as an illustrative example and one could e.g. fiber this $T^3$ over something else to get a good string background. To start, take $(x,y,z)$ as the coordinates on the $T^3$, each with period 1. Additionally, put $N$ units of $H$-flux on the torus, such that 
\begin{equation}
\int_{T^3} H_3 = N,
\end{equation}
where we have set a pre-factor of $1/(2\pi)^2 \alpha^\prime =1$ for convenience. To ensure that this quantization condition is satisfied, we can now pick a gauge where $B_{xy} = Nz$, with $N \in {\bf Z}$. We have introduced an explicit dependence on the coordinate $z$. One can view this space not only as a $T^3$, but also as a $T^2$ in the $(x,y)$ directions fibered over an $S^1$ in the $z$ direction, where the K\"{a}hler modulus $\rho = ( \int B) + iV$ of the $T^2$ undergoes $\rho \rightarrow \rho + N$ as $z \rightarrow z+1$. 

Nothing depends on the coordinates $x$ and $y$, so we can feel free to do a T-duality in either of those directions. T-dualizing on the $x$ direction yields the background
\begin{equation}
ds^2 = (dx - N z \, dy)^2 + dy^2 + dz^2 ; \qquad B=0.
\end{equation}
This is exactly the metric 
\begin{equation}
ds^2 = (dx - f^x_{yz}\,  z  \, dy)^2 + dy^2 + dz^2.
\label{ttmetric}
\end{equation}
with $f^x_{yz} = N$, so we see that this background is a twisted torus. In order to make this metric globally well-defined, we need to identify $(x,y,z) \sim (x + Ny, y, z+1)$. Thus, we see that a T-duality takes 
$
H_{xyz} \stackrel{T_x}{\longrightarrow} f^x_{yz}$.

One can easily picture this space as a $T^2$ in the $(x,y)$ directions fibered over an $S^1$ in the $z$ direction. As one goes around the $S^1$ base, the fiber $T^2$ undergoes a shift in complex structure $\tau \rightarrow \tau + f^x_{yz}$. If we want to end up with an equivalent fiber after traversing the $S^1$, we need to ensure that this is an $SL(2,{\bf Z})$ transformation, so we require $f^x_{yz} \in {\bf Z}$. 
This is as expected: one T-duality has switched the complex structure and 
K\"{a}hler moduli. 

There is a very useful way of thinking about the number $f^x_{yz}$. Define the globally invariant one-forms
\begin{eqnarray}
\eta^x & = & dx - f^x_{yz} \,  z \, dy \\
\eta^ y & = & dy \\ 
\eta^z & = & dz. 
\end{eqnarray}
Clearly, $ d \eta^ y = d \eta^z = 0$, but 
\begin{equation}
d \eta^x  = f^x_{yz} dy \wedge dz= f^x_{yz} \eta^y \wedge \eta^z.
\end{equation}
The $f^x_{yz}$ are just components of the spin connection, by Cartan's structure equations. Additionally, they are the structure constants of a Lie group, which show up as above when the Lie group is viewed as a manifold. This can be easily generalized by
considering a manifold with a basis of globally defined one-forms $e^a$. The generalization of the above construction is that we can write
\begin{equation}
\label{fdef}
de^a = f^a_{bc} \, e^b \wedge e^c,
\end{equation}
with the $f^a_{bc}$ are all constant. Note that the requirement that $f^a_{bc}$ be constant is a nontrivial constraint on the manifold. Additionally, the $f^a_{bc}$ must also obey a constraint:
\begin{equation}
d^2 e^a = 2 f^a_{b [ c} f^b_{de]} e^d e^e e^c = 0.
\end{equation}
Therefore,  the $f^a_{bc}$ obey a Jacobi identity $f^a_{b [ c} f^b_{de]} =0$, as the structure constants of a Lie algebra should. 
For compact spaces one has to require that 
$f^a_{ab} = 0$ (no sum); this comes from requiring $d( \alpha e^1 \wedge ... \wedge e^{d-1}) = 0$.
$f^a_{bc}$ that form a nilpotent algebra automatically satisfy this condition.  Such manifolds are called nilmanifolds, or twisted tori, as we will shortly discuss. 

Before we come to the next example, let us also mention that T-duality not always lead
to a geometrical background \cite{Shelton:2005cf,Dabholkar:2005ve,Shelton:2006fd} (for a review on
non-geometrical backgrounds see \cite{Wecht:2007wu}). 
This can be seen as follows. Starting from the metric (\ref{ttmetric})
we still have another T-duality we can do here, since nothing depends on the $y$ direction. The Buscher rules now give 
\begin{equation}
ds^2 = \frac{1}{1+N^2 z^2} (dx^2 + dy^2 ) + dz^2 ; \qquad B_{xy} = \frac{Nz}{1+N^2 z^2}.
\label{qflux}
\end{equation}
One can check, by examining the K\"ahler modulus
\begin{equation}
\frac{1}{\rho} = Nz -i,
\end{equation}
that $z \rightarrow z+1$ just takes $1/\rho \rightarrow 1/\rho + N$. This is an $SL(2,{\bf Z})$ transformation on $\rho$, so once again we see the fiber $T^2$ shifting its K\"{a}hler modulus as we go around the base circle. 
This is indeed an example of a non-geometric background, since the transformation $1/\rho \rightarrow 1/\rho + N$ mixes the metric and the B-field. More precisely, this background is locally geometric, since the metric and B-field are defined at every point, but it is not globally a manifold. Upon going around a cycle, the metric and B-field mix by an $SL(2,{\bf Z}) \subset O(2,2;{\bf Z})$ transformation. As with the previous two backgrounds, this one is characterized by an integer $N$. Writing
\begin{equation}
H_{xyz} \stackrel{T_x}{\longrightarrow} f^x_{yz}
\stackrel{T_y}{\longrightarrow} Q^{xy}_z,
\end{equation}
we will say that this background is characterized by the non-geometric flux $Q^{xy}_z$. One can show that this object $Q^{xy}_z$ behaves like a one-form, as expected.

Let us study another example, how T-duality acts on a geometrical NS and R flux background,
which is actually
a solution of the supersymmetry conditions (in addition, one needs also some orientifold charges).
We start with a massive  IIA solution that is obtained by taking the internal  manifold  again to be a six-dimensional torus.
All torsion
classes vanish in this case. Note, however, that there are non-vanishing $H$ and $F_{4}$
fields given by eq.(\ref{ltsol})
\begin{eqnarray}
H & =& \frac{2}{5} e^{\Phi} m \left(e^{246}-e^{136}-e^{145}-e^{235}\right) \, , \\
F_4 & = &\frac{3}{5} m \left( e^{1234} + e^{1256} + e^{3456} \right) \, .
\end{eqnarray}
In addition there is a non-vanishing Romans' mass a parameter $m$,
which can be seen as a 0-form flux (see also section (4.4)):
\begin{equation}
F_0  = \frac{2}{5} m^2  e^{2\Phi} \, .
\end{equation}

This solution is related, via a single T-duality, to the type IIB superstring on the
so-called nilmanifold. 
 Indeed, let us perform a T-duality on the $X_6$ coordinate of the six-torus example.\footnote{Note that
it does not matter along which direction one performs the T-duality since
all six perpendicular directions are equivalent.}  After rescaling and relabeling the left-invariant forms
we find the nilmanifold 5.1 described by the following left-invariant basis
of viel-beins:
\begin{eqnarray}
\d e^a&=&0,~~a=1,\dots, 5 \, ,  \nonumber\\
\d e^6&=&e^{12}+e^{35}~.
\end{eqnarray}
The metric is now given by $g=\text{diag} (1,1,1,1,\beta^2,\beta^2)$, and
for the fluxes we have
\begin{eqnarray}
H & =& - \beta \left( e^{235} + e^{145}\right) \, , \nonumber \\
e^{\Phi} F_1 & =& \frac{5}{2} \, \beta^2 e^6 \, , \nonumber \\
e^{\Phi} F_3 & = &\frac{3}{2} \, \beta \left( e^{135}-e^{245}\right) \, , \nonumber\\
e^{\Phi} F_5 & =& \frac{3}{2} \, \beta^2 e^{12346} \, .
\end{eqnarray}
$\beta$ is related to the mass parameter of the torus example via $\beta=\frac{2}{5} m e^{\Phi}$.

Finally, one can perform a second T-duality along the $x_5$ direction,
leading to another geometrical type IIA  geometric background, namely the Iwasawa manifold.

At the end of this section we want to discuss the question 
what is the proper  mathematical description of supersymmetric flux backgrounds including
T-duality.
In particular we have seen that T-duality can lead to backgrounds, which do not allow any more
for a description in terms of standard Riemannian geometry, but are rather non-geometrical string
backgrounds. One of their key properties is that the 
transition functions are not anymore diffeomorphisms, but rather T-duality transformations. That means
local patches of a T-fold are not glued together by coordinate transformations, but rather by
(discrete) T-duality transformations \cite{Dabholkar:2005ve,Hull:2006va}. However, since T-duality is supposed to be a symmetry of
string theory, T-fold constitute a class of consistent non-geometrical string backgrounds, 
like asymmetric orbifolds or general covariant lattice models. 

The geometric flux backgrounds as well
as the non-geometric flux backgrounds can be best described in terms of so-called
generalized geometry, which generalize the $SU(3)$ group structures of the geometric
backgrounds to the generalized $SU(3)\times SU(3)$ group structures 
\cite{Jeschek:2005ek,Gmeiner:2006ni,Grana:2006is,Grana:2006hr,Gmeiner:2007jn,Grana:2008yw,Lust:2008zd}.\footnote{D-branes
and calibrated sources in generalized geometries were described in \cite{Martucci:2006ij,Koerber:2006hh,Evslin:2007ti,Koerber:2007hd,Koerber:2007jb};
warped flux compactifications were recently discussed in \cite{Martucci:2009sf}.}
To understand this
recall that in the world sheet approach the closed string is characterized by independent
left and right moving coordinates $X_L(\sigma-\tau)$ and $X_R(\sigma+\tau)$. 
The standard space-time coordinates are given by the sum of the left and right moving coordinates,
$X=X_L+X_R$, whereas the dual coordinates are given in terms of their difference:
$X^*=X_L-X_R$. Bosonic T-duality transformations on a 6-dimensional background space
are given in terms of $SO(6,6;{\bf Z})$ transformations, which mix the left and right moving degrees
of freedom. In fact, the basic T-duality transformation just exchanges $X_L\leftrightarrow X_R$
resp. the position space coordinates with the dual coordinates, i.e. $X\leftrightarrow X^*$.
Therefore each closed string excitation not only forms representations of the tangent space of $X$,
called $T$, but also of the dual tangent space, denoted by $T^*$.
This fact strong suggest to enlarge the space by combining the tangent and the cotangent bundles
in a single bundle, $T\oplus T^*$, with generalized structure group $SO(6,6)$.
Now each (internal) string excitation transforms as a representation of the group $SO(6,6)$.

In type II superstrings we have left and right moving internal spinors:
$\eta^{(1)}_\pm$ for the left moving spinors and $\eta^{(2)}_\pm$ for the right moving spinors, where
the subscripts $\pm$ denote the two different in-equivalent spinor representations
of $SO(6)$, denoted by ${\bf 4}_S$ and ${\bf f}_C$.
Following our strategy to build proper representations of the enlarged frame rotation group
$SO(6,6)$ it is useful the combine the left and right moving spinors into a single spinor
of $SO(6,6)$, which is often called pure spinor.
Specifically,
the supersymmetry generators $\eta^{(1)}$ and $\eta^{(2)}$ from
eq.(\ref{spinansatz}) are  collected into two spinor bilinears, which using the
Clifford map, can be associated with two polyforms of definite degree
\begin{equation}{\label{polys}
{\Psi}_+ = \frac{8}{|a||b|}\eta^{(1)}_+ \otimes \eta^{(2)\dagger}_+ \, , \qquad
{\Psi}_- = \frac{8}{|a||b|}\eta^{(1)}_+ \otimes \eta^{(2)\dagger}_- \, .
}
\end{equation}
The subindices
plus and minus in $\Psi_{\pm}$ denote the 
Spin(6,6) chirality: positive corresponds
to an even form, and negative to an odd form.

A generalized Calabi-Yau 
is a manifold on which a closed pure spinor $\Psi$ exists:
\begin{equation}
d\Psi  = 
 0\, .
\end{equation}
The requirement of having two invariant closed spinors ensures that ${\cal N}=2$ space-time supersymmetry for type II strings 
and
reduces the structure $SO(6,6)$ to $SU(3)\times SU(3)$.
$\Psi_\pm$ define therefore an $SU(3)\times SU(3)$ group structure on $T\oplus T^*$, similarly
as the supersymmetry condition defines an $SU(3)$ group structure on $T$ for the case
of geometrical background spaces.
However the existence of an   $SU(3)\times SU(3)$ group structure is obviously a more general
property of supersymmetric string backgrounds, and a $SU(3)$ group structure does not always exist.
T-duality transformations act linearly on the pure spinors $\Psi_\pm$, hence
$SU(3)\times SU(3)$ group structures provide a useful mathematical framework for dealing with
T-duality.

In case an $SU(3)$ group structure exist one can relate the pure spinors
with the basic geometric objects $J$ and $\Omega$ (up to a possible phase):
\begin{equation} \label{Fierza} 
\Psi_+ = e^{i J}
\ , \qquad
\Psi_- =  {\Omega} \ .
\end{equation}
Examples of Generalized Calabi-Yau
manifolds are symplectic manifolds and complex 
manifolds with trivial torsion class ${\cal W}_5$ (i.e., if
${\cal W}_1={\cal W}_2=0$, and $\bar {\cal  W}_5=\bar \partial f$ -
then $\Psi= e^{-f} \Omega$ is closed).  
T-duality basically rotates $\Psi_+$ into $\Psi_-$, and one can show that the mirror symmetry
for flux compactifications exchanges symplectic manifolds with complex  manifolds, i.e.
mirror symmetry acts as:
\begin{equation}
\Omega \longleftrightarrow e^{iJ}\, .
\end{equation}

Finally, in order to obtain the pure spinors in IIA and IIB one redefines
\begin{equation}{\label{A/B}
\Psi_1 = \Psi_{\mp} \, , \qquad \Psi_2 = \Psi_{\pm} \, ,
}\end{equation}
with upper/lower sign for IIA/IIB.

\vskip0.2cm
\noindent
{\it Effective supergravity action:}

\vskip0.2cm
\noindent
The superpotential $W$ and K\"{a}hler potential $K$ of the effective $\mathcal{N}=1$ supergravity action
for compactification on spaces with $SU(3)\times SU(3)$ or $SU(3)$ group structures
have been derived in various ways in 
\cite{Grimm:2004uq,Grimm:2004ua,Grana:2005ny,Benmachiche:2006df,Grana:2006hr,Cassani:2007pq,Koerber:2007xk} 
(based on earlier work of \cite{Gukov:1999ya,Taylor:1999ii}).

The part of the effective four-dimensional action containing the graviton and
the scalars reads:
\begin{equation}{
S =  \int \d^4 x \sqrt{-g_4}  \left( \frac{M_P^2}{2} R  - M_P^2{K}_{i \bar{j}} \partial_{\mu} \phi^i \partial^{\mu}  \bar{\phi}^{\bar{j}} - V (\phi,\bar{\phi}) \right)\, ,
}
\end{equation}
where $M_P$ is the four-dimensional Planck mass.
The scalar potential is given in terms of the superpotential via:\footnote{In \cite{Cassani:2008rb,Lust:2008zd} the scalar potential
was for general type II SU(3)$\times$SU(3) compactifications directly derived from dimensional reduction of the action.}
\begin{equation}\label{ftermscp}
V(\phi,\bar{\phi}) = M_P^{-2} e^{{K}} \left( {K}^{i\bar{\jmath}} D_i {W} D_{\bar{j}} {W}^* - 3 |{W}|^2 \right) \, ,
\end{equation}
where the superpotential in the Einstein frame ${W}$ reads
\begin{equation}{
\label{suppoteinstein}
{W}  = \frac{-i}{4 \kappa_{10}^2} \int_M\langle \Psi_2,F+i\,\d_H(\Re \mathcal{T})\rangle\ ,
}
\end{equation}
and $\langle \cdot, \cdot \rangle$ indicates the Mukai pairing, $\Re \mathcal{T}= e^{-\Phi} \Im \Psi_1$, and $\Psi_1$ and $\Psi_2$ are the pure spinors describing the geometry. We can rewrite this
as
\begin{equation}{
{W} = \frac{-i}{4 \kappa_{10}^2} \int_M\langle \Psi_2 e^{\delta B},{F}+i\,\d_{{H}}(e^{\delta B} \Re \mathcal{T}-i\delta C)\rangle\ , .
}
\end{equation}
This shows how the fields organize in complex multiplets $\Psi_2 e^{\delta B}$ and $\Re \mathcal{T}-i\delta C$, which will
be clearer in concrete examples.

The K\"ahler potential reads
\begin{equation}{
\label{kahlereinstein}
{K}  =  - \ln i \int_M \langle \Psi_2, \bar{\Psi}_2 \rangle - 2 \ln i \int_M \langle t , \bar{t} \rangle +3 \ln(8 \kappa_{10}^2 M_P^2)\, ,
}
\end{equation}
where we defined $t=e^{-\Phi} \Psi_1$.
Note that  $\Re t$ should be thought of as a function of $\Im t$ so that $t$ can be seen as (non-holomorphically)
dependent on $\mathcal{T}$. 

Our aim in the next sections will be to find supersymmetric extrema of the scalar potential $V$. We must
therefore impose
\begin{equation}\label{ftermmin}
F_i(\phi_{\mathrm{min}})=e^{K/2}(\partial_{\phi_i}W+W\partial_{\phi_i}K)|_{\rm min}=0 \ \forall  i\, .
\end{equation}

\subsection{Type IIB flux compactifications -- the KKLT scenario}

We will start without geometrical fluxes; then the tree-level 3-form flux
superpotential in type IIB on a Calabi--Yau 3-fold $X$ is of the
standard form \cite{Gukov:1999ya,Taylor:1999ii,Mayr:2000hh,Curio:2000sc,Giddings:2001yu}. It gets two kinds
of contributions, namely from Ramond and Neveu--Schwarz 3-form
fluxes through 3-cycles of the CY space:
\begin{eqnarray}\label{tv}
W_{\mathrm{IIB}}&=&W_H+W_F=\int_X\Omega\wedge \left( F^{\mathrm{R}}_3+SH^{\mathrm{NS}}_3\right)\nonumber \\
&=&e_0+ie_iU_i+i m_0F_0(U)+m_iF_i(U)\nonumber
\\ &&+iS(a_0+ia_iU_i+i b_0F_0(U)+b_iF_i(U))\, .
\end{eqnarray}
Here $\Omega$ is the holomorphic 3-form on the CY space, and
$F^{\mathrm{R}}_3$ ($H^{\mathrm{NS}}_3$) is the Ramond
(Neveu--Schwarz) 3-form field strength field. The $U$-dependent
function $F(U)\equiv F_0(U)$ is the holomorphic prepotential and
the $F_i(U)$ are its first derivatives. The $e_I,m_I$ comprise the
Ramond 3-form fluxes, whereas the $a_I,b_I$ correspond to the
Neveu--Schwarz 3-form fluxes ($I=0,\dots , h^{2,1}$). The
superpotential $W$ depends on the complex-structure moduli fields
$U_i$ ($i=1,\dots ,h^{2,1}$) and on the dilaton $S$, whereas it is
independent of the K\"ahler moduli $T_m$ ($m=1,\dots ,h^{1,1}$).

The supersymmetry conditions eq.(\ref{ftermmin}) now translate into the conditions
on the complex fluxes 3-form fluxes $G_3=F_3^R+SH_3^{NS}$.
First with applying the supersymmetry condition 
respect to the complex structure moduli $U$ one gets
 \begin{equation}\label{uftermmin}
F_U(\phi_{\mathrm{min}})=e^{K/2}(\partial_{U}W+W\partial_{U}K)|_{\rm min}=0 \quad \Rightarrow\quad
G_{(1,2)}=0\, ,
\end{equation}
where $G_{(1,2)}$ is the (1,2)-Hodge component of the complex 3-form fluxes $G_3$.
Similarly supersymmetry with respect to the dilaton $S$ requires
 \begin{equation}\label{sftermmin}
F_S(\phi_{\mathrm{min}})=e^{K/2}(\partial_{S}W+W\partial_{S}K)|_{\rm min}=0 \quad \Rightarrow\quad
G_{(3,0)}=0\, ,
\end{equation}
Finally the supersymmetry condition with respect to the K\"ahler moduli $T$
implies
 \begin{equation}\label{tftermmin}
F_T(\phi_{\mathrm{min}})=e^{K/2}(\partial_{T}W+W\partial_{T}K)|_{\rm min}=0 \quad \Rightarrow\quad
G_{(0,3)}=0\, ,
\end{equation}
In summary, applying all three supersymmetry conditions it follows that the 3-form
flux must be a self-dual (2,1)-form in the supersymmetric minimum of the potential:
\begin{equation}
G_{(2,1)}|_{\rm min}\neq 0
\end{equation}

In type IIB the fluxes generate a $C_4$ tadpole given by
\begin{equation}\label{iibtadpole}
N_{\rm flux}=\int H_3\wedge F_3 =
\sum_{I=0}^{h^{2,1}}a_Im_I+b_Ie_I\, .
\end{equation}
This flux number is equivalent to the Ramond charge of D3-branes, and has to be cancelled by external sources, namely by the  orientifold
O3-planes and an appropriate number of D3-branes. Specifically in addition
to the above supersymmetry conditions one gets the following tadpole
constraints on the 3-form fluxes:
\begin{equation}
N_{\rm flux}+N_{D_3}=N_{O_3}\, .
\end{equation}
This condition can be reformulated in F-theory in a more geometrical way,
 where the O3-planes are related to the
Euler number $\chi$ of an underlying elliptic Calabi-Yau 4-fold $X_4$.
In F-theory the corresponding Gukov-Vafa-Witten superpotential is given in terms
of a 4-form flux $G_4$:
\begin{equation}
W=\int_{X_4}G_4\wedge\Omega\, .
\end{equation}
Now the flux has to satisfy the following tadpole cancellation condition
\begin{equation}
L\equiv \h\int_{X_4}G_4\wedge G_4={\chi(X_4)\over 24}-N_{D-3}\, .
\end{equation}
This gives an upper bound on the fluxes,
\begin{equation}
L\leq L_*\, ,
\end{equation}
with $L_*=\chi(X_4)/24$. 

For a given Calabi-Yau 3-fold like an orbifold (or in F-theory for a given 4-fold) it is possible
to construct many concrete examples of supersymmetric type IIB flux vacua.
E.g. consider a superpotential  of the form \cite{Curio:2000sc}, as it occurs in toroidal or
orbifold compactifications:
\begin{equation}
W_{\mathbb{IIB}}=(p+iqSU_1)(l_2-il_1U_2+in_1U_3-n_2U_2U_3)\, .
\end{equation}
$p,q,l_1,l_2,n_2,n_2$ parametrize the flux quantum numbers that are constrained by the
tadpole condition.
For fixed flux quantum numbers there is a unique solution of the supersymmetry condition
with zero vacuum energy:
\begin{equation}
SU_1=-{p \over q}\, , \quad U_2=\sqrt{l_1l_2\over n_1n_2} 
\, ,\quad U_3=\sqrt{l_2n_1\over l_1n_2} \, .
\end{equation}

Moreover the above tadpole conditions are also useful to estimate the number of maximally
possible flux vacua on a certain background space. Following \cite{Bousso:2000xa,Ashok:2003gk,Denef:2004ze}
this number can be estimated by the following formula:
\begin{equation}\label{nsusy}
N_{\rm SUSY}\simeq{L_*^{2h^{2,1}+2}\over(2h^{2,1}+2)!}\, .
\end{equation}
where the Hodge number $h^{2,1}$ counts of complex structure moduli $U$.
Typical numbers for $h^{2,1}$ and $L_*$ lead to a large number for $N_{\rm SUSY}$:
\begin{equation}
N_{\rm SUSY}={\cal O}(10^{500})\, .
\end{equation}
This is indeed a very huge landscape of flux vacua, from which one can possibly argue
that there is a good chance to find vacua (after proper uplift to positive vacuum energy)
with a tiny positive cosmological constant of order $\Lambda\simeq 10^{-120}M_{\rm Planck}^4$.
In fact, the vast proliferation of string vacua opens the possibility to explain the smallness
of the cosmological constant via the anthropic principle \cite{Weinberg:1987dv}. 
Combing this flux vacua statistics with the intersecting D-brane statistics of section (2.4) (see also
next subsection)
would possibly also lead to an anthropic explanation of the SM and its parameters.
Whether the anthropic principle is really the proper way to understand the landscape
is highly debated among theorists (see e.g. \cite{Susskind:2003kw,Schellekens:2008kg}), and the outcome
of this discussion still has to waited for.
Just note however that in case the SM is realized by D3-branes or by D7-branes
with F-flux, $L$ is considerably lower that $L_*$, such that the actual number of flux vacua is
much smaller than the number that follows  from eq.(\ref{nsusy}).

For
geometrical CY spaces with $h^{1,1}>0$, 
the flux superpotential  (\ref{tv}) fixes the complex moduli $U$ and the dilaton $S$, 
however not the K\"ahler moduli $T$. They are still left as flat direction
of the potential. Moreover, the condition eq.(\ref{tftermmin}) implies that the superpotential
and hence also the scalar potential vanish in the minimum: $W|_{\rm min}=V|_{\rm min}=0$.
So the supersymmetric ground state forms a 4-dimensional Minkowski vacuum, i.e.
 supersymmetry is not compatible with negative
vacuum energy.  
The problem of the K\"ahler-moduli perturbative independence can be
in principle resolved either by introducing geometrical fluxes,
i.e. torsion, hence abandoning the CY structure, or in backgrounds
which are non-geometrical and do not possess \emph{at all}
K\"ahler moduli (i.e. $h^{1,1}=0$ in the framework of CY). 
In addition, the inclusion of non-perturbative effects in the effective superpotential also
can fix the K\"ahler moduli \cite{Kachru:2003aw}.
This is the so-called KKLT scenario, where the total superpotential contains the contributions
from the 3-form fluxes (see eq.(\ref{tv})) plus a non-perturbative contribution that depends 
on the K\"ahler moduli fields (and also due to threshold effects on the complex
structure moduli):
\begin{equation}\label{kklt}
W_{KKLT}(T,U,S)=W_{\rm 3-form}(U,S)+W_{\rm n.p.}(T,U)
\end{equation}
Now, applying the supersymmetry conditions eq.(\ref{ftermmin}) to $W_{KKLT}$ allows also
for non-vanishing 3-form flux component $G_{(0,3)}$. In addition,
now $W|_{\rm min}\neq 0$, $V|_{\rm min}<0$, and hence the flux  vacuum is anti-de Sitter like.
In the second step of the KKLT scenario, a positive contribution, possibly due to anti-D3-branes
or other effects, is added to scalar potential, which leads to a positive vacuum energy, i.e. a
4-dimensional de Sitter vacuum with broken supersymmetry. The typical KKLT potential 
as a function of the overall K\"ahler modulus $T$ before and
after the uplift is shown in figure 13.
 \begin{figure}
\begin{center}
  \includegraphics[width=0.55\textwidth]{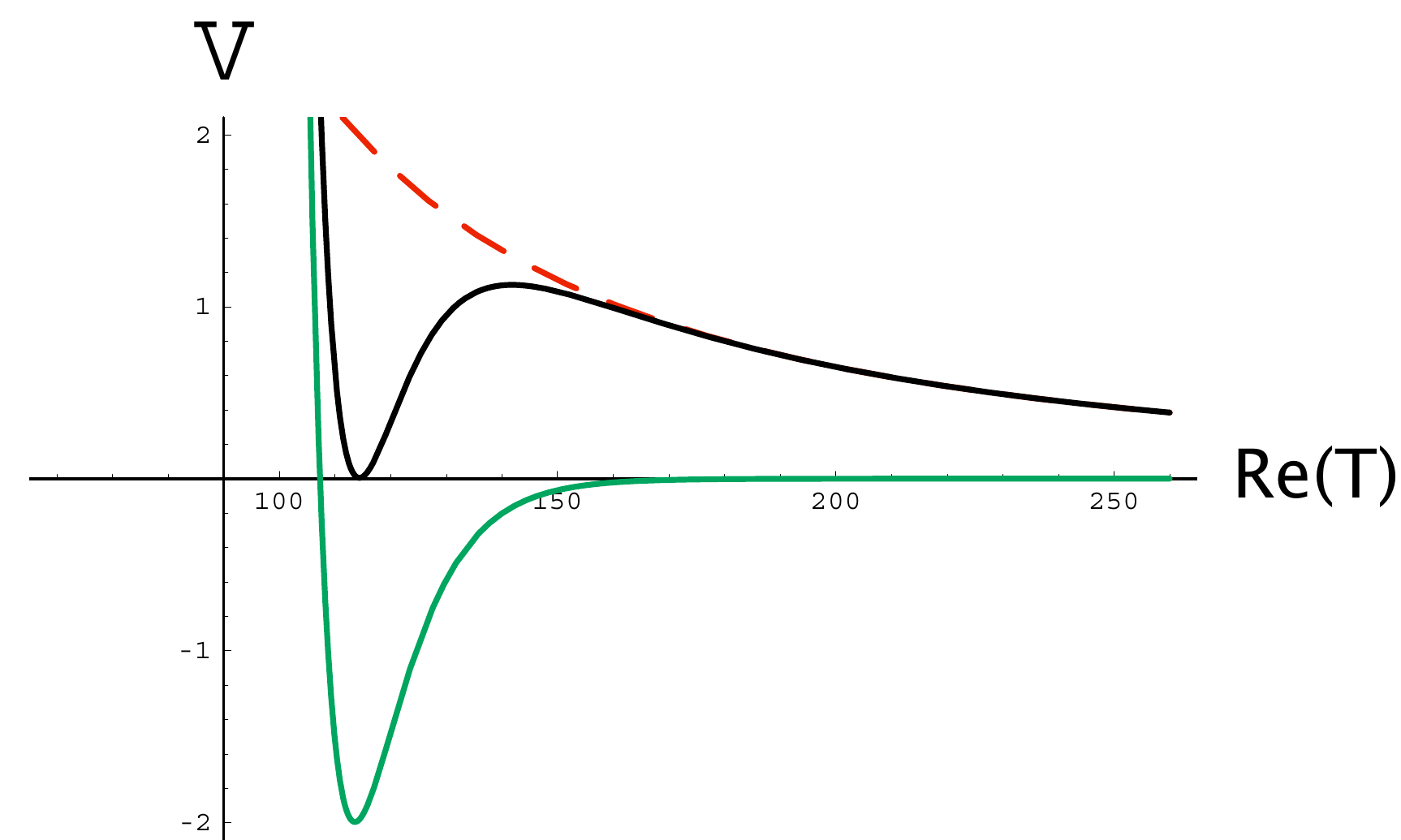}
\end{center}
\caption{The KKLT potential before the uplift (in red) and after the de Sitter uplift (in black).}
\end{figure}
It is then possible to analyze the pattern of the soft SUSY breaking mass parameters
in the effective supergravity action with D3/D7-branes, which arises in type IIB flux vacua after
supersymmetry breaking
\cite{Camara:2003ku,Grana:2003ek,Lust:2004fi,Camara:2004jj,Lust:2004dn,Choi:2005ge, Conlon:2005ki, Conlon:2006us,Conlon:2006wz}.

The non-perturbative part of the KKLT superpotential  is provided by
Euclidean D3-instantons \cite{Witten:1996bn}, 
which are wrapped around 4-cycles (divisors)
$D$ inside ${\cal M}_6$,
and/or gaugino condensations in hidden gauge group sectors
on the world volumes of D7-branes,
which are also wrapped around certain
divisors $D$. 
Both give rise to terms in the superpotential of the form
\begin{equation}
W_{\rm n.p.}(T,U)\sim g_i(U)\Phi^ne^{-a_iV_i(T)}\, ,
\end{equation}
where $V_i$ is the volume of the divisor $D_i$, depending on the
K\"ahler moduli $T$, and $g_i(U)$ is a pre-factor, which generically depends on the complex
structure moduli $U$.
The fields $\Phi$ are matter fields in bifundamental representations that are located at the intersections of
space-time filling D7-branes, which are at the same time also
intersected by the D3-instantons, resp. the D3-instantons lie
on top of the D7-branes.

For gaugino condensation in a hidden gauge group, $V_i$ is the 
(holomorphic) gauge coupling constant of $G_{\rm hidden}$, and
$W_{\rm n.p.}$ corresponds to the field theory ADS/TVY superpotential 
\cite{Taylor:1982bp,Affleck:1983rr}.
Here the number of matter fields in  $W_{\rm n.p.}$ is determined by the number of
colors and flavors of $G_{\rm hidden}$. In the simplest case with $G_{\rm hidden}=SU(N_C)$
and $N_F=N_C-1$, $W_{\rm n.p.}$ is induced by a single D3-instanton that is wrapping
the same 4-cycle as the $N_c$ gauge D7-branes.
Using these techniques, the moduli stabilization in the KKLT scenario 
with non-perturbative superpotential was investigated in several orbifold compactifications
and their blow-up variants \cite{Lust:2005dy,Denef:2005mm,Lust:2006zg,Lust:2006zh}.

\subsection{Combing type IIB flux compactifications and D-brane model building -- large volume
compactifications}

In type IIB orientifolds we assume that the D7-branes are wrapped around
4-cycles inside a CY-orientifold.
 In the string frame\footnote{In the Einstein frame the
K\"ahler moduli $t_k$ are multiplied by the factor $e^{-\h\phi_{10}}$.
Therefore, in the Einstein frame the CY volume reads $V_6=\fc{1}{6}e^{-\fc{3}{2}\phi_{10}}\
\kappa_{ijk}\ t_it_jt_k$.}, the volume~$V_6$ of a CY space
$X$ is given by
\begin{equation}
V_6={1\over 3!}\int_XJ\wedge J\wedge J={1\over 6}\ \kappa_{ijk}\ t_it_jt_k\ ,
\end{equation}
with $t_i$ ($i=1,\dots , h^{1,1}$) the (real) K\"ahler moduli
in the string basis and $\kappa_{ijk}$ the triple
intersection numbers of $X$. The K\"ahler form $J$ is expanded
w.r.t. a base $\lbrace\hat D_i\rbrace$ of the cohomology
$H^{1,1}(X,{Z})$ as $J=\sum\limits_{i=1}^{h^{1,1}}{t}_i\
\hat D_i$. Without loss of generality we restrict
to orientifold projections with $h^{1,1}_-=0$, $h^{1,1}_+=h^{1,1}$.
On the other hand, the real parts of the physical K\"ahler moduli $T_i$ correspond to the
volumes of the CY homology four-cycles $D_k$ and
are computed from the relation:
\begin{equation}\label{physT}
T_i={1\over 2}\ \int_{D_i}J\wedge J={\partial V_6\over\partial{t}_i}=\h\
\kappa_{ijk}\ t_jt_k\ .
\end{equation}
It follows that the volume $V_6$ of $X$
becomes a function of degree 3/2 in the K\"ahler moduli~$T_i$:
\begin{equation}
V_6= {1\over 3!}\ \int_XJ\wedge J\wedge J={\cal O}(T_i^{3/2})\ .
\end{equation}

For D7--branes wrapped around the four-cycle $D_k$, the corresponding gauge 
coupling constant takes the form\footnote{On the other hands, for (space--time filling)
D3-branes the corresponding gauge coupling constant is given by:
\begin{equation}
g^{-2}_{D3}=(2\pi)^{-1}e^{-\phi_{10}}\equiv S\ .
\end{equation} 
In the case of magnetic F-fluxes on the D7-brane world--volume
the gauge couplings (\ref{gaugedseven})
receive an additional $S$-dependent contribution, cf. \cite{Lust:2004cx,Lust:2005bd}.} 
\begin{equation}\label{gaugedseven}
g^{-2}_{D7_{k}}=(2\pi)^{-1}\ \alpha'^{-2}\ T_k\ ,
\end{equation}

Now we consider CY manifolds which allow for large
volume compactification \cite{Balasubramanian:2005zx,Conlon:2005ki,Conlon:2007xv}.
  Here
one assumes that a set of four-cycles $D^b_{\alpha}$ ($\alpha=1,\dots ,
h^{1,1}_b$) can be chosen arbitrarily large while
keeping the rest of the four-cycles $D^s_{\beta}$ ($\beta=1,\dots
,h^{1,1}-h^{1,1}_b$) small, i.e. $T^b_{\alpha}\gg T^s_{\beta}$.
Since we want the gauge couplings of the SM gauge groups to have finite,
not too small values, we must assume that the SM gauge bosons originate from
D7-branes wrapped around the small 4-cycles $D^s_{\beta}$.
This splitting of the four-cycles into big and small cycles
is only possible, if the CY triple
intersection numbers form a specific pattern.
In addition, the Euler number of the CY space
must be negative, i.e. $h^{2,1}>h^{1,1}>1$.
For a simple class of CY spaces with this property the overall volume
$V_6$ is controlled by
one big four-cycle $T^b$, and the volume has to take the  form
\begin{equation}
V_6\sim  (T^b)^{3/2}-h(T^s_\beta)\ ,
\end{equation}
where $h$ is a homogeneous function of the small K\"ahler moduli $T^s_\beta$ of
degree 3/2.
E.g. one may consider the following more specific volume form:
\begin{equation}
V_6\sim (T^b)^{3/2}-\sum_{\beta=1}^{h^{1,1}-1}(T^s_\beta)^{3/2}\ .
\end{equation}
Looking from the geometrical point of view, these models have a "Swiss cheese"
like structures, with holes
inside the CY-space given by the small four-cycles.

The simplest example of a Swiss cheese example is the CY manifold
${\bf P}_{[1,1,1,6,9]}[18]$ with $h^{1,1}=2$.
In terms of the 2-cycles the volume is given by
\begin{equation}
V_6=6\ ({t}_1^{3}+{t}_2^{3})\ .
\end{equation}
According to eq.(\ref{physT}) the corresponding 4-cycle volumes become:
\begin{eqnarray}
T^b & =&{\partial V_6\over\partial {t}_1}=18\ 
{t}_1^2\quad\Longleftrightarrow\quad {t}_1=
{\sqrt{T^b}\over 3\sqrt 2}\ ,\nonumber\\
T^s & =&{\partial  V_6\over\partial {t}_2}=18\ 
{t}_2^2\quad\Longleftrightarrow\quad {t}_2=
-{\sqrt{T^s}\over 3\sqrt 2}\ .
\end{eqnarray}
Then the volume can be written  in terms of the 4-cycles as
\begin{equation}
V_6={1\over 9\sqrt 2}\ \biggl[\ (T^b)^{3/2}-(T^s)^{3/2}\ \biggr]\ .
\end{equation}

So far only discussed the algebraic structure of the CY spaces that allow for large volume compactifications.
The next step is then to show that minima with large 4-cycle volumes can be indeed found
in the effective potential. This problem was addressed in \cite{Conlon:2005ki,Conlon:2007xv}. They
used the standard KKLT 3-form flux superpotential plus the non-perturbative D3-instanton contribution,
as given in equation (\ref{kklt}).
In addition, in order to get large four-cycle volumes, perturbative $\alpha'$ corrections have to be included
into the tree level K\"ahler potential, which then reads:
\begin{equation}
K=-2\ln\biggl(V_6+{\xi\over 2g_s^{3/2}}\biggr)-\ln\biggl(S+\bar S\biggr)-\ln\biggl(-i\int_{CY}\Omega\wedge\bar
\Omega\biggr)\, .
\end{equation}
Here the parameter $\xi$ is related to the Euler number $\chi$ of the CY-space as follows:
\begin{equation}
\xi=-{\zeta(3)\chi\over 2(2\pi)^3}\, .
\end{equation}
Analyzing the structure of the effective potential one can show the minima with large overall
volume can indeed occur. Many of the phenomenological
properties of large volume compactifications were discussed. Assuming that supersymmetry is
spontaneously at an intermediate string scale, $M_{\rm string}={\cal O}(10^{11}{\rm GeV})$
in the hidden sector of the theory, the pattern  of the supersymmetry breaking soft terms
in the observable sector were discussed in \cite{Conlon:2006wz,Conlon:2007xv}.
In \cite{Blumenhagen:2007sm} a toy model with gauge
group $G=U(N)\times Sp(2N)$ and with 
with matter fields on wrapped D7-branes around small cycles and large
overall volume was investigated. This model is based 
on the Swiss cheese example CY manifold ${\bf
P}_{[1,3,3,3,5]}[15]$ with $h^{1,1}=3$. 
Here the conditions for getting a non-perturbative superpotential due to D3-brane
instantons and the possibility to have chiral fermions from open strings were discussed in detail.
However the matter content of this model was not very realistic, and also not all complex structure
moduli could be fixed by the 3-form flux superpotential.
Hence, certainly a larger class of large volume CY-spaces has to be investigated in order
to find consistent SM realizations by D-branes.

\subsection{Type IIA flux compactifications}

\subsubsection{Type IIA $AdS_4$ vacua}

Now we want to switch from type IIB flux vacua to type IIA flux compactifications to 4-dimensional
anti-de Sitter vacua 
\cite{Behrndt:2000tr,Behrndt:2004km,Derendinger:2004jn,Lust:2004ig,Villadoro:2005cu,DeWolfe:2005uu,Camara:2005dc}.
The main motivations to view into the type IIA, $AdS_4$ flux landscape are the following:

\begin{itemize}
\item
In type IIA, $AdS_4$ flux vacua all moduli are generically fixed. As we discuss
in the following, one can construct several
explicit examples with stabilized moduli in the context of massive type IIA supergravity.
\item After a proper uplift to a de Sitter vacuum, type IIA flux compactifications may serve
as a good basis for string inflation or obtaining a small cosmological constant.
Some aspects of this will be discussed in section (5.3.1).
\item Some of the $AdS_4$ flux vacua can be realized in terms of branes. As we
discuss in section 4.5., these branes provide
interpolating domain wall solutions between $AdS_4$ and flat 4-dimensional Minkowski space-time,
which may induce transitions between different flux vacua (see section 4.6).
\item
Type IIA, $AdS_4$ flux vacua are conjectures to be dual to 3-dimensional Chern-Simons
gauge theories, which are conformal field theories in the three dimensions. One known
example is type IIA on the background $AdS_4\times CP^3$ \cite{Aharony:2008ug}.
In would be interesting to determine the 3-dimensional Chern-Simons theories that are dual
to massive $AdS_4$ supergravity, discussed in the following.

\end{itemize}

In this section we give a short review on the form of the $AdS_4$ flux vacua
in massive IIA supergravity.
Up to now all  explicit ten-dimensional examples of ${\cal N}=1$ supersymmetric compactifications to AdS$_4$ fall within the class of type IIA SU(3)-structure compactifications and T-duals thereof.\footnote{Recently $AdS_4$ compactifications with $SU(3)\times SU(3)$
group structure were discussed in \cite{Lust:2009zb}.}
The most general form of $\mathcal{N}=1$ compactifications of IIA supergravity to AdS$_4$ with the
ansatz $\eta^{(1)} \propto \eta^{(2)}$ for the internal supersymmetry generators
(the strict SU(3)-structure ansatz) was given  in \cite{Lust:2004ig}. These vacua must have
constant warp factor and constant dilaton, $\Phi$. Setting the warp factor to one, the solutions of 
\cite{Lust:2004ig} are given by:
\begin{eqnarray}
\label{ltsol}
H &=&\frac{2m}{5} e^{\Phi}\Re \Omega \, ,\\
F_2&=&\frac{f}{9}J+F'_2 \, , \\
F_4&=&f\mathrm{vol}_4+\frac{3m}{10} J\wedge J \, , \\\label{21d}
W e^{i \theta} &=&-\frac{1}{5} e^{\Phi}m+\frac{i}{3} e^{\Phi }f \, .
\end{eqnarray}
where $H$ is the NSNS three-form, and $F_{n}$ denote the RR forms.
As before, ($J$, $\Omega$) is the SU(3)-structure of the internal six-manifold.
$f$, $m$ are constants parameterizing the solution: $f$ is the Freund-Rubin parameter, while $m$
is the mass of Romans' supergravity \cite{Romans:1985tz}, which can be identified with the type IIA flux $F_0$. $e^{i\theta}$ is a phase associated with the internal supersymmetry generators:
$\eta^{(2)}_+ = e^{i \theta} \eta^{(1)}_+$.
The constant $W$ is defined by the following relation for the AdS$_4$ Killing spinors, $\zeta_\pm$,
\begin{equation}
\label{defW}
\nabla_{\mu} \zeta_- = \frac{1}{2} W \gamma_{\mu} \zeta_+ \, ,
\end{equation}
so that the radius of AdS$_4$ is given by $|W|^{-1}$. The two-form $F'_2$ is the primitive part of
$F_2$ (i.e.\ it is in the $\bf{8}$ of SU(3)). It is constrained by the Bianchi identity:
\begin{equation}
\d F'_2=(\frac{2}{27}f^2-\frac{2}{5}m^2 ) e^{\Phi} \Re \Omega - j^{6} ~,\label{ltsolb}
\end{equation}
where we have added a source, $j^6$, for D6-branes/O6-planes on the right-hand side.

Finally, the only non-zero torsion classes of the internal manifold are ${\cal W}^-_1,{\cal W}^-_2$:
\begin{equation}
{\cal W}^-_1=-\frac{4i}{9} e^{\Phi} f  \, , \qquad
{\cal W}^-_2=-i e^{\Phi} F'_2 \,  .
\label{ltsolc}
\end{equation}
For a given geometry to correspond
to a vacuum without orientifold sources, one finds that the following bound on
(${\cal W}_1^-,{\cal W}_2^-$) has to be satisfied
\begin{equation}
\frac{16}{5} e^{2\Phi} m^2 = 3|{\cal W}^-_1|^2-|{\cal W}^-_2|^2\geq 0
~.
\label{condition}
\end{equation}

The constraint (\ref{condition}) can however be relaxed by allowing
for an orientifold source, $j^{6}\neq 0$. As a particular example, let us consider:
\begin{equation}
j^{6}=-\frac{2}{5}e^{-\Phi}\mu \Re\Omega~,
\label{jo}
\end{equation}
where $\mu$ is an arbitrary, {\it discrete}, real parameter of dimension (mass)$^2$, so that $-\mu$  is proportional to
the orientifold/D6-brane charge
($\mu$ is positive for net orientifold charge  and negative for net D6-brane charge).
The addition of this source term  was also considered in \cite{Acharya:2006ne,Villadoro:2007tb}.
Eq.~(\ref{jo}) above guarantees that the calibration conditions, which
for D6-branes/O6-planes read
\begin{equation}
\label{calcond}
j^6 \wedge \Re \Omega = 0 \, , \qquad j^6 \wedge J = 0 \, ,
\end{equation}
are satisfied and thus the source wraps supersymmetric cycles.
In fact, using the supersymmetry conditions in the presence of calibrated sources together with the Bianchi
identities one can prove a useful integrability theorem, namely that the equations of motion are satisfied in the
presence of sources \cite{Koerber:2007hd}.
The bound (\ref{condition}) changes to
\begin{equation}\label{boundtra}
e^{2\Phi} m^2=\mu+\frac{5}{16}\left(3|{\cal W}^-_1|^2-|{\cal W}^-_2|^2\right) \ge 0 \, .
\end{equation}
Since $\mu$ is arbitrary the above equation
can always be satisfied, and therefore no longer imposes any constraint on the
torsion classes of the manifold.

 The corresponding supersymmetric solutions are all of the form $AdS_4\times X_6$, where
$X_6$ is a certain space which possesses an $SU(3)$ group structure. Recently constructed
examples include flat 6-dimensional tori, Nilmanifolds (twisted tori) and several homogenous
coset spaces  
\cite{Aldazabal:2007sn,Tomasiello:2007eq,Koerber:2008rx,Caviezel:2008ik}.
In addition to the mass parameter $m$, several other internal
fluxes, like $H_3$, $F_2$ and $F_4$, are also needed in order to preserve supersymmetry, 
to satisfy all Bianchi identities and to get everywhere regular solutions with finite scalar fields.
First, the  table 3 shows the coset spaces solutions for massive type IIA supergravity.\footnote{These coset spaces also appeared
as backgrounds for heterotic string compactifications already some time ago in \cite{Lust:1986ix,Castellani:1986rg}.}
\begin{table}[h!]
\begin{center}
\begin{tabular}{|c|c|c|c|c|c|}
\hline
& \rule[1.2em]{0pt}{0pt} $\frac{\text{G}_2}{\text{SU(3)}}$& $\frac{\text{Sp(2)}}{\text{S}(\text{U(2)}\times \text{U(1)})}$ & $\frac{\text{SU(3)}}{\text{U(1)}\times \text{U(1)}}$ & SU(2)$\times$SU(2) & $\frac{\text{SU(3)}\times \text{U(1)}}{\text{SU(2)}}$\\
\hline
Light fields & 4 & 6 & 8 & 14 & 8 \\
Unstabilized & 0 & 0 & 0 & 1 & 0 \\
Decouple KK & no & yes & yes & yes & no\\
$R<0$ possible & no & yes & yes & yes & yes\\
\hline
\end{tabular}
\caption{Results for the coset spaces}
\label{cosettable}
\end{center}
\end{table}
In the massless limit, the $\mathbb{CP}^3$ coset reduces to the 
ABJM case, and the supersymmetry is enhanced discontinuously from 
${\cal N}=1$ to ${\cal N}=6$.

The other class solutions for massive $AdS_4$ supergravity require besides the fluxes
also orientifold sox-planes due to some otherwise uncanceled tadpoles:
\begin{table}[h!]
\begin{center}
\begin{tabular}{|c|c|c|}
\hline
IIA & IIB & IIA\\
\hline
T$^6$ & nilmanifold 5.1 & Iwasawa \\
D4/D8/NS5& D3/D5/D7/NS5/KK & D2/D6/KK\\
\hline
\end{tabular}
\caption{Brane picture}
\label{branepic}
\end{center}
\end{table}
It is interesting to note that these flux geometries also allow for an equivalent brane
interpretation of intersecting branes, where the branes act precisely as the sources for
the non-vanishing fluxes. In this case, the Romans mass parameter $m$ always corresponds to the 
presence of $m$ D8-branes. As we will discuss in section (4.5),  the flux geometries $AdS_4\times X_6$
always arise as the near horizon geometry of the corresponding intersecting brane configurations.

A specific massive  IIA solution is the first example in table 4, where the internal  manifold to be a six-dimensional torus.
All torsion
classes vanish in this case. Note, however, that there are non-vanishing $H$ and $F_{4}$
fields given by (\ref{ltsol})
\begin{eqnarray}
H & =& \frac{2}{5} e^{\Phi} m \left(e^{246}-e^{136}-e^{145}-e^{235}\right) \, , \\
F_4 & = &\frac{3}{5} m \left( e^{1234} + e^{1256} + e^{3456} \right) \, .
\end{eqnarray}
{} From (\ref{boundtra}) we find that there is an orientifold source of the type (\ref{jo})
with $\mu=e^{2\Phi} m^2$, which corresponds to smeared orientifolds along $(1,3,5)$, $(2,4,5)$, $(2,3,6)$
and $(1,4,6)$. The corresponding orientifold involutions are
\begin{eqnarray}
\label{torusinv}
O6: &  &\qquad e^2 \rightarrow -e^2 \, , \quad e^4 \rightarrow -e^4 \, , \quad e^6 \rightarrow -e^6 \, , \\
O6: & & \qquad e^1 \rightarrow -e^1 \, , \quad e^3 \rightarrow -e^3 \, , \quad e^6 \rightarrow -e^6 \, , \\
O6: &  &\qquad e^1 \rightarrow -e^1 \, , \quad e^4 \rightarrow -e^4 \, , \quad e^5 \rightarrow -e^5 \, , \\
O6: & &\qquad e^2 \rightarrow -e^2 \, , \quad e^3 \rightarrow -e^3 \, , \quad e^5 \rightarrow -e^5 \, .
\end{eqnarray}

\subsubsection{Type IIA effective flux potentials}

Now we turn to the 4-dimensional effective action description of type IIA, $AdS_4$ flux vacua.
It was shown for several of the above examples that the spectrum obtained from
the effective action matches 
matches the results from direct dimensional reduction on these spaces \cite{Caviezel:2008ik}.

The type IIA effective superpotential receives three kinds of
contributions (see e.g.
\cite{Derendinger:2004jn,Villadoro:2005cu,DeWolfe:2005uu,Camara:2005dc,Villadoro:2007yq,Kounnas:2007dd}):
\begin{equation}\label{IIAfull}
W_{\mathrm{IIA}}=W_H+W_F+W_{\mathrm{geom}}\, .
\end{equation}
The first term is due to the Neveu--Schwarz 3-form fluxes and
depends on the dilaton $S$ and the type  IIA complex-structure
moduli $U_m$ ($m=1,\dots ,\tilde h^{2,1}$):
\begin{equation}W_H(S,U)=\int_Y\Omega_c\wedge H_3=i\tilde a_0S+i\tilde c_mU_m\, ,
\end{equation}
where in type IIA the 3-form $\Omega_c$ is defined by
$\Omega_c=C_3+i\mathrm{Re}(C\Omega)$. Second, we have the
contribution from Ramond 0-, 2-, 4-, 6-form fluxes (the 0-form
flux corresponds to the mass parameter $\tilde m_0$ in massive IIA
supergravity):
\begin{eqnarray}
W_F(T) &=&\int_Y \mathrm{e}^{J_c}\wedge F^{\mathrm{R}}\nonumber
\\ &=&\tilde m_0\frac{1}{6}\int_Y\left(J_c\wedge J_c\wedge
J_c\right) +\frac{1}{2}\int_Y\left(F_2^{\mathrm{R}}\wedge
J_c\wedge J_c\right)+
\int_YF_4^{\mathrm{R}}\wedge J_c+\int_YF_6^{\mathrm{R}}\nonumber \\
&=&i\tilde m_0F_0(T)-\tilde m_iF_i(T)+i\tilde e_iT_i+\tilde
e_0\,\label{IIAF} .
\end{eqnarray}
Here $F(T):= F_0(T)$ is the type IIA prepotential, which depends
on the IIA K\"ahler moduli $T_i$ ($i=1,\dots ,\tilde h^{1,1}$) and
$F_i(T):=\partial F_0/\partial T_i$. We use the notation $J_c$ for
the complexified K\"{a}hler metric $J_c:=B+iJ$. Finally we have
the contribution of the geometrical (metric) fluxes, which
captures the non-Calabi--Yau property of $Y$:
\begin{eqnarray}
W_{\mathrm{geom}}(S,T,U)=i\int_Y\Omega_c\wedge \mathrm{d}J=-\tilde
a_iST_i-\tilde d_{im}T_iU_m\, ,\label{IIAg}
\end{eqnarray}
where the metric fluxes $\tilde a_i$, $\tilde d_{im}$ parameterize
the non-vanishing of $\mathrm{d}J$.

The type IIA Ramond tadpole follows from the equation of motion of
the field $C_7$. Specifically it is of the form
\cite{Camara:2005dc}:
\begin{equation}\label{iiatadpole}
\tilde N_{\rm flux}=\int\left(C_7\wedge
\mathrm{d}F_2+C_7\wedge(\tilde a_0H_3+\mathrm{d}\bar F_2)\right)\,
,
\end{equation}
where $G_2=\mathrm{d}C_1+\tilde a_0B_2+\bar F_2$ and
$^*F_2=F_8=\mathrm{d}C_7$. The metric fluxes $\tilde a_i$
contribute to $\mathrm{d}\bar F_2$, and one gets for non-vanishing
fluxes $\tilde a_I$ and $\tilde m_I$ that
\begin{equation}\label{naflux}
\tilde N_{\rm flux}=\sum_{I=0}^{\tilde h^{1,1}}\tilde a_I\tilde
m_I\, .
\end{equation}
This non-vanishing flux tadpole, which corresponds to a
non-vanishing D6-brane charge,  must be cancelled by the orientifold O6-planes and an appropriate
number of  D6-branes:
\begin{equation}
\label{tadpoleiia1} \tilde N_{\rm
flux}+N_{\mathrm{D6}}=2N_{\mathrm{O6}}\, .
\end{equation}

We now come to the crucial point of generating supersymmetric
$\mathrm{AdS}_4$ ground states in type IIA with all main moduli
stabilized. 
The superpotential
must depend on all chiral fields for the vacuum energy to be
negative with unbroken supersymmetry. Following
\cite{Derendinger:2004jn, DeWolfe:2005uu, Camara:2005dc} we will
concentrate on the case without metric fluxes, i.e. $\tilde
a_i=\tilde d_{im} =0$. Furthermore, the 6-form fluxes as well as
the 2-form fluxes can be shown to be gauge dependent and hence can
be set to zero: $\tilde e_0=\tilde m_i=0$ \cite{DeWolfe:2005uu,
Camara:2005dc}. The fluxes $\tilde m_0$ and $\tilde a_0$ must be
non-zero for $W$ to be kept non-vanishing. Finally, we combine
$\tilde e_i$ and $\tilde m_0$ as
\begin{equation}
\gamma_i=\tilde m_0\tilde e_i\, .
\end{equation}

We will now assume that the internal space $Y$ is simply given by the product of three 2-tori or
an orbifold of it. This space has
 a simple (toroidal) cubic prepotential $F=T_1T_2T_3$, and the
superpotential has the generic form:
\begin{eqnarray}
W_{\mathrm{IIA}}&=&W_F+W_H= \tilde m_0\int_Y(J\wedge J\wedge J)+
\int_YF_4^{\mathrm{R}}\wedge J+
\int_Y\Omega_c\wedge H_3\nonumber\\
&=&i\tilde e_iT_i+i\tilde m_0T_1T_2T_3+i\tilde a_0S+i\tilde
c_mU_m\, .\label{IIAU}
\end{eqnarray}
Using Eq. (\ref{naflux}), the D6-tadpole of corresponding fluxes
is simply given by
\begin{equation}\label{nafluxa}
\tilde N_{\rm flux}=\tilde a_0\tilde m_0\, .
\end{equation}
According to Eq. (\ref{tadpoleiia1}) this number has to be
balanced by the D6-branes and the O6-planes.

We may also consider the generalization to the case where the
prepotential is given by $F=\frac{1}{6}c_{ijk}T_iT_jT_k$. In CY
compactifications, the $c_{ijk}$ would be the classical triple
intersection numbers and the corresponding superpotential would
read:
\begin{eqnarray}
W_{\mathrm{IIA}}&=&W_F+W_H =i\tilde e_iT_i+i\tilde
m_0\frac{1}{6}c_{ijk}T_iT_jT_k+i\tilde a_0S+i\tilde c_mU_m\,
.\label{2.18}
\end{eqnarray}
However, in order to keep the algebra simple we will focus in the
following on the toroidal prepotential with $c_{ijk}=1$.

Coming back to the  superpotential (\ref{IIAU}),
and K\"ahler potential $K= - \log (S+\bar S) \prod_{i=1}^3(T_i+\bar
T_i)\prod_{i=1}^3(U_i+\bar U_i)$, the equations $F_{\phi_i}=0$
admit the following solution:
\begin{equation}\label{solutionIIAg}
|\gamma_i|T_i =\sqrt{\frac{5|\gamma_1\gamma_2\gamma_3|}{3 \tilde
m_0^2}}\, ,\quad S=-\frac{2}{3 \tilde m_0\tilde a_0}\gamma_iT_i\,
,\quad \tilde c_m U_m=-\frac{2}{3 \tilde m_0}\gamma_i T_i\, .
\end{equation}
This solution corresponds to supersymmetric $\mathrm{AdS}_4$
vacua.
The vacuum energy, i,.e. the $AdS_4$ cosmological constant is given
by the  following expression, which entirely depends on that quantized flux
quantum numbers;
\begin{equation}
\Lambda_{AdS}=-3e^K|W|^2= -{3^7\sqrt{3\over 5}\over 100}{|\tilde a_0\tilde c_1\tilde c_2\tilde c_3|(|\tilde m_0\tilde e_1\tilde e_2 \tilde e_3|)^{5/2}\over(\tilde e_1\tilde e_2\tilde e_3)^4}M_P^4\, .   \label{lambdaads}
\end{equation}

Let us end the present section by discussing some T-dual/mirror
transforms of the of the IIA models. 
T-duality will in general transform the NS-fluxes into geometrical
fluxes. We can for instance investigate within the toroidal models
the T-duality transformation in the internal directions $x^1$ and
$x^2$, acting as $T_1\rightarrow 1/T_1$, $T_{2,3}\rightarrow
T_{2,3}$. Then the T-dual superpotential of Eq. (\ref{IIAU})
becomes
\begin{equation}
W_{\mathrm{IIA}}=\tilde e_1 + \tilde e_2 T_1T_2+ \tilde m_0 T_2T_3
+ \tilde e_3 T_3 T_1 + \tilde a_0 S T_1 + \tilde c_mT_1U_m\,
.\label{IIAUtdual}
\end{equation}
The fluxes $\tilde a_0$ and $\tilde c_m$ become now geometrical.
The corresponding $\mathrm{AdS}_4$ ground states can be simply
obtained by replacing $T_1$ by $1/T_1$ in Eq.
(\ref{solutionIIAg}).

Alternatively let us consider the IIB mirror transform of the
superpotential (\ref{IIAU}), which is obtained by applying
T-duality transformations in the three directions $x^1$, $x^3$ and
$x^5$. This exchanges the IIA K\"ahler moduli by the IIB
complex-structure moduli and vice versa. In the presence of the
IIA NS-fluxes $\tilde c_m$ as in (\ref{IIAU}), the type IIB mirror
superpotential will necessarily contain geometrical fluxes:
\begin{equation}
W_{\mathrm{IIB}} = i\tilde e_i U_i + i \tilde m_0 U_1 U_2 U_3 + i
\tilde a_0 S + i \tilde c_m T_m\, .\label{IIBG}
\end{equation}
In this case, T-duality takes the system away from the original CY
framework.

\subsection{$AdS_4$ domain wall solutions}

As we discussed in the previous sections, supersymmetric
$\mathrm{AdS}_4$ can appear as ground states of type IIA flux
compactifications. Our aim here is precisely to characterize the sources that
generate the fluxes necessary for the compactifications under
consideration. This complementary, or dual picture, gives another
perspective to the emergence of $\mathrm{AdS}_4$. The latter
appears as near-horizon geometry of a certain distribution of
intersecting/smeared branes and calibrated sources that act as domain walls,
 connecting $\mathrm{AdS}_4$ to an asymptotically
flat region. (Domain wall solutions and flow equations were also investigated within group structure manifolds and generalized geometry 
in \cite{Koerber:2008rx,Mayer:2004sd,Smyth:2009fu,Haack:2009jg}.)
The brane picture is the first step towards
the counting of microscopic states. From the viewpoint of
four-dimensional gauged supergravity, one could presumably go
further and consider the attractor equations and the macroscopic
entropies. This is outside the scope of the present paper.

As we will see explicitly, the appearance of
$\mathrm{AdS}_4$ as near-horizon geometry requires that all
branes have two common spatial directions in non-compact
four-dimensional space-time, i.e. they have the geometry of a
domain wall in four dimensions. Moreover, depending on their
dimensionality, the branes will fill part of the internal space
$M_6$. Keeping this structure in mind, let us summarize, the
relations between the various fluxes and the corresponding source
branes \cite{Kounnas:2007dd}.
 \begin{itemize}
 \item
 For a Neveu--Schwarz 3-form flux $H_3$ through a 3-cycle $\Sigma_3$ inside
 $M_6$ the sources are NS5-branes wrapped around the dual 3-cycle $\tilde\Sigma_3$.
 \item
 In the Ramond sector we have fluxes of the Ramond field strengths $F_n^{\mathrm{R}}$ through
 some internal $n$-dimensional cycles $\Sigma_n$. The desired domain-wall configuration in space-time,
 requires that these fluxes be generated by magnetic brane sources, namely by D($8-n$)-branes,
 wrapped around internal cycles $\tilde\Sigma_{6-n}$
 dual to $\Sigma_n$.
 \item For  geometrical fluxes we expect to have Kaluza--Klein monopoles as sources.
 In fact, performing T-duality to directions orthogonal to the NS5-brane, one obtains a KK-monopole.
 \end{itemize}

The fluxes are quantized and this reflects that the number of
branes is not arbitrary. 

Any self-consistent system of space-time-filling branes must obey
the tadpole cancellation condition, which is a consequence of the
(generalized) Gauss law. Alternatively, in supersymmetric
configurations, this condition can be thought of as arising from
the integrated Bianchi identities. Specifically for 
the following D4/D8/NS5 example , we will need the D6/O6
tadpole cancellation condition:
\begin{equation} \frac{1}{2\pi\sqrt{\alpha^{\prime}}}\int_{\Sigma} F_0~ H_3+
N_{\mathrm{D6}}-2N_{\mathrm{O6}}=0~, \label{tadp}
\end{equation}
where  $N_{\mathrm{D6}}$, $N_{\mathrm{O6}}$ is the total number
space-time-filling D6-branes, O6-planes wrapping a three-cycle
$\Sigma$ in the internal space.

 For these
  cases it is significant that whenever stabilization is complete,
  the values of the moduli found by minimizing the scalar
  potential are recovered by a careful analysis of the space-time
  background fields near the horizon. In particular the dilaton approaches a
  finite constant in this limit. 

  The examples under consideration here are the following:
\begin{itemize}
  \item Configuration with D4/D8/NS5 branes. This model contains in particular
 four stacks of intersecting NS5-branes. The background is a IIA ground state
  of the superpotential (\ref{IIAU}) with \emph{all} terms non-vanishing:
  $W_{\mathrm{IIA}}=i\tilde e_iT_i+i\tilde m_0T_1T_2T_3+i\tilde a_0S+i\tilde c_mU_m$.
  This allows for full moduli stabilization (Eqs. (\ref{solutionIIAg})).
  \item The next model is obtained by performing a T-duality along one
  direction (say $x^1$). This is a type IIB model with
  D3/D5/D7/NS5/KK-branes/monoples. Its superpotential, generated by $F_1, F_3, F_5, H_3$ and geometric fluxes,
  reads $W_{\mathrm{IIB}} = i(\tilde{e}_1 U_1 + \tilde{c}_2 U_2 + \tilde{c}_3 U_3  + \tilde{c}_1 T_1 +
  \tilde{e}_2 T_2 + \tilde{e}_3 T_3 + \tilde{m}_0 U_1 T_2 T_3 + \tilde{a}_0 S)$,  and it exhibits
  an $\mathrm{AdS}_4$ vacuum with all moduli stabilized. The latter appears as the
  near-horizon geometry of the brane/monopole configuration at hand.
   \end{itemize}

Specifically, the first example is given by the
following system of intersecting D4/D8/NS5-branes \cite{Kounnas:2007dd}:
%
\begin{center}
  \begin{tabular}{|c||c|c|c|c|c|c|c|c|c|c|}
    \hline
    & $\xi^0$ & $\xi^1$  & $\xi^2$  & $y$ & $x^1$  & $x^2$  & $x^3$  & $x^4$  & $x^5$  & $x^6$ \\
    \hline
    \hline
    $\mathrm{D}4$ & $\bigotimes$ & $\bigotimes$  & $\bigotimes$ &  &$\bigotimes$  &  $\bigotimes$ &   &
                &   &  \\

    \hline
    $\mathrm{D}4^{\prime}$ & $\bigotimes$ & $\bigotimes$  & $\bigotimes$ &  &  &  & $\bigotimes$  &
               $\bigotimes$  &   &  \\

    \hline
    $\mathrm{D}4^{\prime\prime}$ & $\bigotimes$ & $\bigotimes$  & $\bigotimes$ &   &   &  &   &
                 & $\bigotimes$  & $\bigotimes$ \\
\hline
    $\mathrm{NS}5$ & $\bigotimes$ & $\bigotimes$  & $\bigotimes$ &   &  $\bigotimes$  &   &  $\bigotimes$ &
                 & $\bigotimes$  & \\
\hline
    $\mathrm{NS}5^{\prime}$ & $\bigotimes$ & $\bigotimes$  & $\bigotimes$ &   &  $\bigotimes$  &  &   &  $\bigotimes$
                 &  & $\bigotimes$ \\
 \hline
    $\mathrm{NS}5^{\prime\prime}$ & $\bigotimes$ & $\bigotimes$  & $\bigotimes$ &   &  & $\bigotimes$ &   &
               $\bigotimes$    &  $\bigotimes$  & \\
\hline
    $\mathrm{NS}5^{\prime\prime\prime}$ & $\bigotimes$ & $\bigotimes$  & $\bigotimes$ &   &  & $\bigotimes$ &  $\bigotimes$     &
               &  &  $\bigotimes$ \\
 \hline
    $\mathrm{D}8$ & $\bigotimes$ & $\bigotimes$  & $\bigotimes$ &   &  $\bigotimes$ & $\bigotimes$ & $\bigotimes$  &  $\bigotimes$
                 & $\bigotimes$  & $\bigotimes$ \\
\hline
   \end{tabular}
\end{center}
%
These generate a supersymmetric solution of IIA supergravity in
the presence of supersymmetric (calibrated) sources. The tadpole cancellation condition induces in
addition O6-planes and/or D6-branes. These brane distributions
should thus be referred to as D4/D8/NS5/O6/D6.

In the following we will list a few important facts about this intersecting brane solution:

\begin{itemize}

\item
The solution is supersymmetric. It can be explicitly given in terms of almost harmonic functions
that describe a smearing of the branes over the transverse space.

\item The near horizon form of the solution is $AdS_4\times T^6$. At the other spatial limit
the solution just describes $R^{1,3}\times T^6$. Hence this solution is an interpolating domain
wall solution between these spaces (see figure 14).
\begin{figure}
\begin{center}
  \includegraphics[width=0.5\textwidth]{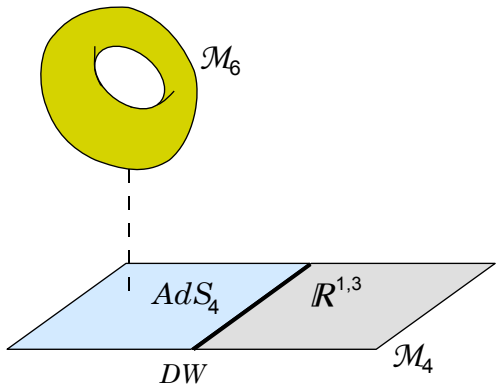}
\end{center}
\caption{A domain wall separating a region of $AdS_4$ from a region of $R^{1,3}$. The internal
manifold ${\cal M}_6$ is fibered over ${\cal M}_4$.}
\end{figure}
\item
At the horizon we recover that the scalar fields precisely take those fixed 
values, which were derived for the moduli,in the effective-superpotential description above.

\end{itemize}

Let us now perform a T-duality along $x^1$.  This results in the
following flux superpotential:
\begin{equation}
 W_{\mathrm{IIB}}=i\tilde{e}_1U_1+
i\tilde{c}_2 U_2+i\tilde{c}_3 U_3+i\tilde{a}_0
S+i\tilde{c}_1T_1+i\tilde{e}_2 T_2+i \tilde{e}_3T_3+i\tilde{m}_0
U_1T_2T_3\, ,
\end{equation}
 where the last four terms are the geometrical-flux
contributions and guarantee all-moduli stabilization around the
type IIB $\mathrm{AdS}_4$ vacuum, as in the type IIA mirror
situation.

The corresponding brane configuration is now \cite{Kounnas:2007dd}:
\bigskip
\begin{center}
  \begin{tabular}{|c||c|c|c|c|c|c|c|c|c|c|}
    \hline
    & $\xi^0$ & $\xi^1$  & $\xi^2$  & $y$ & $x^1$  & $x^2$  & $x^3$  & $x^4$  & $x^5$  & $x^6$ \\
    \hline
    \hline
    $\mathrm{D}3$ & $\bigotimes$ & $\bigotimes$  & $\bigotimes$ &  &  & $\bigotimes$  &   &
                &   &  \\
    \hline
    $\mathrm{D}5$ & $\bigotimes$ & $\bigotimes$  & $\bigotimes$ &   &  $\bigotimes$   &  & $\bigotimes$  &  $\bigotimes$
                 &   & \\
    \hline
    $\mathrm{D}5^{\prime}$ & $\bigotimes$ & $\bigotimes$  & $\bigotimes$ &  &$\bigotimes$   & &   &
                 & $\bigotimes$    &$\bigotimes$    \\

    \hline
    $\mathrm{D}7$ & $\bigotimes$ & $\bigotimes$  & $\bigotimes$ &   &   & $\bigotimes$ & $\bigotimes$    &$\bigotimes$
                 & $\bigotimes$  & $\bigotimes$ \\
\hline
    $\mathrm{NS5}$ & $\bigotimes$ & $\bigotimes$  & $\bigotimes$ &   &  $\bigotimes$  &  &  $\bigotimes$ &
                 & $\bigotimes$  & \\
\hline
    $\mathrm{NS5}^{\prime}$ & $\bigotimes$ & $\bigotimes$  & $\bigotimes$ &   &  $\bigotimes$  & &   &  $\bigotimes$
                 &  & $\bigotimes$ \\
 \hline
    $\mathrm{KK}$ & $\bigotimes$ & $\bigotimes$  & $\bigotimes$ &   &$\bullet$  & $\bigotimes$ &   &
               $\bigotimes$    &  $\bigotimes$  & \\
\hline
    $\mathrm{KK}^{\prime}$ & $\bigotimes$ & $\bigotimes$  & $\bigotimes$ &   & $\bullet$ & $\bigotimes$ &  $\bigotimes$     &
               &  &  $\bigotimes$ \\
\hline
   \end{tabular}
\end{center}
\bigskip
Here $KK$ denotes the Kaluza-Klein monopoles that are T-dual to NS5-branes. Note that the near
horizon geometry is not anymore a six-dimensional torus, but the T-dual near horizon
geometry is that of a {\em twisted torus}, or a {\em nilmanifold}, i.e. $AdS_4\times N_{5,1}$.

\subsection{Transitions in the flux landscape}

In order to get transitions between vacua with different flux quantum numbers,
one needs non-perturbative, gravitational configurations which are coupled
to the flux background fields, and which interpolate between different flux
vacua. These are given in terms of BPS or nearly BPS domain walls (membranes) 
(for earlier work see e.g. \cite{Cvetic:1991vp,Cvetic:1992bf,Cvetic:1996vr,Behrndt:2001mx,Louis:2006wq})
in four-dimensional
space time that are coupled to the scalar moduli fields.
The profile of the domain wall is such that it separates spatial regions with different flux quantum
numbers from each other. For the case that the domain wall is interpolating between two 
supersymmetric vacua, the interpolating solutions is describing a BPS domain wall.
Of course, eventually we are interested in the decay of a non-supersymmetric flux
vacuum with positive cosmological constant (our vacuum) and broken space-time supersymmetry
into another (supersymmetric) flux vacuum, which cam have either positive, zero or also
negative cosmological constant ($AdS_4$) vacuum. The formation of an $AdS_4$ domain wall
is particularly interesting, since $AdS_4$ are very common in the string landscape.
In this case our universe would be decaying into a contracting space, which at first sight seems
to be problematic. Nevertheless the corresponding transition amplitude 
from $dS_4$ to $AdS_4$ is expected to be non-vanishing,
as it was  discussed in \cite{Ceresole:2006iq}.

Now let us consider the corresponding the domain wall solution which interpolates 
between the above $AdS_4$ flux vacuum and flat Minkowski space-time with vanishing
fluxes. As discussed in the previous section it is given in terms of interesting D4,- D8- and NS 5-branes.
In addition one also needs orientifold 6-planes (O6) in order to cancel the induces
D6-brane charge from the fluxes. 
The
four dimensional part of the metric is such of an interpolating domain wall, where the
intersecting branes are smeared in the direction transversal to the domain wall.
Specifically, this 4-dimensional part of the metric can be written as 
\begin{equation}
{\rm d}s^2=a(r)^2(-{\rm d}t^2+{\rm d}x^2+{\rm d}y^2)+{\rm d}r^2\, .\label{domainmetric}
\end{equation}
For $r\rightarrow 0$ this metric approaches the metric of $AdS_4$, and the scalar fields
are fixed to the values determined by the non-vanishing fluxes, as given in eq.(\ref{solutionIIAg}).
For $r\rightarrow \infty$, the function $a(r)$ becomes a constant, and the eq.(\ref{domainmetric})
become the metric of flat Minkowski space. 

The tension $\sigma$ of the domain wall can be computed by introducing a central function
$Z(r)$ which is defines as
\begin{equation}
Z(r)={a'(r)\over a(r)}\, .
\end{equation}
By comparison with the exact metric of  \cite{Kounnas:2007dd} one obtains
\begin{equation}
Z(r)|_{r=0}=e^{K/2}|W|\, ,\quad \Lambda_{AdS}=-3|Z(r)|^2_{r=0}\, .
\end{equation}
The (membrane) tension $\sigma$ of the domain wall is then given by the following expression:
\begin{equation}
\sigma\simeq (|Z|_{r=\infty}-|Z|_{r=0})\, .
\label{tension}
\end{equation}

Now let us determine the decay amplitude of the Minkowski vacuum with vanishing fluxes
into the $AdS_4$ vacuum with non-vanishing fluxes. 
The decay of the Minkowski vacuum occurs due to the creation of the domain wall, which sweeps
through space-time until the entire universe is in the new $AdS_4$ vacuum. This is similar
but not completely equal to the creation of a bubble via the Coleman/De Luccia instanton \cite{Coleman:1980aw}.
In fact in order to be realistic, one
should break supersymmetry and uplift the Minkowski vacuum by a small amount to
obtain a de Sitter vacuum which decays into the $AdS_4$ vacuum. 
Neglecting the problem of supersymmetry breaking and the uplift, the decay amplitude
of the Minkowski (de Sitter)  vacuum is then given by the
following expression:
\begin{equation}
\Gamma \simeq M_P\exp\biggl( -{8\pi^2M_P^4C\over \sigma^2}\biggr)=
M_P\exp\biggl( {24\pi^2M_P^4C\over \Lambda_{AdS}}\biggr)\, .
\end{equation}
The constant $C$ depends on the details of the domain wall solution.

As also discussed in \cite{Ceresole:2006iq}, 
the corresponding decay amplitude is independent of the de Sitter cosmological constant $\Lambda
=V_0$, but  only depends on the value of $\Lambda_{AdS}$.
In order to avoid too fast decay of our vacuum, $|\Lambda_{AdS}|$ must not be too large. 
E.g. if $|\Lambda_{AdS}|\simeq m_{3/2}^4$, the life-time of our universe is long enough.
However $AdS_4$ vacua with $|V_1|\sim M_P^4$  create too much decay of our vacuum.
Using the known expression
for $\Lambda_{AdS}$ in
eq.(\ref{lambdaads}),  this constraint can be translated into the following restriction on the flux 
quantum numbers:
\begin{equation}
{3^7\sqrt{3\over 5}\over 100}{|\tilde a_0\tilde c_1\tilde c_2\tilde c_3|(|\tilde m_0\tilde e_1\tilde e_2 \tilde e_3|)^{5/2}\over(\tilde e_1\tilde e_2\tilde e_3)^4}<<1\, .
\end{equation}

\section{String and brane inflation}

\subsection{General remarks}

\subsubsection{Inflation from scalar fields: Slow roll conditions -- F- and D-term inflation}

Recent astrophysical experiments have provided an enormous amount of high precision
data about the early history of our universe. In fact, we know now that the universe
is spatially flat, i.e. $\Omega=1$, and the latest CMB data from WMAP5 
\cite{Komatsu:2008hk} agree with an
almost scale invariant spectrum with spectral scalar index:
\begin{equation}
n_s=0.96\pm 0.013\,.
\end{equation}
(Note that this value assumes that there is basically no contribution from cosmic strings.
In case cosmic strings contribute to the formation of large scale structures, the
allowed value for $n_s$ might be increased \cite{Bevis:2007gh,Pogosian:2008am}, see also the later discussion.)
Second there can be only small tensor perturbations, the relevant ration $r$ of
tensor to scalar perturbations is bound as follows:
\begin{equation}
r\equiv {\Delta_T^2\over \Delta_\phi^2}<0.30\, .
\end{equation}
These limits will be further improved within the next years, most notably by the mission of
the Planck satellite. 

All these data can be nicely explained by an epoch of cosmic inflation in the early universe (for a nice text book on
cosmology and inflation see \cite{Mukhanov:2005sc}).
Hence the goal will be to use the data from the CMB to find or to probe the fundamental
theory of gravity and matter in the early universe.  This strategy works best focussing on correlated signatures, e.g. on the correlation between $n_s$ and the possible abundance of cosmic strings.
The most common method for this is to use effective scalar field theories wit a scalar potential
$V(\phi)$, where $\phi$ is the inflaton field that drives inflation. Of course, 
a huge collection of effective field theories exist. So we would like to constrain the viable
effective field theories as much as possible, first from the experimental data.
Second, we also like to ask the question, which effective field theory can be consistently embedded
into quantum gravity and into string theory. This will also give us some further constraints
on inflationary effective theories, as we will discuss in the next subsection.

From observations we know that the potential $V(\phi)$ of inflation must be
sufficiently flat. Specifically, the following slow roll conditions have to be satisfied:
\begin{eqnarray}\label{etaep}
\epsilon&=&-{\dot H\over H^2}={M_P^2\over 2}\Biggl({V'(\phi)\over V(\phi)}\Biggr)^2<<1\, ,
\nonumber \\
\eta&=&{\dot \epsilon\over \epsilon H}={M_P^2}\Biggl({V''(\phi)\over V(\phi)}\Biggr)^2<<1\, .
\end{eqnarray}
The smallness of the two slow parameters follows, since the scalar spectral index
is related to $\epsilon$ and $\eta$ as
\begin{equation}
n_s=1-6\epsilon+2\eta \, .
\end{equation}

Slow roll inflation can be roughly achieved in two different ways:
\begin{itemize}
\item Large field inflation: $\phi\geq M_P$.

The prime example for large field inflation is chaotic inflation \cite{Linde:1983gd} with
only a mass term in the inflaton potential:
\begin{equation}
V(\phi)={1\over 2}m^2\phi^2\, .
\end{equation}
Large field inflation generically leads to gravitational waves, i.e. to a large $r$ parameter.
Then the Lyth bound \cite{Lyth} relates the field range of $\phi$ to $r$,
\begin{equation}
r=0.01\Biggl({\Delta\phi\over M_P}\Biggr)^2\, ,
\end{equation}
and hence observations tell us that $\Delta\phi\leq (5-6)M_P$.

\item
Small field inflation: $\phi\leq M_P$.

Here there are no constraints from gravitational waves, but, as we will discuss, the scenario
generically requires large fine tuning of parameters for eqs.(\ref{etaep}) to be satisfied.
In particular, the effective, F-term supergravity scalar potential (\ref{ftermscp}) is not flat at all.
So, assuming a canonical K\"ahler potential for the inflaton field, $K=\phi\bar\phi$, 
the effective supergravity potential generically behaves as
\begin{equation}
V\sim e^{|\phi|^2}\, .
\end{equation}
As we will see, a possible way out is D-term inflation \cite{Binetruy:1996xj,Halyo:1996pp}
 (hybrid inflation  \cite{Linde:1993cn}) where the theory possesses a $U(1)$ shift symmetry, such that the D-term scalar potential does not at all depend on
the inflaton field, but a potential is only generated by Coleman-Weinberg loop effects.

\end{itemize}

\subsubsection{Constraints from black hole decays}

It is usually assumed that from the knowledge of low-energy perturbative physics (e.g., such as, the particle spectrum, and their couplings) in our vacuum, one cannot draw any conclusion about the physics in other vacua  on the landscape, without knowing  the non-perturbative structure of underlying high scale theory.  This belief  is based on the intuition, that different vacua correspond to different non-perturbative solutions of the high energy theory, largely separated by the expectation values 
 of the classical order parameters (e.g., vacuum expectation values (VEVs) of the scalar fields),  
 whereas low energy perturbative  physics only accounts for small fluctuations about this solutions. 
As a result, even in the neighboring vacua, physics may be arbitrarily different and unpredictable for a low energy observer in our vacuum.    
 We  wish to  show that  black hole (BH) physics can provide a powerful guideline for overcoming this obstacle. 
Among, the expected enormity  of the vacuum landscape, there is a large subset that shares common gravitational physics.  In these vacua, the classical black hole physics is also common and imposes 
the same consistency constraints on perturbative particle physics.  

 In particular,  by incorporating the consistency bounds, that BH physics imposes on number and masses of particle species \cite{Dvali:2007hz},  we can derive non-trivial constraints not only on our vacuum, but on any quasi-stationary state, which can be obtained by a continuous deformation  of it.   Under {\it continuous deformation}, we mean a change of expectation values that preserves invariant characteristics of the vacuum (such as,   the number of species, 
their chirality, and possibly other topological characteristics).   
 In a certain well-defined sense, to be made precise below,  BH physics allows us to ``see" through the landscape.  
 In this part of the discussion, the key tool in our consideration will be a BH constraint on number of particle species and their masses.  This bound can be derived from the flat space thought experiment, with BH formation and evaporation.   In this experiment, an observer  forms a classical BH and later  
detects its evaporation products.  In each case, when the lifetime of a BH is less than the lifetime of the species, a powerful bound follows.  For example, in the simplest case  the number of stable species of mass $M$ cannot exceed 
\begin{equation}
\label{nmax}
N_{max} \, \equiv \, {M_P^2 \over M^2}\, .
\end{equation}
 This consistency constraint must be satisfied in every vacuum of the theory.   This fact automatically  limits the number of possible deformations of our vacuum, which 
from perturbative physics alone  one would never guess.  For example,  in our vacuum, {\it a priory}, we may  have a very large  number of massless  species coupled to a modulus $\phi$.   Naively,  nothing forbids existence of another vacuum,  obtained by giving an arbitrary VEV to the modulus $\phi$.  However, since such a deformation of the vacuum gives masses to the species coupled to $\phi$, only 
deformations permitted by the BH bound are possible. Thus, BH physics, automatically constraints 
physics in such vacua.  The vacua in question does not  have to be degenerate with ours, or even be stationary.  Below we shall generalize BH bound for such vacua.   Primary target of this study will be 
the de Sitter and quasi de Sitter vacua, that may be connected to ours by a continuous deformation of some  scalar VEVs.  The phenomenological importance of this study is obvious. 
Existence of such vacua is suggested by the strong cosmological evidence that our Universe  underwent a period of inflation, which is responsible for solving the flatness and the horizon problems, and creating the spectrum of density perturbations.  Knowing that we, most likely,  rolled down from another vacuum, we wish  to understand constraints on such states by using BH physics, and whatever knowledge of perturbative physics we have in our present vacuum. 
The bounds from BH physics, which we discuss in this paper, 
set  powerful criteria about
 what is the class of effective string actions, which can
be consistently coupled to quantum gravity, and eventually capture string physics, which might
have been lost in the effective action approach. Those effective field theories
or vacua which cannot fulfill this criterion are called swampland \cite{Vafa:2005ui} (see also
\cite{ArkaniHamed:2006dz}).

The  generalization of the BH bound to the de Sitter and quasi de Sitter vacua  relies on  certain relations between the Schwarzschild radius  and the lifetime of a ``test"  BH, and  the Hubble radius and the lifetime of the corresponding  (quasi) de Sitter vacuum respectively \cite{Dvali:2008sy}.  
  Shortly, for a given number and masses of species, there is an upper limit on the lifetime and the Hubble size of the vacuum, or else the BH bound (\ref{nmax}) must be satisfied.    In the other words, a given vacuum can only invalidate this  BH bound on species, by becoming more curved and/or  shorter lived. For the slow-roll inflationary vacua,  this implies constraints on the slow-roll parameters, and subsequently, on the allowed number of the inflationary e-foldings. 
  
We now generalize the BH proof of the bound to the de Sitter and inflation \cite{Dvali:2008sy}.  Let $M$ be the mass of the species, and let $H$ be the Hubble parameter in de Sitter. 
 Consider a slow roll inflation driven by a single inflaton field $\phi$.  The equation for the spatially-homogeneous time-dependent field is, 
  \begin{equation}
\label{infeq}
\ddot{\phi}\, + \, 3H\, \dot{\phi} \, + \, V(\phi)'\, = \, 0\, ,   
\end{equation}
where, prime stands for the derivative with respect to $\phi$.      
   The main idea of the slow roll inflation is, that for certain values of  $\phi$, the potential $V(\phi)$ is sufficiently flat, so that the friction term dominates and this  allows $\phi$ to roll slowly.  
 The energy density is then dominated by the slowly-changing potential energy.  The 
 Hubble parameter is approximately given by $H^2 \, \simeq \, V(\phi) /3M_P^2$, and can be regarded as constant on the time scales $\sim \, H^{-1}$.   Obviously, the inflationary region of the potential must be away from  todays minimum with almost zero vacuum energy.  In any inflationary scenario the value of the inflaton field during inflation is very different from its todays expectation value $\phi_0$ corresponding 
 to the minimum of $V(\phi)$, which without loss of generality we can put at $\phi_0 \, = \, 0$.  
   
 Soon after the end of the inflationary period, inflaton oscillates about its true minimum $\phi_0$, and reheats 
 the Universe.   For this to happen,  inflaton should necessarily interact with the standard model particles and possibly with the other fields.  Let us consider an inflaton coupled to $N$ species, with masses 
 $M_j$.   For the efficient reheating, the masses of the the particles  
about the minimum $\phi_0$,  must be less than the inflaton mass about the same minimum. 
That is,  $M_j \, \ll \, V''(\phi_0)$.   Due to coupling to the inflaton field, the masses of species 
are functions of its expectation value,  $M_j(\phi)$, and it is very common that these masses change substantially during inflation.   The key point that we are willing to address now, is that the masses of these species are subject to the BH bound, and give useful restriction on the inflationary trajectory.  Thus, knowing the 
couplings of the inflaton in {\it our} vacuum,  one can get an non-trivial information about
the much remote inflationary vacua of the same theory. 

  For simplicity, we shall assume the universality of the species masses  $M_j(\phi) \, = \, M(\phi)$.  
 During the slow-roll inflation, Universe is in a quasi-de-Sitter state, in which the inflationary Hubble parameter sets the size of the causally-connected event horizon $H^{-1}$.   However, the difference 
 from the stationary de Sitter vacua, is that in realistic inflationary scenarios the slow roll phase 
 (in any given region)  is not exponentially long lived,  and lasts for several Hubble times.
 So $H^{-1}$ sets the time scale on which parameters can be regarded as constant. 
 
  Thus, a hypothetical observer  located within a given  causally-connected 
  inflationary patch can perform a sensible experiment with BH formation and evaporation.    In such a case, the black hole
  bound can be directly applied, and we arrive to the bound \cite{Dvali:2008sy}
 \begin{equation}
\label{massbound}
M(\phi) \, < \, {M_P \over (H^{-1}(\phi)M_P)^{{1\over 3}}}.
\end{equation}
  All the information that this bound implies for a given inflationary scenario, is encoded in the functions 
$M(\phi)$  and  $H(\phi)$. We shall now illustrate this on some well known examples.

 \subsection{Several scenarios}

 \subsubsection{Chaotic Inflation}
  
   Let us consider the example of Linde's chaotic inflation \cite{Linde:1983gd}. This is based on  a single scalar field 
   with a mass $m$ and no self-coupling
   \begin{equation}
\label{chao}
V(\phi) \, = \, {1 \over 2} m^2 \phi^2 \, + \,  g \phi \bar{\psi}_j\psi_j\, .
\end{equation}  
The last term describes the coupling to $N$-species, which for definiteness we assume to be fermions, 
and $g$ is the interaction constant.   As said above, the coupling of the inflaton to the species is crucial
for the reheating. 

 The above theory has a Minkowski vacuum, in which $\phi \, = \, 0$ and  all the species are massless. 
 Due to the latter fact, in this vacuum  the BH bound on the  number and mass of the species is satisfied.  However, as we shall see,  the same bound, puts non-trivial restriction  on the inflationary epoch, since 
 during inflation  $\phi \,  \neq \, 0$ and species are massive.  

 Ignoring for a moment the coupling to the species, the logic  in the  standard Chaotic inflationary scenario goes  as follows.  The expectation value of the field $\phi$ can be arbitrarily large, as  long as the 
 energy density remains sub-Planckian, that is 
 \begin{equation}
\label{conditionaa}
m^2\phi^2 \,  \ll \, M_P^4\, .  
\end{equation}   
 The equation (\ref{infeq}) then can be applied and takes the form 
  \begin{equation}
\label{cheq}
\ddot{\phi}\, + \, 3H\, \dot{\phi} \, + \, m^2\phi \, = \, 0\, ,  
\end{equation}
where $H^2 \, = \, {m^2\phi^2 \, + \, \dot{\phi}^2 \over 6M_P^2}$. As long as $H \, \gg \,  m$, the friction dominates and $\phi$ rolls slowly.  This implies (up to a factor of order one)  
\begin{equation}
\label{slow}
\phi \,  \gg \, M_P,
\end{equation}
which is compatible with (\ref{conditionaa})  as long as $m \, \ll \,  M_P$. 
If the above is satisfied, $\phi$ rolls slowly, and Universe undergoes the exponentially fast expansion. 
Let us now see how the coupling to the species restricts the above dynamics. During inflation the mass of the species is $M \, = \, g\phi$ and they are subject to the BH bound.  To see what this bound implies we can simply insert the current values of $M(\phi)$ and $V(\phi)$  in  (\ref{massbound}), and we get  
 \begin{equation}
\label{massbound1}
g\phi \, \lsim \, M_P\left ({m\phi  \over M_P^2}\right )^{{1\over 3}}.
\end{equation}
Non-triviality of the above constraint is obvious. For example, the standard argument assumes  that 
inflation could take place for arbitrary $m \, \ll \, M_P$, and from arbitrarily large values of $\phi$ satisfying
(\ref{condition}), irrespective to the number of species to which inflaton is coupled.
The above expression tells us that in the presence of species,  this is only possible, provided, 
$g \, \lsim \, (M_P/\phi)^{2/3} (m/M_P)^{1/3}$. 

For the practical reasons of solving  the flatness and the horizon problems, in the standard Chaotic scenario, last $60$ e-foldings happen for $\phi \,  \lsim \,  10 M_P$, whereas 
from density perturbation we have  $m \, \sim  10^{12}$GeV or so.   This implies, 
$g \, < \, 10^{-3}$. This constraint can be easily accommodated by the adjustment of couplings, however  it is remarkable that no fine tuning can make $g \sim 1$ consistent.

 \subsubsection{Hybrid Inflationary Vacua} 
 
 The essence of the  hybrid inflation \cite{Linde:1993cn}  is that inflationary energy density is not dominated by the potential 
 of the slowly-rolling inflaton field $\phi$, but rather by a false vacuum energy of other scalar fields, $\chi_j$. These fields are  trapped in a temporary minimum, created 
due to large positive mass$^2$-s, which  they acquire  from the coupling to the inflaton field.  The slowly rolling inflaton then acts as a clock,  which at some critical point triggers the transition that liberates the trapped fields, and converts their false vacuum energy into radiation.  However, usually  Inflation ends before this transition, because 
of breakdown of the slow-roll.  Thus, in hybrid inflation, the presence of fields with inflaton-dependent masses is essential not only for the  reheating, but for the inflation itself.  
\begin{figure}\label{azero}
\begin{center}
  \includegraphics[width=0.55\textwidth]{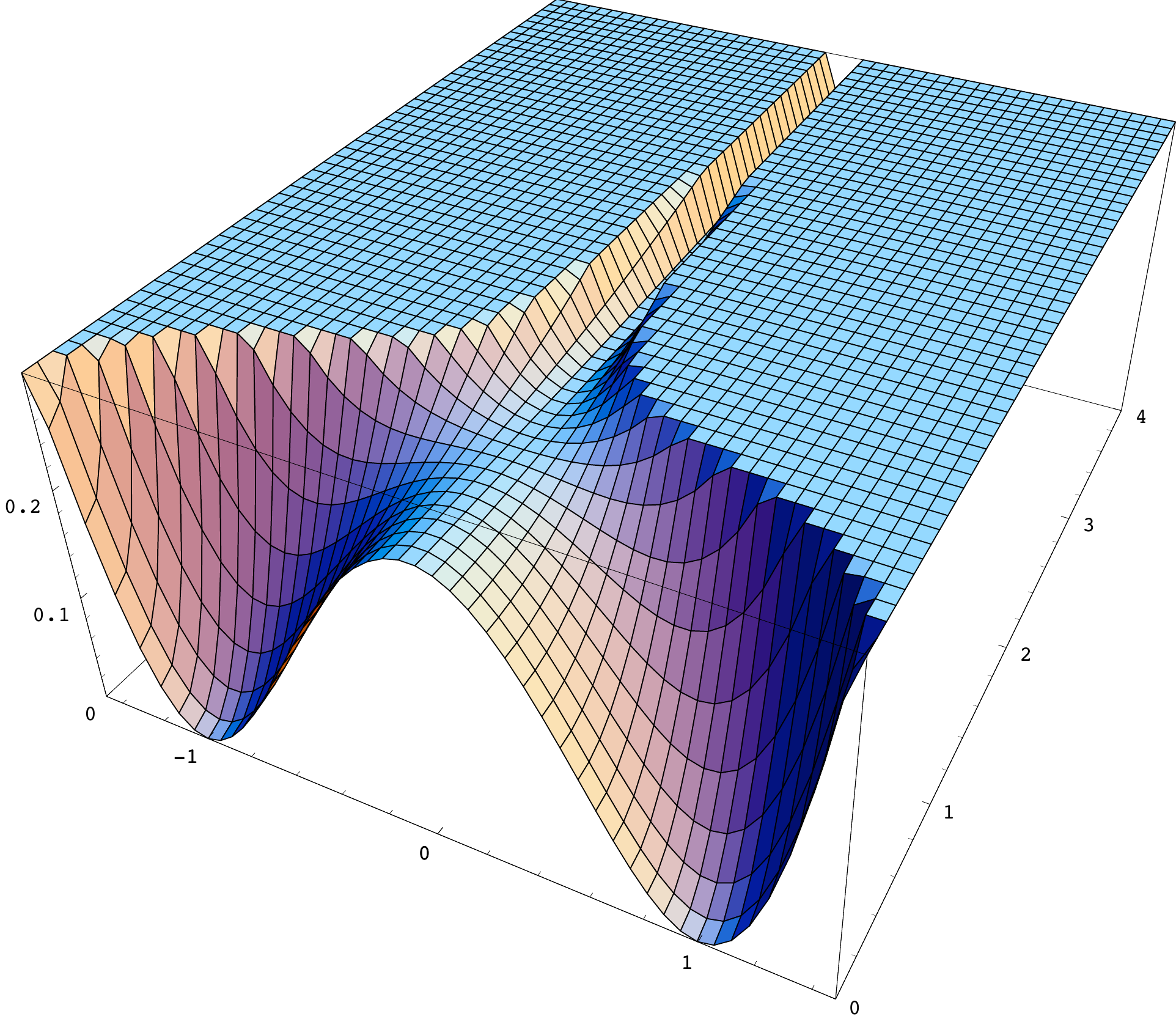}
\end{center}
\caption{The potential for hybrid inflation as a function of the inflaton $\phi$ (in the $y$-direction)and the tachyon field
$\chi$ (in the $x$-direction).}
\end{figure}
The simplest prototype model 
realizing this idea is  
  \begin{equation}
\label{hybrid}
V\, =\, \lambda^2 \, \phi^2 \chi_j^2 \,  + \, \left ({g \over 2} \,  \chi_j^2 \, - \, \mu^2 \right )^2,
\end{equation}
where $\lambda$ and $g$ are constants. 
Then,  for $|\phi| \, > \, \phi_t\, \equiv \, \mu \sqrt{{g \over \lambda^2}}$,  the effective potential for 
$\chi_j$ is minimized at $\chi_j = 0$, and the false vacuum energy density is a $\phi$-independent  constant, $\mu^4$.  Thus, in the classical treatment of the problem, starting at arbitrary initial value $\phi \, \gg \, \phi_t$ and with zero initial velocity, $\phi$ would experience zero driving force and system would inflate forever.  One could slightly lift this flat direction by adding an appropriate self interaction potential 
for $\phi$ (e.g., such as a positive mass term $m^2\phi^2$) which would drive $\phi$ towards the small values.  In such a picture inflation ends abruptly after $\phi$ drops to its critical value $\phi_t$, for which $\chi_j$ becomes tachyonic, and system rapidly relaxes into the true vacuum. This is shown in
figure 15. 
However, the above story is only true classically, and quantum mechanical corrections 
are very important and always generate potential for $\phi$. Because of to these corrections, typically, inflation ends way  before the phase transition, due to breakdown of the slow-roll.  Existence of supersymmetry cannot change the latter fact, however, supersymmetry  does make
the corrections to the potential finite and predictive. 

 The simple supersymmetric realizations of the hybrid inflation idea have been suggested in form of $F$-term \cite{Copeland:1994vg,Dvali:1994ms}
 and  $D$-term 
 \cite{Binetruy:1996xj,Halyo:1996pp}
  inflationary models.   As a result of supersymmetry, in $F$-term inflation $\lambda\, = \, g$. Due to renormalization 
of the K\"ahler function via $\chi_j$ loops, the non-trivial inflaton potential is inevitably generated, which for $\phi \gg \phi_t$ 
has the following Coleman-Weinberg form
\begin{equation}
\label{oneloop}
V(\phi) \, \simeq \, \mu^4 \left [ 1 \, + \, {N g^2\over 16 \pi^2} {\rm ln}{g|\phi| \over Q}  \right ] \,, 
\end{equation}
where, $Q$ is the renormalization scale.  Notice, that  this potential cannot be fine tuned away by addition of some local counter terms.  The condition of the slow roll is that  $V'' \ll H^2$, implying that  
\begin{equation}
\label{slowroll}
N\, g^2 \, \ll \, {\phi^2 \over M_P^2}\, . 
\end{equation}
Because of the logarithmic nature, the slope flattens out for large $\phi$. 
However, even if one tries to ignore any other correction to the potential,  
nevertheless, the slow-roll condition will eventually run in conflict with the black hole bound,
which implies that 
\begin{equation}
\label{conflict}
N\, g^2 \, \lsim \,  {M_P^2 \over \phi^2} \,.  
\end{equation} 
This fact indicates, that even if the theory is in seemingly-valid perturbative regime (that is, 
$  {N g^2\over 16 \pi^2} {\rm ln}{g\phi \over Q}  \, \ll  \, 1$), nevertheless, the perturbative corrections 
to the K\"ahler cannot be the whole story, and theory has to prevent growth of $\phi$, by consistency with 
the black hole physics.

\subsection{String inflationary Vacua}

String theory provides an interesting and also promising microscopic framework to
realize cosmic inflation (for reviews on
string inflation and string cosmology see\cite{HenryTye:2006uv,cline06,kallosh07,burgess07,silverstein07}).
The 4D effective potential for the inflaton typically arises after
compactification from ten to four space-time dimensions. Specifically
consider the following effective potential as a function of the moduli scalar fields $M$:
\begin{eqnarray}
 V(M)=V_0(M')+V'(M',\phi)\quad\Longrightarrow \quad V(\phi)=V_0+V'(\phi)\,.
 \end{eqnarray}
 here we have assumed that we can slit the moduli fields $M$ into one inflaton field $\phi$ and
 the remaining fields $M'$. The potential $V(M')$ can be very steep and fixed all
 moduli fields $M'$ to some particular values, where we have to assume that the minimum $V_0$ of
 this part of the potential is positive, i.e. it forms a de Sitter minimum.
 The remaining potential after stabilization of all $M'$-fields is then the inflaton potential, which has
 to meet the slow roll conditions discussed before.
 However often, the typical string compactification does not give rise to vanilla potentials of this type, as one can in general not separate out a stabilizing  potential
$V_{0}(M')$ that does not depend on the putative inflaton $\phi$. Instead one usually has moduli stabilization potentials $V_{0}(M',\phi)$ that also depend on $\phi$ and hence generically interfere with slow-roll inflation. Another potential problem, which will be the most relevant for the present paper, is that the value $V_{0}$ at which all orthogonal moduli $M'$ can be stabilized might be constrained to be negative due to the peculiar structure of the scalar potential $V_{0}(M')$.
Then one might always have at least one steep direction whenever the potential is positive. Potentially dangerous moduli of this type are, in particular, the overall volume modulus of a compactification manifold or the dilaton, as these often enter as steep directions in many contributions to the scalar potential.

 Now we can imagine two scenarios for string inflation:
 
 \begin{itemize}
 \item Closed string inflation: the inflaton is a closed string modulus.
 
 Here the effective potential is due to the potential energy from fluxes, including geometrical fluxes, 
 and/or (Euclidean) branes wrapped around internal cycles.  
 
\item Open string inflation: inflaton is a open string modulus, the distance between D-branes.

Here the effective potential is due to the attractive or repulsive forces between (non-BPS) branes.

\end{itemize}

 Let us discuss the mechanism of brane inflation \cite{Dvali:1998pa,Burgess:2001fx} in more detail.
  In this picture the role of the inflaton field $\phi$ is played by the brane-separation field. A simplifying but crucial assumption of the original brane inflation model, is that compactification moduli are all fixed, with the masses being at least of order of the inflationary Hubble parameter, so that 
branes can be considered to be moving in a fixed external geometry, weakly affected by the brane motion.  In the same time, the  $4d$ Hubble volume must be larger than the size of the compact extra dimensions. These conditions allow us to apply 
 the  power of the effective four-dimensional supergravity reasoning. 
 Below we shall focus on the case of D-brane inflation, based on the motion and subsequent annihilation of branes an anti-branes.  This picture 
 from the four-dimensional perspective can be understood as the hybrid inflation, in which
 $\phi$ is a brane distance field, and role of $\chi$ is played by the open string tachyon.  
  In this picture, the supersymmetry breaking by a non-BPS
brane-anti-brane system  corresponds to the spontaneous supersymmetry
breaking via FI  D-term. 


When branes are far apart, there  is a light  field $\phi$, corresponding
 to their relative motion. This mode is a
combination of the lowest lying scalar modes of the open strings that are
attached to a brane or anti-brane only. 
We are
interested in the combination that corresponds to the relative radial
motion of branes.

\begin{equation}
\label{X} \phi \, = \, M_{\rm string}^2 r\, .
\end{equation}

 In the simplest case of a single brane-anti-brane pair,  we have the two gauged $U(1)$-symmetries.
 One of these two  provides a non-vanishing
D-term.  The tachyon ($\chi$) is an open string state
that connects the brane and the anti-brane.  The mass of this stretched
open string is $M_s^2 r$. In $4d$ language,  the tachyon as
well as other open string states get mass from the coupling to $\phi$.

 The energy of the system is given by the
D-term energy, which is constant at the tree-level, but not at one-loop
level. At one-loop level the gauge coupling depends on $\phi$.  $g^2$ gets renormalized,
because of the loops of the heavy $U(1)$-charged states, with $\phi$-dependent masses.  
 For instance,
there are one-loop contributions  from the $\chi$ and $\bar\chi$ loops.
More precisely there is a renormalization of $g^2$ due to one-loop open
string diagram, which are stretched between the brane and anti-brane.
Since the mass of these strings depend on $\phi$, so does the renormalized
$D$-term energy
\begin{equation}
\label{DX} V_D \, = \, {g^2(\phi)\over 2}D^2  \, = \, {g^2_0 \over 2} \left ( 1 \,
+ \, g^2_0f(\phi)\right ) \xi^2\,,
\end{equation}
where $g_0^2$ is the tree-level gauge coupling, and $f(\phi)$ is the
renormalization function.  For example, for $D_3$-$D_7$ system, discussed in section (5.3.2) at the intermediate distances 
($M_{\rm string}^{-1} \ll r \ll R$, where $R$ is the size of two transverse extra dimensions), 
this takes the form (\ref{oneloop}). 


 We shall now see, why  at  least in the simplest $D$-brane setup, the $U(1)$ symmetry must be 
 Higgsed throughout the inflation. 
Let us again think about the  process of  D$_{3
+ q}$-$\bar{\mbox{D}}_{3+q}$ driven inflation, with the subsequent brane annihilation.  We assume that $q$
dimensions are wrapped on a compact cycle, and relative motion takes place in $6 - q$ remaining
transverse dimensions. 

The low energy gauge symmetry group is $U(1)\times U(1)$, one linear
superposition of which is Higgsed by the tachyon VEV.
The crucial point is, that this Higgsed $U(1)$
gauge field is precisely the combination of the original $U(1)$'s that
carries a non-zero RR-charge (the other combination is neutral).  The
corresponding gauge field strength $F_{(2)}$ has a coupling to the closed string RR
 $2+q$-form $C_{(2\, + \, q)}$ via the WZ terms,
\begin{equation}
\label{fccoupling} \int_{3 + 1 + q} F_{(2)}\wedge C_{(2 \, +\, 
q)}\,,
\end{equation}
where, since we are interested in the effective $4d$ supergravity
description,  we have to integrate over extra $q$-coordinates, and only keep the $4d$ zero mode component of the RR field. This then becomes an effective $2$-form, $C_{(2)}$. 

The connection with the $4d$ supergravity  $D$-term  language,
is made by  a
dual description of the $C_{(2)}$-form in terms of an axion ($a$),
\begin{equation}
\label{cduala}
d {\rm C}_{(2)} \rightarrow *\, d {\rm a}\,,
\end{equation}
where star denotes a $4d$ Hodge-dual. Under this duality transformation
we have to replace
\begin{equation}
\label{c-a} 
(d {\rm C}_{(2)})^2  \, + \, {\xi \over M_P^2} F_{(2)}\wedge {\rm C}_{(2)}\, ~~
 \rightarrow \,~~ M_P^2 (d{ \rm a \, - \,g Q_a
W})^2\,,
\end{equation}
where $Q_a \, = \, {\xi \over M_P^2} $ is the axion charge under $U(1)$.
As it should, this charge vanishes as the compactification volume goes to infinity, and
$4d$ supergravity approaches the rigid limit.
  We thus see that the $U(1)$ gauge field ($W_{\mu}$) acquires a mass $m_{W}^2 \, \gsim \xi^2 /M_P^2$.

 The D-term (hybrid) leads to a very specific prediction, namely to the appearance of
 cosmic strings as non-trivial
 topological defects after the spontaneous breaking of the $U(1)\times U(1)$ gauge symmetry \cite{Dvali:2003zh}.
 Since cosmic strings contribute to the energy density of the
 universe, to the generation of the CMB and can be also
 visible by cosmic lensing, there are stringent experimental bounds on the abundance
 of cosmic strings in the universe. These bounds are pretty dangerous
 for open string D-term inflation, resp. can be used to give strong constraints on these kind of models,
 as we will further discuss in section (5.3).

\subsubsection{The search for type IIA inflation and de Sitter vacua with positive cosmological constant}

In the  section (4.4), we have derived part of the low energy effective action for the type IIA,
$AdS_{4}$ compactifications. Here we would briefly like
to discuss the question if some of these IIA 
vacua can be used for inflation \cite{Caviezel:2008tf,Caviezel:2008ik}.  Specifically, we want to address the question if the scalar
potential in the closed string moduli sector can be flat enough in order to allow inflation by one
of the closed string moduli. Therefore
the parameter $\epsilon$ must be small enough in
some region of the closed string scalar potential. In addition, this analysis is also relevant for open string inflation
on these IIA vacua, since in this case we have to find closed string minima of the scalar potential,
i.e. $\epsilon=0$ somewhere in the closed string moduli space.
Extending the earlier work
\cite{Hertzberg:2007ke}, the authors of \cite{Hertzberg:2007wc} proved a no-go theorem against 
small $\epsilon$, i.e. against a period of slow-roll inflation
in type IIA compactifications on Calabi-Yau manifolds
with standard RR and NSNS-fluxes, D6-branes and O6-planes at large volume and with small string coupling. More precisely, they show that the slow-roll parameter $\epsilon$ is at least $\frac{27}{13}$ whenever the potential is positive, ruling out slow-roll inflation in a near-de Sitter regime, as well as meta-stable dS vacua.
As emphasized in \cite{Hertzberg:2007wc}, however, the inclusion of other ingredients such as NS5-branes,
geometric fluxes and/or non-geometric fluxes evade the assumptions that underly this no-go theorem. In fact in \cite{Silverstein:2007ac} de Sitter vacua in type IIA were found using some of these additional ingredients. Furthermore a concrete string inflationary model on nilmanifolds with D4-branes and large
field inflation was presented in \cite{Silverstein:2008sg,McAllister:2008hb}. The coset models of section
(4.4) could thus be candidates for circumventing the no-go theorem as they all have geometric fluxes. So let us study this in some more detail.

The proof of this no-go theorem is remarkably simple and uses only the scaling properties of the scalar potential with respect to
the volume modulus
\begin{equation}{
\rho = \left(\frac{\text{Vol}}{\rm vol}\right)^{1/3} \, ,
}\end{equation}
where ${\rm vol}=|\int e^{123456}|$ is a standard volume,
and the dilaton modulus
\begin{equation}{
\tau = e^{-\Phi} \sqrt{\rm {vol}} \, ,
}\end{equation}
 as well as the signs of the various contributions to the potential.

 Classically, the four-dimensional scalar potentials of such compactifications may receive contributions from the NSNS $H_3$-flux, geometric fluxes $f^i_{jk}$, O6/D6-branes and the RR-fluxes $F_p,\,p =0,2,4,6$: 
\begin{equation}\label{eq:pot}
V= V_3 + V_f + V_{O6/D6} + V_0 + V_2 + V_4 +V_6,  
\end{equation}
where
$V_3, V_0, V_2, V_4, V_6 \geq 0$,
and
$V_f$ and $V_{O6/D6}$ can a priori have either sign.

In \cite{Hertzberg:2007wc} the authors studied the dependence of this scalar potential on the volume modulus $\rho =(\textrm{Vol})^{1/3}$ and the four-dimensional dilaton $\tau = e^{-\phi} \sqrt{\textrm{Vol}}$. Using only this $(\rho,\tau)$-dependence, they could derive a no-go theorem in the absence of metric fluxes that puts a lower bound on the first slow-roll parameter, 
\begin{equation}
\epsilon \equiv \frac{K^{A\bar{B}}\partial_{A}V \partial_{\bar{B}}V}{V^2} \geq \frac{27}{13}, \quad \textrm{whenever }  V>0, \label{Bound1} 
\end{equation}
where $K^{A\bar{B}}$ denotes the inverse K\"{a}hler metric, and the indices $A,B,\ldots$ run over all moduli fields.
This then not only excludes slow-roll inflation but also de Sitter vacua (corresponding to $\epsilon=0$). 

The lower bound (\ref{Bound1}) follows from the observation that 
a linear combination of the derivatives with respect to $\rho$ and $\tau$ is always greater than a  certain positive multiple of the  scalar potential $V$. More precisely, the
 general scalings
\begin{eqnarray}\label{scalings}
V_3\propto \rho^{-3}\tau^{-2}, \qquad V_{p}\propto \rho^{3-p}\tau^{-4}, \qquad V_{O6/D6}\propto \tau^{-3}, \qquad V_{f}\propto \rho^{-1}\tau^{-2}
\end{eqnarray}
imply,
for the scalar potential (\ref{eq:pot}),
\begin{equation}{\label{nogoKachru}
  -\rho\frac{\partial V}{\partial \rho} -3\tau \frac{\partial V}{\partial \tau}=9V+\sum_{p \, \epsilon \, {2,4,6}}pV_{p} -2V_{f}.
}\end{equation}
Hence, whenever the contribution from the metric fluxes is zero or negative, the right hand side in (\ref{nogoKachru}) is at least equal to $9V$, which can then be translated to the above-mentioned lower bound $\epsilon \geq \frac{27}{13}$ \cite{Hertzberg:2007wc}. 
Avoiding this no-go theorem without introducing any new ingredients would thus require $V_f > 0$. Since $V_f \propto -R$, where $R$ denotes the internal curvature scalar, this is equivalent to demanding that the internal space have  negative curvature. Since all terms in $V$ scale with a negative power of $\tau$ we see from (\ref{eq:pot}) and (\ref{scalings}) that we would also need $V_{O6/D6} <0$ to avoid a runaway.

 In summary, if geometric fluxes \emph{alone} are to circumvent this no-go theorem, they can do so at most if they are positive:
\begin{equation}
\label{avoid}
 V_{f}>0 \qquad \textrm{(Necessary condition for evading the no-go theorem)}.
\end{equation}

In fact, we can immediately find the geometric part of the potential from the Einstein-Hilbert term in the ten-dimensional action:
\begin{equation}{
V_f = - \frac{1}{2} M_P^4 \kappa_{10}^2 e^{2 \Phi} \text{Vol}^{-1} R = - \frac{1}{2} M_P^4 \kappa_{10}^2 \tau^{-2} R \, ,
}\end{equation}
where $R$ is the scalar curvature of the internal manifold. For cosets/group manifolds $R$ can be explicitly calculated.
This expression has indeed the expected scaling behavior since $R \propto g^{-1} \propto \rho^{-1}$.
It follows that the condition (\ref{avoid}) for avoiding the no-go theorem can be rephrased as
\begin{equation}{
R < 0 \, .
}\end{equation}
Let us display the scalar curvature for some of our coset models in section 4.4:
\begin{eqnarray}
\frac{\text{G}_2}{\text{SU(3)}}: & & \qquad R  = \frac{10}{k_1} \, , \nonumber\\
\frac{\text{Sp(2)}}{\text{S}(\text{U(2)}\times \text{U(1)})}: &  &\qquad R = \frac{6}{k_1} + \frac{2}{k_2} - \frac{k_2}{2(k_1)^2}  \, , \nonumber \\
\frac{\text{SU(3)}}{\text{U(1)}\times \text{U(1)}}: & &\qquad R = 3 \left(\frac{1}{k_1}+\frac{1}{k_2}+\frac{1}{k_3} \right) - \frac{1}{2} \left(\frac{k_1}{k_2k_3}+\frac{k_2}{k_1k_3}+\frac{k_3}{k_1k_2} \right)\, , \nonumber \\
\frac{\text{SU(3)}\times \text{U(1)}}{\text{SU(2)}}: & &\qquad
R = \frac{1}{\sqrt{1+\rho^2}} \left(\frac{6}{k_1} - \frac{3 \rho k_2}{4(1+\rho^2)k_1^2}\left|\frac{U_2}{U_1}\right| \right)\, ,
\end{eqnarray}
where $k_i>0$ are the K\"ahler moduli and $U_i$ the complex structure moduli that enter the expansion of $J$ and $\Im \Omega$. 
We see that for $\frac{\text{G}_2}{\text{SU(3)}}$ the curvature is always positive, so inflation is still excluded, however for
the other models there are values of the moduli such that $R<0$. For SU(2)$\times$SU(2) we did not display the curvature, because taking generic values
of the complex structure and K\"ahler moduli, its expression is quite complicated and not very enlightening. However, also in that case it is possible
to choose the moduli such that $R<0$.

Note that this does not yet guarantee that the $\epsilon$ parameter is  indeed small, it just says that the theorem that requires it to be at least $27/13$ no longer applies. Hence, a logical next step would be to calculate $\epsilon$ in this region, ideally by taking also all other moduli into account
(see the general expression  and try to make $\epsilon$ small or zero. These would
be necessary conditions for, respectively, inflation or de Sitter vacua. They are not sufficient however,
because for inflation, we would also need the $\eta$ parameter to be small and further obtain a satisfactory inflationary model
which could end in a meta-stable de Sitter vacuum etc. For a meta-stable de Sitter vacuum, on the other hand, one would also have to check that
the matrix of second derivatives only has negative eigenvalues.

\subsubsection{Type IIB D3/D7-brane inflation}
Except the large field inflation model in type IIA on nilmanifolds of \cite{Silverstein:2008sg}
most of the inflationary string inflationary models are so far in within type IIB
compactifications. In particular, a well studied case is given in terms
of warped inflation on a IIB Calabi-Yau, with D3($\overline{\rm D3}$)-branes located at the tip
of a long throat in the internal geometry 
\cite{Kachru:2003sx,Baumann:2006th,Baumann:2007np,Krause:2007jk,Baumann:2007ah,Pajer:2007zt, Pajer:2008uy,Baumann:2008kq}.
This class of models realizes F-term inflation with relatively
small inflaton field, being the distance between
D3- and ${\overline {\rm D3}}$ brane on the throat.
Considerable fine tuning of parameters is necessary in warped inflation in order
to meet the constraints of slow roll inflation.
Here we like to discuss another working example
of type IIB inflation, namely the IIB orientifold on $K3\times T^{2}/\mathbb{Z}_{2}$ 
with D3- and D7-branes
\cite{Herdeiro:2001zb, Dasgupta:2002ew,Hsu:2003cy,Burgess:2006cb,Haack:2008yb,Burgess:2008ir}.
As we will see this model provides a concrete realization 
of D-term inflation with a $U(1)$ shift symmetry, which ensures the flatness of the
potential. However quantum corrections destroy the shift symmetry and also generate
a F-term contribution to the effective potential. So again some amount of fine-tuning
is required. As we will discuss it is nevertheless possible to directly confront this
model
with experimental data, which makes the model interesting in itself.

As said already, the model we would like to study here
is D3/D7-brane inflation on the background
$K3\times T^{2}/\mathbb{Z}_{2}$. The resulting model is a stringy version of a hybrid D-term inflation
model discussed above \cite{Binetruy:1996xj,Kallosh:2003ux,Dvali:2003zh,Binetruy:2004hh}
with a waterfall stage at the end in which a charged scalar field
condenses.\footnote{This condensing field corresponds to a particular
state of the strings stretching between the D3- and D7-brane, which
becomes tachyonic at a certain  critical interbrane  distance due to
the world volume flux on the D7-brane. The D3-brane is then dissolved
on the D7-brane as an instanton, and $\mathcal{N}=1$ supersymmetry
becomes restored \cite{Dasgupta:2002ew}.}
As a D-term inflation model,  D3/D7-brane
inflation, a priori, does not suffer from the generic supergravity
eta-problem of F-term inflation models. The main problem of D-term
inflation  is instead the cosmic string production during the
waterfall stage,  when  the spontaneous breaking of the underlying
$U(1)$-symmetry takes place and the D-flatness condition is restored.

One of the reasons  to study the D3/D7-model on
$K3\times T^{2}/\mathbb{Z}_{2}$ is its high computability. 
Type IIB string theory compactified on $K3\times T^{2}/\mathbb{Z}_{2}$ is related to M-theory compactified on $K3\times K3$  
\cite{Tripathy:2002qw,Lust:2005bd,Aspinwall:2005ad} and is associated with 4D, ${\cal N}=2$ supergravity specifically described in  \cite{Angelantonj:2003zx,D'Auria:2004qv,D'Auria:2004td}.
Bulk moduli  stabilization in these models was studied in a series of  papers,
and it is one of the best understood string theory models with
stabilization of all bulk moduli. In its simplest
incarnations this model does not contain the
D-branes necessary to describe the Standard
Model of  particle physics at low energies.
Therefore, the D3/D7-system studied here
should be regarded as a brane/flux module, which is responsible
for inflation and
moduli stabilization, and which has to be complemented by
additional D-branes in order
to obtain realistic Standard Model phenomenology at lower energies.

In the D3/D7-brane inflationary model, an attraction between a D3- and a D7-brane is
triggered by a non-self-dual world volume flux on a D7-brane,  which
we will henceforth call the Fayet-Iliopoulos (FI)  D7-brane.
If both branes are space-time-filling, and the D7-brane wraps
the $K3$-factor, the transverse interbrane distance on
$T^{2}/\mathbb{Z}_{2}$ plays the role of the inflaton.
A distinguishing feature of this model (as compared, e.g.,
with $D3/\overline{D3}$-brane inflation) is that the supersymmetry breaking
during the slow-roll de Sitter phase is spontaneous, and hence well-controlled.
More precisely, the supersymmetry breaking can be understood in terms
of a two-step process: Certain \emph{bulk} three-form fluxes on
$K3\times T^{2}/\mathbb{Z}_{2}$ may
spontaneously break the original $\mathcal{N}=2$ supersymmetry
preserved by the geometry to $\mathcal{N}=1$. In the resulting
effective $\mathcal{N}=1$ theory, the \emph{world volume}
fluxes on the D7-brane
then give rise to a D-term potential. Assuming the volume
modulus of the K3-factor to be fixed, this D-term potential
is non-zero for sufficiently large D3-D7-distance, breaking supersymmetry
spontaneously to $\mathcal{N}=0$.
This final spontaneous supersymmetry breaking induces a Coleman-Weinberg type
one-loop correction to the scalar potential that drives the D3-brane
towards the FI D7-brane. This motion corresponds to the phase of slow-roll inflation.

\vskip0.2cm
\noindent
{\sl 5.3.2.1 The effective K\"ahler potential}
\vskip0.2cm
\noindent
The NSNS- and RR-two-forms with one leg along the non-compact
directions and one
along the torus give rise to four vector fields in 4D. One linear
combination of these four vectors corresponds to the 4D graviphoton,
whereas the other three enter three vector multiplets. The three
complex scalars of these vector multiplets
are\footnote{Unlike in section 4.2. we are using the standard notation of \cite{Angelantonj:2003zx}
for the moduli of $K3\times T^2$.}
\begin{eqnarray}\label{notation}
s&=&C_{(4)}-i \textrm{Vol}(K3)\ ,\label{sdef1}\\
t&=&\frac{g_{12}}{g_{11}}+i\frac{\sqrt{\det g}}{g_{11}}\ , \label{t} \\
u&=& C_{(0)}-ie^{\varphi} \label{udef}
\end{eqnarray}
which denote, respectively, the $K3$-volume modulus with its
axionic RR-partner $C_{(4)}$, the $T^2$ complex structure
modulus and the axion-dilaton.

The position moduli of the 16 D7-branes on the torus are denoted by $y_7^{k}$ $(k=1,\ldots,16)$. Depending on where one chooses the origin of these coordinates, they could obviously be defined in various ways. A very convenient way to define them for brane configurations close to the orientifold limit is to use $y_{7}^{1,2,3,4}$ for the complex positions of branes number 1-4 with respect to fixed point number 1, and similarly, to use $y_{7}^{5,6,7,8}$ to  denote the positions of the branes number
5-8 with respect to fixed point number 2,  and so forth. In this notation,
$y_{7}^{k}=0$ for all $k=1,\ldots,16$ thus would mean that there are
four D7-branes sitting on top of each O7-plane, and we are at the
orientifold limit with constant dilaton.

Classically, the moduli space of the vector multiplet sector
is described by the special K\"{a}hler manifold
\begin{equation}
\mathcal{M}_{V}\cong \left( \frac{SU(1,1)}{U(1)}\right)_{s}\times
\frac{SO(2,18)}{SO(2)\times SO(18)}\ , \label{SO(18)coset}
\end{equation}
where the first factor is parametrized by $s$, and the remaining scalars
$(t,u,y_{7}^{k})$ span the second factor.
This geometry can be obtained from the following cubic prepotential:
\begin{equation}\label{prep}
\mathcal{F}(s,t,u,y_7^k)=stu-\frac{1}{2}sy_7^k y_7^k\ .   \label{prepnoD3}
\end{equation}
In F-theory language, the dilaton, $u$, corresponds to the complex structure of the
elliptic fiber of a second $K3$ factor, which we
will denote by $\widetilde{K3}$. In this picture, the $SO(2,18)/(SO(2)\times SO(18))$
factor of $\mathcal{M}_{V}$ describes the complex structure moduli space of
$\widetilde{K3}$. It should be noted that, far away from the orientifold limit,
the convenient separation of the scalars into closed and open string moduli is in general no longer possible, and the 10D meaning of e.g.\ $y_{7}^{k}$ as  brane positions is less clear \cite{Lust:2005bd}.

\begin{equation}\label{kinscalars}
K = -\ln \Big[-8\,({\rm Im}(s)\,{\rm Im}(t){\rm
Im}(u)-\frac{1}{2}\,{\rm Im}(s)\,({\rm
Im}(y_7^k))^2-\frac{1}{2}\,{\rm Im}(u)\,({\rm
Im}(y_3^r))^2)\Big] \ . \label{KP}
\end{equation}

The remaining moduli of the original $K3$-factor, as well as the torus volume and the remaining
axions from the  RR-four-form with two legs along K3 and two legs along
$T^{2}/\mathbb{Z}_{2}$ live in altogether 20 hypermultiplets and
parametrize, at tree-level, the quaternionic K\"{a}hler manifold
\begin{equation}
\mathcal{M}_{H}=SO(4,20)/(SO(4)\times SO(20))\ .
\end{equation}
This manifold has 22 translational isometries
along  the 22 real axionic directions, $C^{I}$ $(I=1,\ldots,22)$,
which descend  in the above-mentioned way from the RR-four-form.
These 22 axions transform in the vector representation of
$SO(3,19)\subset SO(4,20)$, and hence decompose into an $SO(3)$
triplet $C^{m}$ $(m=1,2,3)$ and an $SO(19)$-vector $C^{a}$ $(a=1,\ldots,19)$.

\vskip0.2cm
\noindent{\sl 5.3.2.2. Volume stabilization due fluxes and due to a non-perturbative superpotential}
\vskip0.2cm
\noindent
Three-form fluxes on $K3\times T^{2}/\mathbb{Z}_{2}$ lead to a superpotential
of the form
\begin{eqnarray}\label{tvinf}
W_{\rm flux}=W_H+W_F=\int_{K3\times T^2}\Omega\wedge \left( F_3+uH_3\right)\, ,
\end{eqnarray}
where $H_{3}$ and
$F_{3}$ denote the NSNS and RR three-form field strengths, respectively.
These fluxes generically stabilize
the moduli $(t,u,y_{7}^{r})$ and may lead to spontaneous partial
supersymmetry breaking $\mathcal{N}=2\rightarrow\mathcal{N}=1$.   In
an $\mathcal{N}=1$ vacuum, 
one of the two $\mathcal{N}=2$ gravitini (together with some of the other fields)
gains a mass. 
In addition the fluxes have to satisfy the tadpole condition
\begin{equation} \label{tadpol}
 \frac{1}{2}N_{\rm flux}+N_{D3}=24\ ,
\end{equation}
where
\begin{equation}
 N_{\rm flux}=\frac{1}{(2\pi)^4(\alpha^{\prime})^{2}}\int_{K3\times T^{2}}
H_{3}\wedge F_{3}
\end{equation}
with the integral being evaluated on the covering torus (which
explains the factor of $1/2$ in front of $N_{\rm flux}$ in (\ref{tadpol})).

Volume stabilization of the K3 space is finally  is achieved by a non-perturbative
F-term potential
due to either Euclidean D3-brane instantons or gaugino condensation
on stacks of D7-branes, which may arise after
spontaneous breaking of supersymmetry to ${\cal N}=1$. Here, we only focus on the volume of the $K3$-factor (the other K\"{a}hler moduli
could be stabilized by Euclidean D3-brane instantons \cite{Aspinwall:2005ad}\footnote{These Euclidean D3-instantons necessarily wrap the $T^{2}/\mathbb{Z}_{2}$-factor.
As the only open string dependence of the resulting superpotentials is via
the transverse distance between the space-time filling
D3-brane and the corresponding D3-instanton, these
superpotentials are independent of the
D3-brane position along $T^{2}/\mathbb{Z}_{2}$
and, hence, the inflaton.}). Moreover, we restrict ourselves to
the mechanism of gaugino condensation. This implies a  constraint
on the charged matter spectrum of the brane setup, which has to
allow for the presence
of a non-perturbative superpotential from gaugino condensation
(for the case of Euclidean D3-branes analogous constraints were discussed in
\cite{Witten:1996bn,Gorlich:2004qm,Aspinwall:2005ad}). 

Thus,
in order to comply with our notation from eq.(\ref{notation}) and,
we define the ${\cal N}=1$ gauge kinetic function $f_{D7}$
as
\begin{equation} 
f_{D7}= i s
\end{equation}
so that the gauge coupling is given by the real part of $f_{D7}$.

In order to ensure the appearance of a non-perturbative superpotential
one has to require that the quantity
\begin{equation} \label{c}
c = \sum_j T(r_j) - T({\rm adj})
\end{equation}
be negative. In (\ref{c}), the sum runs over the light (charged)
${\cal N}=1$ chiral multiplets in the representation $r_j$ of the gauge group.
In particular, no
adjoint matter is allowed in the light spectrum of the ${\cal N}=1$
gauge theory. We assume that the charged matter content of the D7-brane
gauge theory is such that it fulfills $c<0$,
for example by giving mass to unwanted matter via fluxes
\cite{Gorlich:2004qm,Cascales:2004qp,Lust:2005bd}.

\begin{equation}\label{ias}
W=W_{\rm flux}+A_0 \exp\Big( {\frac{8 \pi^2 f}{c}}\Big) = W_{\rm flux}+A_0 e^{\frac{i8s \pi^2}{c}
}\ , 
\end{equation}

The D7-brane stack on
  which gaugino condensation takes place should be at a different
  position on $T^{2}/\mathbb{Z}_{2}$ than the D7-brane on which
  world-volume flux is supposed to attract the D3-brane (see, e.g.,
  Fig. 16  for a possible realization). Otherwise, the
  $K3$-volume is destabilized after inflation. 
In our model, the role of the attracting anti-D3-brane is played by the
  D7-brane with the world-volume flux on it, i.e., by the  FI  D7
  brane. It should thus likewise be placed away from the stack of the volume
 stabilizing D7's so as to avoid the destabilization of the volume at the exit from inflation.
 \begin{figure}
\begin{center}
  \includegraphics[width=0.55\textwidth]{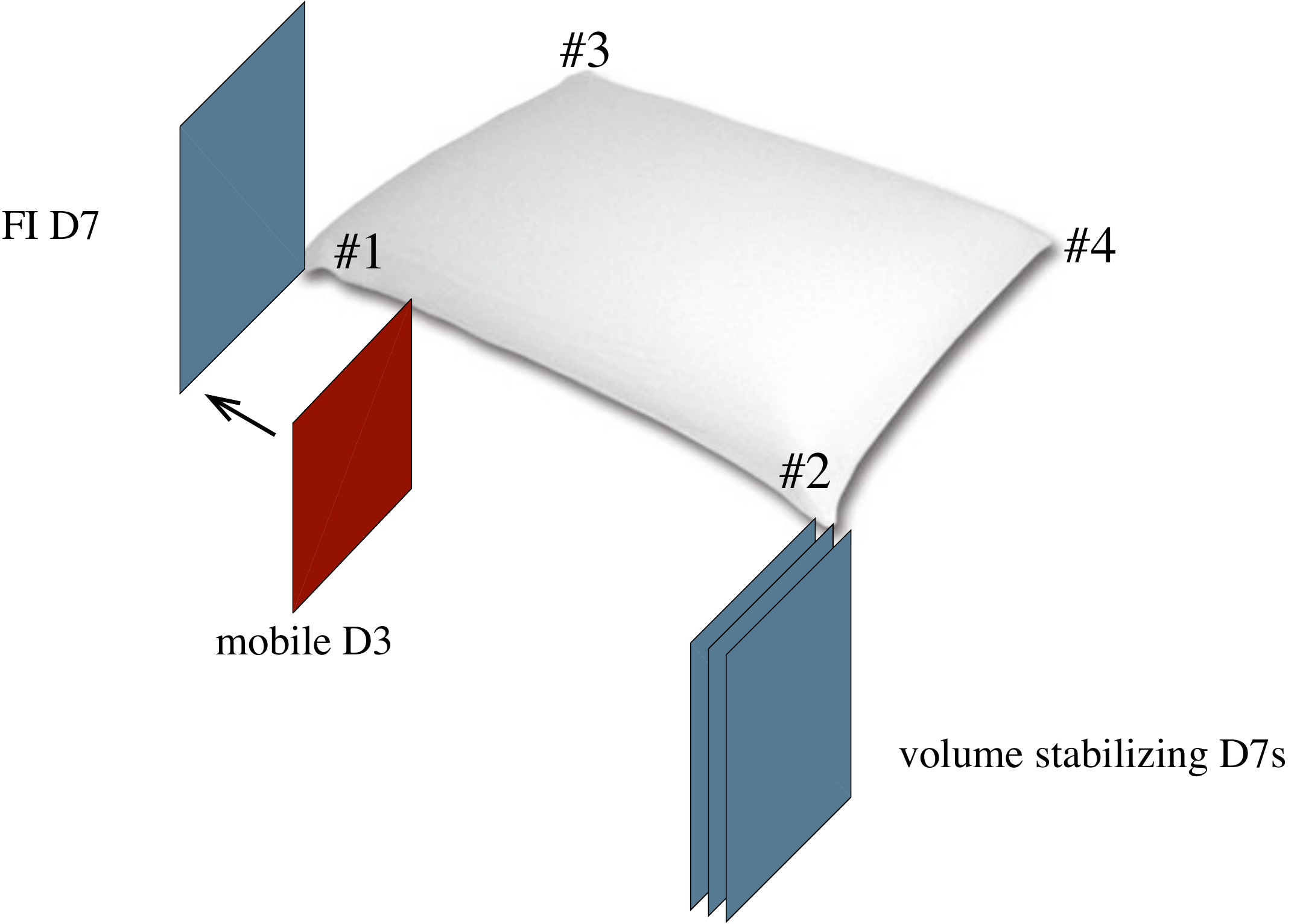}
\end{center}
\caption{Brane realization for D3/D7-inflation. The $K_3$ volume is stabilized
by a stack of D7-branes at fixed point No. 2; supersymmetry breaking by the the world volume
flux occurs at the D7-brane at fixes point No. 1, and the D-term, inflaton potential is
generated by the force between the mobile D3-brane and the FI D7-brane.}
\end{figure}

\vskip0.2cm
\noindent{\sl 5.3.2.3 Inflationary D-term potential}
 \vskip0.2cm
\noindent
 In our model, the role of the attracting anti-D3-brane is played by the
  D7-brane with the world-volume flux on it, i.e., by the  FI  D7
  brane. It should thus likewise be placed away from the stack of the volume
 stabilizing D7's so as to avoid the destabilization of the volume at the exit from inflation.
 The inflaton potential is then generated by  spontaneous supersymmetry breaking
due to a non-selfdual world volume flux on another D7-brane, which then triggers
an attraction of a nearby D3-brane towards that D7-brane. In 4D, the supersymmetry
breaking due to the world-volume fluxes can be attributed to a non-vanishing
D-term potential.
It is important that  the D7-brane with the world volume
flux is different from the D7-branes on which gaugino condensation
takes place and that both types of D7-branes are at different locations on
$T^{2}/\mathbb{Z}_{2}$. The reason for this is that the function $A(y_{3},\ldots)$ (see next section)
entering the non-perturbative superpotential  (\ref{ias})
vanishes if the D3-brane sits on top of the D7-branes responsible for the
gaugino condensation \cite{Ganor:1996pe}. If the gaugino condensation D7-branes and those with
world-volume flux were the same, this would lead to volume destabilization at the end
of inflation, when the D3-brane dissolves as an instanton on the D7-branes.
The situation is thus similar to the setup described in
\cite{Baumann:2007np,Baumann:2007ah}, where the mobile D3-brane
also moves away from the volume stabilizing D7-branes and approaches the
anti-D3-brane at the tip of the throat. The analogue of the anti-D3-brane
would then be the D7-brane with world volume flux in our
setup.

The $\mathcal{N}=2$ theory with prepotential (\ref{prep})
features a shift symmetry for the K\"{a}hler potential and the D7-brane gauge kinetic function along the real parts of the D3-brane position
moduli, $y_{3}^{r}$.
 If we assume that the D7-brane with the non-self-dual
world-volume flux sits at $y_7=0$ (we are from now on suppressing the indices
$k$ and $r$ of the D7- and D3-brane coordinates wherever it
does not cause confusion), the attractive force it exerts on
a mobile D3-brane only depends on the
absolute value, $|y_{3}|$, of that D3-brane's position \cite{Dasgupta:2002ew}.
Hence, if we assume that the initial position of the D3-brane
has $\textrm{Im}(y_{3})=0$, the D3-brane is attracted towards
the flux D7-brane along the $\textrm{Re}(y_{3})$ direction,
which is unaffected by the non-perturbative F-term
potential.
It should be noted that if the K\"{a}hler potential and the relevant
gauge couplings had been functions of $|y_{3}|$ instead of
$\textrm{Im}(y_{3})$ (as would be the case, e.g., for a ``canonical''
K\"{a}hler potential $K=|y_{3}|^{2}$),
one would also have had a shift symmetry along the phase of
$y_{3}$.\footnote{This phase is
a compact direction in field space, but so is $\textrm{Re}(y_{3})$
due to the compactness of the torus.}
However, in that case, also the attractive potential between the
D7-brane with world volume flux and the D3-brane  would be independent
of the phase of $y_{3}$, and one would have a completely flat direction
and no inflation. It is thus important that the shift symmetry is along
a direction in field space along which the inflationary potential is not flat.

Now, we will determine this field
range for the real part of the canonically normalized
D3-brane coordinate, $\phi\equiv\textrm{Re}(y_{3}^c)$.
Neglecting quantum corrections to the K\"ahler potential,
the kinetic term of $\textrm{Re}(y_3)$ can be read off from (\ref{kinscalars}).
\begin{equation}
M_P^2 \int d^4 x \sqrt{\det(\tilde{g}_{\mu\nu})} \tilde{g}^{\mu\nu}
\frac{\partial_{\mu}\textrm{Re}(y_{3})\partial_{\nu}
\textrm{Re}(y_{3})}{4\textrm{Im}(t)\textrm{Im}(s)
- 2[\textrm{Im}(y_{3})]^{2}}\ . \label{kinscalars2}
\end{equation}
The canonically normalized field, $\phi$, is therefore
\begin{equation} \label{canonphi}
\phi=\frac{M_{P}\textrm{Re}(y_3)}{\sqrt{2\textrm{Im}(t)\textrm{Im}(s) -[\textrm{Im}(y_{3})]^2}} \ ,
\end{equation}
or 
\begin{equation} \label{phi}
\phi=M_{P}\textrm{Re}(y_{3})\sqrt{-\frac{(2\pi)^{5}g_{s}
(\alpha^{\prime})^{2}}{\textrm{Vol}_{0}(K3)\textrm{Im}(t)}}\ .
\end{equation}

The potential of D-term inflation in the near de Sitter valley where
inflationary perturbations are generated   is given by a constant term and the
Coleman-Weinberg term:
\begin{equation}
V =\frac {g^{2}\xi^{2}}{ 2}\left(1+\frac{g^{2}}{ 16 \pi^{2}} U(x)\right) \ ,
\label{D3D7CW}
\end{equation}
where $x\equiv \frac{\phi}{ \sqrt\xi}$ and
\begin{equation}
U(x)= (x^2+1)^2 \ln (x^2+1) + (x^2-1)^2 \ln (x^2-1)
- 2x^4 \ln (x^2) - 4\ln 2 \ .
\label{U}
\end{equation}
Supersymmetry is broken by the FI parameter $\xi$, which depends on
the 2-form flux on the FI D7-brane.
The last term is added to account for the normalization condition $U(1) = 0$,
but it can be ignored in our subsequent calculations. Indeed, in the approximation
which we are going to use, the corrections to the potential do not affect much its
value, $V \approx  g^{2}\xi^{2}/2$, but these corrections are fully responsible for
the value of its derivative $V'$, which does not depend on the last term in (\ref{U}).

\vskip0.2cm
\noindent{\sl 5.3.2.4 Additional quantum corrections: mixture of F- and D-term inflation}
 \vskip0.2cm
\noindent
As it is well known the gauge coupling constants on the D7-brane will
receive non-vanishing quantum corrections, the so-called threshold corrections \cite{Dixon:1990pc}.
They will also affect the non-perturbative superpotential due to gaugino
condensation and hence also eventually the inflaton potential.
The quantum corrections to the D7-brane gauge coupling break the shift symmetry of the
real part of $y_3$, and in general the real part of $y_3$ is no longer a distinguished direction.
These threshold corrections to gauge coupling constant in orientifold compactifications
were computed in 
\cite{Bachas:1996zt,Antoniadis:1999ge,Lust:2003ky,Berg:2004ek,Berg:2004sj,Akerblom:2007np,Akerblom:2007uc}.
For the non-perturbative superpotential only 
the real part of a holomorphic function is relevant,  and for the model under consideration
this is given by the modular function $\vartheta_1$ \cite{Berg:2004ek,Berg:2004sj}:
\begin{equation} \label{gs}
g^{-2}= \textrm{Re}(i s) - \frac{1}{(2\pi)^2} \textrm{Re}\, \zeta (y_{3},t)\ , \quad
\zeta (y_{3},t) = \ln \vartheta_{1}(\sqrt{2 \pi} y_{3},t) + \ldots \ ,
\end{equation}
where the gauge kinetic function on the D7-branes has the form
\begin{equation} \label{fd7}
f_{D7}= i s - \frac{1}{8 \pi^2} \zeta(y_3-\mu,t)
- \frac{1}{8 \pi^2} \zeta(y_3+\mu,t) + \ldots \ ,
\end{equation}
This leads to the following non-perturbative superpotential due to 
gaugino condensation:
\begin{equation}
W_{\rm n.p.}=A_0 \exp\Big( {\frac{8 \pi^2 f}{c}}\Big) = A e^{\frac{8 \pi^2}{c}
(i s - \frac{1}{8\pi^2} \zeta (y_{3}-\mu,t)- \frac{1}{ 8\pi^2} \zeta (y_{3}+\mu,t))}\ , \label{W}
\end{equation}
where $A_0$ now might depend on any light charged matter fields
and $A$ incorporates in addition an overall factor independent
of $y_3$ coming from the ellipsis in (\ref{fd7}). Using the explicit
form of the string threshold corrections
eq.\ (\ref{gs}) we derive
\begin{eqnarray}\label{supo2}
W_{\rm n.p.}=A_0 \exp\Big(\frac{8 \pi^2 f(M}{ c}\Big) = A
\biggl(\vartheta_1\Big(\sqrt{2 \pi} (y_3+\mu),t\Big)
\vartheta_1\Big(\sqrt{2 \pi} (y_3-\mu),t\Big)\biggr)^\frac{- 1 }{ c}
e^{8 i \pi^2 s/c}.
\end{eqnarray}
For small values of $y_3-\mu$ (with $y_3+\mu$
staying finite) this becomes
\begin{eqnarray}\label{supo4}
W_{\rm n.p.} = A\,
\biggl(\vartheta_1\Big(\sqrt{2 \pi} (y_3+\mu),t\Big)\biggr)^\frac{- 1}{ c}
\left((2 \pi)^{3/2} \eta(t)^3\right)^\frac{-1}{ c}
(y_3-\mu)^\frac{-1 }{ c} e^{8 i \pi^2 s/c}+ \ldots \ .
\end{eqnarray}

This $y_3$ dependence of the non-perturbative superpotential leads
to a non-vanishing F-term potential $V_F$ for the inflaton field $\phi$ \cite{Haack:2008yb}.
Neglecting possible quantum corrections to the K\'ahler potential, 
and expanding $V_F$ up to quadratic order in $\phi$ we obtain the following new mass
term for the canonically normalized inflaton field:
\begin{equation}
V_{F}=\frac{|Ae^{-ia\tilde{s}}|^{2}\tilde{s}_{2}}{2u_{2}}\Big[\frac{3 a^{2}}{t_{2}} - 2 \phi^{2}\Big(3 a\textrm{Re}(\Delta) + 4t_2 |\Delta|^{2}\Big)\Big] +\mathcal{O}(\phi^{4}).\label{paper2}
\end{equation}
%
%

\begin{equation} \label{s}
s=\tilde{s} +i\lambda [{\rm Re}(y_{3})]^{2}
+\mathcal{O}([{\rm Re}(y_{3})]^{4})
\end{equation}

In order to ensure that the cosmological constant is almost
zero after inflation, the first term in (\ref{paper2}) (a negative
contribution to the vacuum energy), has to be canceled. This might
require an additional uplifting mechanism.

Here we just assume that the $\phi$-independent contribution to the
F-term potential is canceled after inflation ends.
In that case, the correction due to the F-term potential arising from
stringy corrections to the superpotential takes the
form
\begin{equation} \label{msquare}
V_{F}= -\frac {m^2}{2}  \phi^2 \ , \qquad
m^2= \frac{2|A|^2 \tilde{s}_2 e^{2 a \tilde{s}_2}}{u_2} \left[
3 a {\rm Re} (\Delta) + 4 t_2 |\Delta|^2\right]\ .
\end{equation}
The function $m^2$ of (\ref{msquare}) gets a strong suppression from
the exponential pre-factor (note that $\tilde{s}_2$ is negative in our conventions and that
$|\tilde{s}_{2}|$ has to be considerably larger than one in the supergravity
regime). Furthermore, also $|u_2|$ is large in the weak coupling limit. In addition, $m^{2}$
depends on the complex structure
and is thus tunable via a choice of fluxes. Note that even though $t_2,u_2$ and $\tilde{s}_2$ are all negative in our
conventions, $m^2$ is not necessarily positive, because ${\rm Re} (\Delta)$ can have either sign.
In figure 17, we plot the function
\begin{equation} \label{tildem}
\tilde{m}^2 \equiv 3 a {\rm Re} (\Delta) + 4 t_2 |\Delta|^2
\end{equation}
for $a = 8\pi^2/10$ and $\Upsilon=2\pi^3/30$  as a function
of $-t_2$ for the sample value $t_1=0.26$. As $\tilde{m}^2=\gamma m^2$ with $\gamma>0$,
the vanishing of $\tilde{m}^2$ means also a vanishing of $m^2$.
One can show  \cite{Haack:2008yb}  that $m^2$ can be made
 small and positive by tuning the parameters $t_2$ and $s_2$.
 Furthermore one can show that quartic corrections to the inflaton can be kept small compared
 to the mass term considered here.
  \begin{figure}
\begin{center}
  \includegraphics[width=0.55\textwidth]{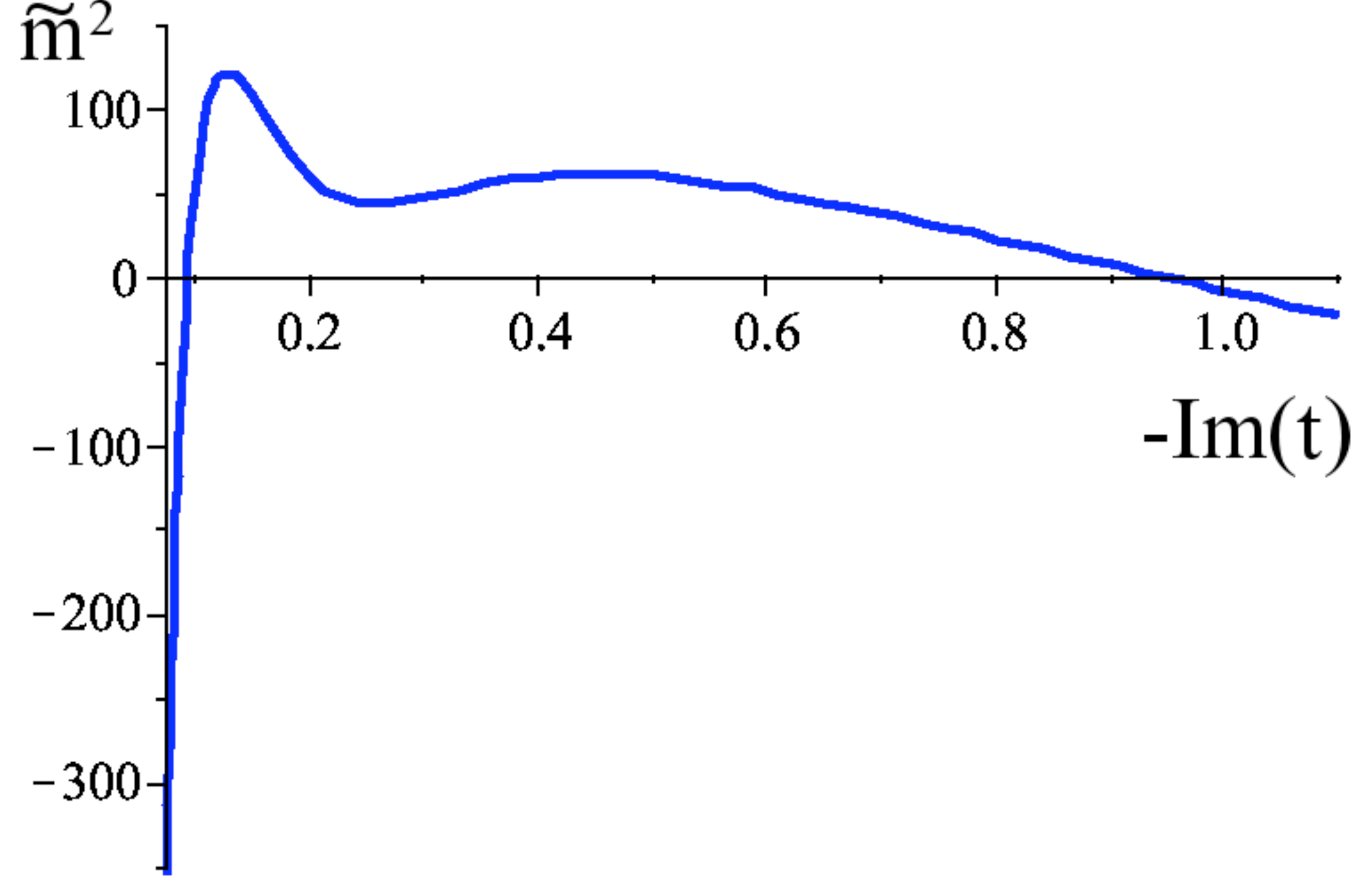}
\end{center}
\caption{Rescaled mass term for the inflaton fields as a function of the complex structure of
the torus $T^2$.}
\end{figure}

Thus, the whole D3/D7-brane inflation model potential at small $\phi$ (i.e.\
in the regime where inflationary perturbations are generated)
in Planck units, and with
account of stringy corrections from the stabilizing $F$-term as
explained above, is (see figure 18)
\begin{equation}
V =\frac {g^{2}\xi^{2}}{ 2}\left(1+\frac {g^{2}}{ 16 \pi^{2}} U\Big(\frac{\phi}{\sqrt{\xi}}\Big)\right) -\frac {m^2}{ 2}\phi^{2} \ , \label{D3D7potentialcorrFull}
\end{equation}
where $U(x)$ is given in (\ref{U}).
 \begin{figure}
\begin{center}
  \includegraphics[width=0.55\textwidth]{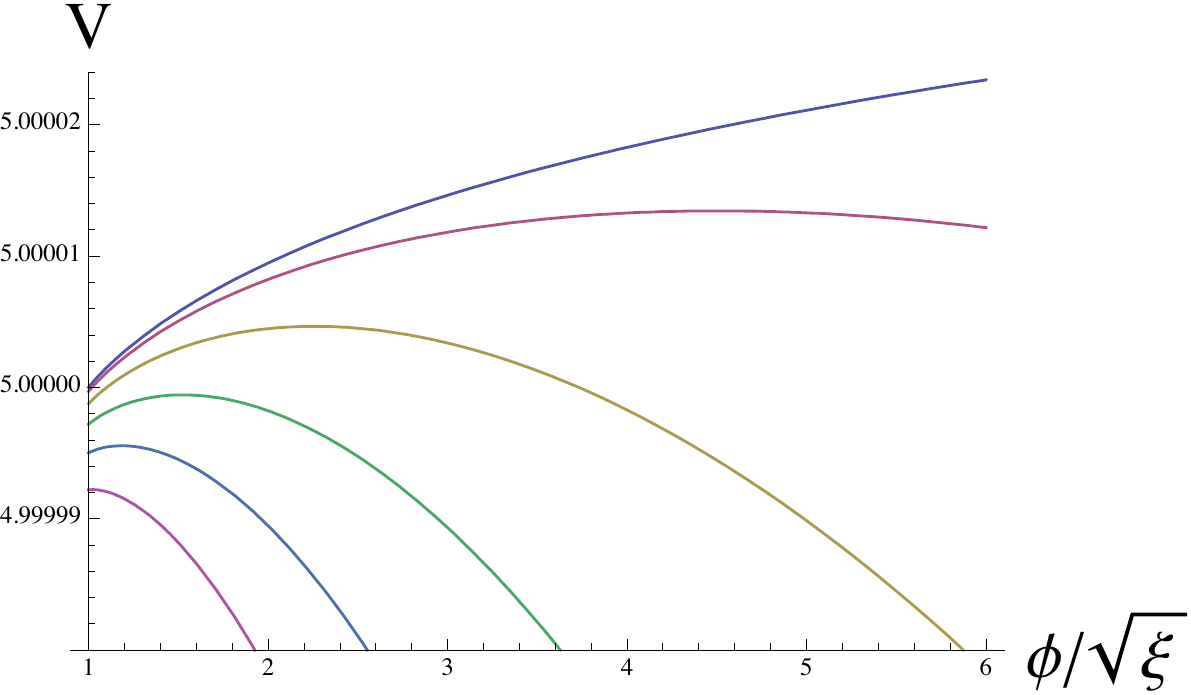}
\end{center}
\caption{The inflaton potential including the F-term mass corrections. The upper curve is
for $m=0$.}
\end{figure}

\newpage
\vskip0.2cm
\noindent{\sl 5.3.2.5  Towards experimental tests of D3/D7 inflation}
\vskip0.2cm
\noindent
Let us first discuss the model without quantum corrections, i.e. $m=0$.
As shown in  \cite{Kallosh:2003ux} to be in agreement with data one needs very small coupling constant.
Namely for larger couplings $g\geq 2\times 10^{-3}$ one gets
\begin{equation}\label{COBE4}
n_s = 1
  -{3}\left(\frac{V'}{V}\right)^2 + 2\frac{V''}{V} \approx 1-\frac{1}{N} \approx 0.98 \ .
\end{equation}
The problem here is that the tension of the cosmic strings
produced after inflation in this model is given by
\begin{equation}
G\mu  =  \frac{ \xi}{ 4} \approx   2.8 \times 10^{-6} \ .
\end{equation}
This is significantly higher than  the current bound on the cosmic string tension.

On the other hand for very small couplings, If $g << 2\times  10^{-3}$,
we can find a solution 
\begin{equation}
n_s= 0.997\ .
\end{equation}
\begin{equation}
G\mu = 7\times 10^{-7}\ , \qquad \xi = 2.8\times 10^{-6}\ .
\end{equation}
That looks
very interesting in view of some recent work on cosmic
strings and the CMB \cite{Bevis:2007gh,Pogosian:2008am}.  According
to \cite{Pogosian:2008am}, the recent puzzle of some
high $l$ excess power in CMB data from the ACBAR experiment, reported
in \cite{Reichardt:2008ay}, might possibly be considered as an evidence
for the existence of cosmic strings with tensions near the observational
bound.

Now consider the case $m^2\neq0$. The main result is that these 
quantum corrections suppress cosmic strings.
Since the analytic solution is known we may try to extract the most important properties of this model concerning the string tension and the spectral index.
We have the value of $\xi$ as follows
\begin{equation}
\xi =  \frac{2.7 \times 10^{-4}\, \sqrt{ \alpha} \,  e^{-\alpha N}}{\pi  \sqrt {1 - e^{-\alpha N}}}\ ,
\end{equation}
where $N$ can be in the range of 50 to 60. In our estimates we will use, for definiteness, $N = 60$. One now finds the following expression for the spectral index
\begin{equation}
n_s=1-\alpha \left (1+\frac{1}{ 1 - e^{-\alpha N}}\right)\ .
\end{equation}
$\alpha$ is related to mass parameter due to quantum corrections:
\begin{equation}
\alpha =\frac{4m^2}{ g^2\xi^2}\ .
\end{equation}
No one can make parametric plots for the values of the string tension versus the spectral index (see figure 19).
  \begin{figure}
\begin{center}
  \includegraphics[width=0.3\textwidth]{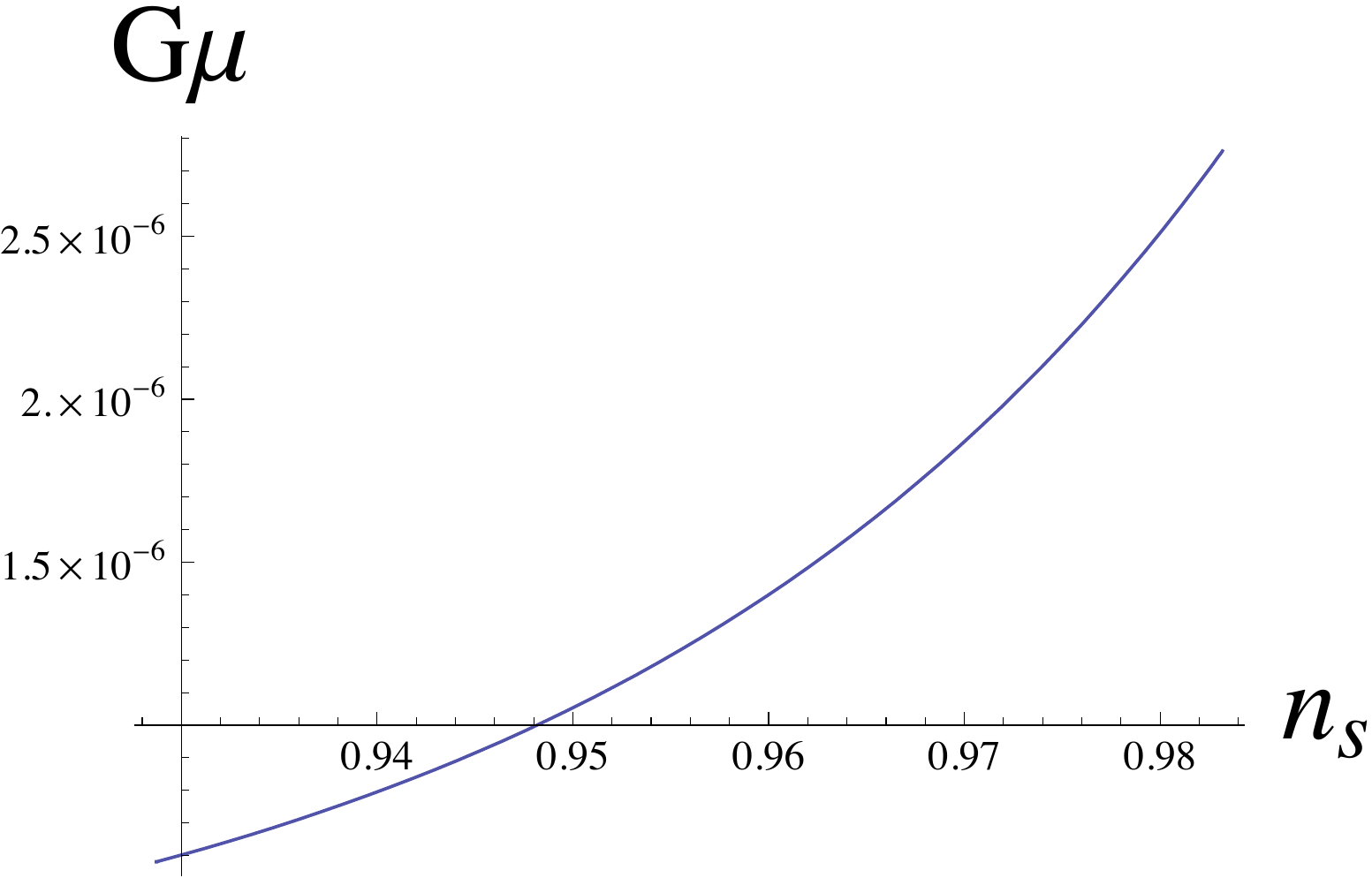}
\end{center}
\caption{The spectral index $n_s$ versus the cosmic string mass density.}
\end{figure}
We see that one can suppress the contribution of the cosmic strings by the quantum corrections.
In particular, in the limit $\alpha \rightarrow 0$, i.e.\ in the absence of stringy corrections, we get back to $G\mu={\xi\over 4}=2.8\times10^{-6}$ and $n_s=0.98$ for $N=60$. In general, a wider range of values for $\xi$ and $n_s$ is possible.
Finally note that tensor modes are highly suppressed in this model, since we are dealing with
small field inflation.

\section{Summary}
In this review article we discussed several aspects of the string landscape.
String theory possesses a huge number of
ground states upon compactification to lower dimensions, and hence it is quite evident
that the landscape exists. So the question is how to handle it. At the first
sight the string landscape leads to an apparent lack of predictive power. Verification as
well as falsification of string theory seems almost impossible. Some part of the community in high
energy physics takes this against string theory. However the actual 
situation is not as bad. Intersecting D-branes  as well as heterotic string
compactifications allow to derive models that come remarkably close to the SM,
which is not a priori granted in a fundamental theory that includes gravity and with many constraints
like string theory. Of course other low energy worlds with different gauge groups and
matter content are also possible. String or D-brane statistics provide some likelihood functions
about how often the SM appears in a given ensemble (e.g. closed string background).
It is the hope that one one finds statistical correlations telling us that every choice of
parameters or gauge groups with associated matter content is possible.
The inclusion of background fluxes allows to fix moduli parameters, which enables
us to compute couplings in a given flux background. Moreover the smallness of the
cosmological constant can be explained by statistical means. Many people object that
string statistics is nothing else that the entrance to the anthropic principle. This might be in fact true,
and the only explanation for the SM might be the anthropic principle.
In other words our universe
is only one out of many universes in the cosmic landscape of a so-called multiversum.
However, contrary to common believe, the anthropic principle is not completely meaningless
in the sense that it still allows for concrete predictions, as predicted by Weinberg \cite{Weinberg:1987dv} for
the cosmological constant or recently discussed by Bousso  \cite{Bousso:2008bu} in other instances. 

Apart these anthropic considerations, classes of string compactifications still allow
for concrete comparisons with experiment. E.g. this is true in large volume compactifications
with intermediate string scale of order $10^{11-12}$ GeV, where one can
make definite predictions on the form of the soft SUSY breaking parameters in the MSSM.
Even more dramatically, in models where the string scale is around 1 TeV
(as an alternative solution of the hierarchy problem to supersymmetry), a large class
of D-brane models lead to model independent predictions about the spectrum 
of string Regge excitations that can be possibly measured at the LHC. 
These measurements,
e.g. in collisions of two gluons into two gluons or into quarks, only see
the Regge excitations of the open strings but are insensitive against any details
of the internal compact space. Hence in this range of parameters the entire landscape
is nullified. Other processes like the scattering of two quarks into quarks provide
additional informations about the KK spectrum of the underlying geometry.
So these processes could provide an image of the internal part of the string landscape.
Of course, observing  a low string scale at the LHC requires also some big portion of
luck, and a low string scale brings many other problems like FCNC's, which have
to discussed in a model dependent way. So, we would not be surprises, if
the string scale is unfortunately high, after all.

Cosmology provides another promising avenue to make contact between string theory
and observations. Here in particular the WMAP5 data about the CMB and
future experiments like the PLANCK mission can be used to compare and to constrain
string compactifications. This is shown in the following plot, taken from the 5-years report of WMAP
\cite{Komatsu:2008hk}. The dots indicate where inflationary models are lying in the $n_s$,$r$-plane,
like chaotic inflation with quadratic or quartic inflaton potential, or the large inflaton
model of type IIA compactification on nilmanifolds. Here, the small field inflation of
type IIB on the $K3\times T^2$ orientifold lies in the low $r$ range, denoted by a 
$\clubsuit$
in this plot.

\newpage

\begin{center}
\begin{picture}(0,0)(0,0)
\put(-145,-145){
   \includegraphics[scale=.30]{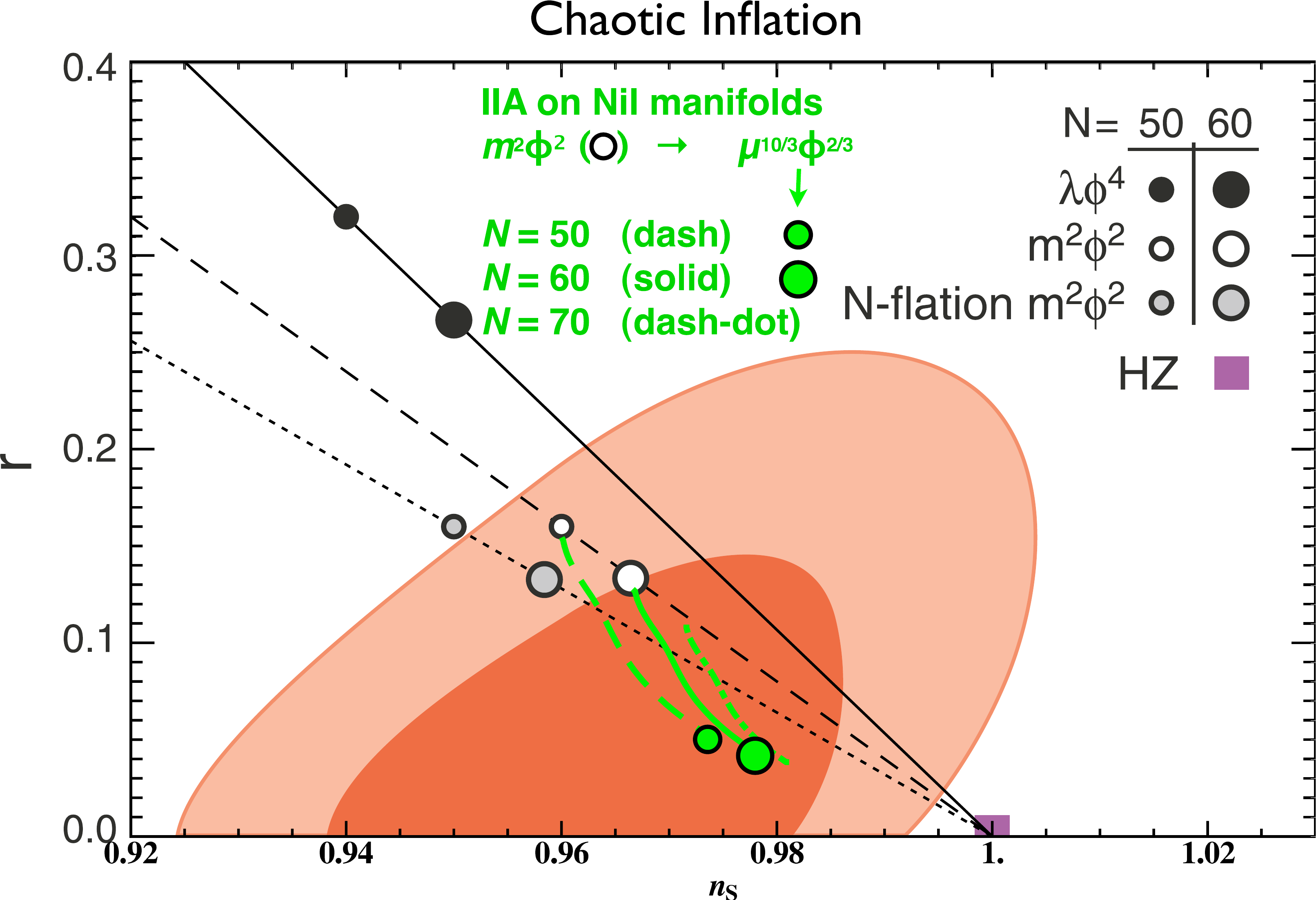}
   }
\end{picture}
\end{center}

\vskip2.6cm

\begin{eqnarray}\clubsuit\nonumber
\end{eqnarray}

\vskip2cm

\vskip0.6cm
\noindent{\bf Acknowledgments:}
I am very much indebted to my collaborators on the work presented here:
N. Akerblom, R. Blumenhagen, M. Cvetic,  F. Gmeiner, L. G\"orlich, G. Honecker, B. K\"ors, P. Mayr, E. Plauschinn,
S. Reffert, R. Richter, M. Schmidt-Sommerfeld, M. Stein, S. Stieberger and T. Weigand on 
D-brane models, D-brane
statistics and D-brane effective action.
L. Anchordoqui, H. Goldberg, S. Nawata, S. Stieberger and T. Taylor on the
string scattering at the LHC.
C. Caviezel, P. Koerber, S. K\"ors, C. Kounnas, L. Martucci, M. Petropoulos,
S. Reffert, E. Scheidegger, W. Schulgin, S. Stieberger, P. Tripathy, D. Tsimpis and M. Zagermann on flux compactifications and moduli stabilization.
G. Dvali on black hole constraints for the string landscape.
C. Caviezel, M. Haack, R. Kallosh, P. Koerber, S. K\"ors, A. Krause, A. Linde, T. Wrase and M. Zagermann on string inflation.


\end{document}